\documentclass[aip,jcp,reprint,noshowkeys,superscriptaddress]{revtex4-2}
\usepackage{graphicx,dcolumn,bm,xcolor,microtype,multirow,amscd,amsmath,amssymb,amsfonts,physics,wrapfig,bbold,siunitx,xspace}
\graphicspath{{../data/plotdata/}}
\usepackage[version=4]{mhchem}
\usepackage{comment}
\usepackage[utf8]{inputenc}
\usepackage[T1]{fontenc}

\usepackage{hyperref}
\hypersetup{
    colorlinks,
    linkcolor={red!50!black},
    citecolor={red!70!black},
    urlcolor={red!80!black}
}

\usepackage{listings}
\definecolor{codegreen}{rgb}{0.58,0.4,0.2}
\definecolor{codegray}{rgb}{0.5,0.5,0.5}
\definecolor{codepurple}{rgb}{0.25,0.35,0.55}
\definecolor{codeblue}{rgb}{0.30,0.60,0.8}
\definecolor{backcolour}{rgb}{0.98,0.98,0.98}
\definecolor{mygray}{rgb}{0.5,0.5,0.5}

\definecolor{sqred}{rgb}{0.85,0.1,0.1}
\definecolor{sqgreen}{rgb}{0.25,0.65,0.15}
\definecolor{sqorange}{rgb}{0.90,0.50,0.15}
\definecolor{sqblue}{rgb}{0.10,0.3,0.60}

\lstdefinestyle{mystyle}{
    backgroundcolor=\color{backcolour},
    commentstyle=\color{codegreen},
    keywordstyle=\color{codeblue},
    numberstyle=\tiny\color{codegray},
    stringstyle=\color{codepurple},
    basicstyle=\ttfamily\footnotesize,
    breakatwhitespace=false,
    breaklines=true,
    captionpos=b,
    keepspaces=true,
    numbers=left,
    numbersep=5pt,
    numberstyle=\ttfamily\tiny\color{mygray},
    showspaces=false,
    showstringspaces=false,
    showtabs=false,
    tabsize=2
  }

\newcolumntype{d}{D{.}{.}{-1}}

\DeclareSIUnit{\au}{a.u.}
\DeclareSIUnit{\debye}{D}

\newcommand{\ie}{\textit{i.e.}}

\usepackage[normalem]{ulem}

\newcommand{\SupMat}{\textcolor{blue}{Supporting Information}\xspace}

\newcommand{\T}[1]{#1^{\intercal}}
\DeclareMathOperator{\sgn}{sgn}

\newcommand{\br}{\boldsymbol{r}}

\newcommand{\bx}{\boldsymbol{x}}

\newcommand{\HF}{\text{HF}}

\newcommand{\etaopt}{\eta_{\text{opt}}}
\DeclareMathOperator*{\argmin}{\arg\min}

\newcommand{\hW}{\Hat{W}}
\newcommand{\hw}{\Hat{w}}
\newcommand{\hH}{\Hat{H}}
\newcommand{\hHcap}{\Hat{H}(\eta)}

\newcommand{\bO}{\boldsymbol{0}}
\newcommand{\bI}{\boldsymbol{1}}
\newcommand{\bA}{\boldsymbol{A}}
\newcommand{\bB}{\boldsymbol{B}}
\newcommand{\bC}{\boldsymbol{C}}
\newcommand{\bCt}{\Tilde{\boldsymbol{C}}}
\newcommand{\bF}{\boldsymbol{F}}
\newcommand{\bW}{\boldsymbol{W}}
\newcommand{\bS}{\boldsymbol{S}}
\newcommand{\bL}{\boldsymbol{L}}
\newcommand{\bX}{\boldsymbol{X}}
\newcommand{\bY}{\boldsymbol{Y}}

\newcommand{\bOmega}{\boldsymbol{\Omega}}

\newcommand{\cmel}[3]{\mel*{#1}{#2}{#3}_\text{c}}
\newcommand{\cbraket}[2]{\braket*{#1}{#2}_\text{c}}

\newcommand{\ii}{\mathrm{i}}

\newcommand{\LCPQ}{Laboratoire de Chimie et Physique Quantiques (UMR 5626), Universit\'e de Toulouse, CNRS, Toulouse, France}

\begin{document}	

\title{Complex Absorbing Potential Green's Function Methods for Resonances}
\author{Loris \surname{Burth}}
	\affiliation{\LCPQ}
\author{F\'abris \surname{Kossoski}}
	\affiliation{\LCPQ}
\author{Pierre-Fran\c{c}ois \surname{Loos}}
	\email{loos@irsamc.ups-tlse.fr}
	\affiliation{\LCPQ}
	
\begin{abstract}
The complex absorbing potential (CAP) formalism has been successfully employed in various wavefunction-based methods to study electronic resonance states. 
In contrast, Green's function-based methods are widely used to compute ionization potentials and electron affinities but have seen limited application to resonances.
We integrate the CAP formalism within the $GW$ approximation, enabling the description of electronic resonances in a Green's function framework. 
This approach entails a fully complex treatment of orbitals and quasiparticle energies in a non-Hermitian setting.
We validate our CAP-$GW$ implementation by applying it to the prototypical shape resonances of \ce{N2^-}, \ce{CO^-}, \ce{CO_2^-}, \ce{C2H2-}, \ce{C2H4-}, and \ce{CH2O-}.
It offers a fast and practical route to approximate both the lifetimes and positions of resonance states, achieving an accuracy comparable to that of state-of-the-art wavefunction-based methods.
\bigskip
\begin{center}
	\boxed{\includegraphics{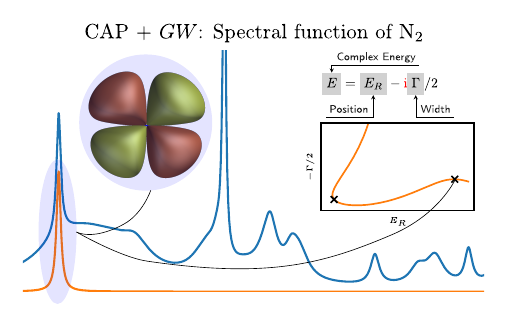}}
\end{center}
\bigskip
\end{abstract}
\maketitle
%=================================================================%
\section{Introduction}
\label{sec:introduction}
%=================================================================%
Anions can be classified as either electronically bound or unbound, based on the sign of the electron affinity (EA) of their corresponding neutral parent state.\cite{Jordan_1987,Jordan_2003,Simons_2008,Simons_2011,Jordan_2014,Simons_2000,Simons_2023}
Bound anions, characterized by positive EAs, can accommodate an additional electron in a square-integrable state.
Two sub-classes exist.
In valence-bound anions, an additional electron occupies a $\pi^*$ or $\sigma^*$ antibonding orbital.
For example, small molecules such as \ce{C2} possess a negative lowest-unoccupied molecular orbital (LUMO) energy and can thus bind an additional electron.
Valence-bound anions are even more prevalent in large $\pi$-conjugated systems, such as DNA, where low-energy electron attachment (\SIrange{0.1}{0.2}{\eV}) can weaken chemical bonds and lead to single- or double-strand breaks.\cite{Boudaiffa_2000,Martin_2004,Simons_2006,Alizadeh_2015}
In dipole-bound anions (like \ce{NC-CH^-_3} or \ce{NCH-}), the extra electron resides in an extremely diffuse orbital stabilised by the long-range electric field of a permanent dipole moment (typically $\mu \ge \SI{2.5}{\debye}$). \cite{Jordan_2003,Simons_2008} 
Here, the electron remains far from the molecular core, and the binding energy is very small, often between a few \si{\milli\eV} to a few hundred \si{\milli\eV}.
Such dipole-bound states are ubiquitous in water clusters, where the cumulative dipole moment is sufficient to weakly bind an electron. \cite{Hammer_2004}

In contrast to bound anions, unbound anions possess negative EAs.\cite{Simons_2023,Clarke_2024} 
These anions are therefore temporary, corresponding to metastable electronic resonances, that is, discrete states embedded within the continuum that decay via electron autodetachment.
Unlike bound states, which have real-valued energies, resonances are characterized by a complex (Gamow-Siegert) energy \cite{Gamow_1928,Siegert_1939}
\begin{equation}
	E = E_R - \ii \Gamma/2
	\label{eq:complex_energy}
\end{equation}
where $E_R$ denotes the resonance position and $\Gamma$ is the resonance width, which is related to the lifetime against autoionization by $\hbar/\Gamma$. \cite{Klaiman_2012}
In the inner molecular region, these so-called Gamow-Siegert states exhibit a large amplitude similar to that of bound states. 
However, in the asymptotic region, their wavefunctions oscillate like those of unbound states. 
The complex energy leads to an exponential decay in time of the state's amplitude, reflecting its metastable character.
As a result, these states are not square-integrable, and standard Hermitian methods are inapplicable (see below).

Resonances are typically classified into two families. 
Shape resonances involve a one-electron process in which the excess electron is temporarily trapped in an antibonding orbital. 
Typical examples are the $\Pi$-shape resonances of \ce{N2-} and \ce{CO-}, or electronically excited states of closed-shell anions. \cite{Zuev_2011,Bravaya_2013,Kunitsa_2015}
Feshbach resonances, \cite{Feshbach_1958,Feshbach_1962,Domcke_1991,Averbukh_2005,Kolorenc_2008,Kolorenc_2011} in contrast, occur when electron attachment is accompanied by an internal electronic excitation of the molecule, resulting in a two-electron transition for detachment. 
A well-known example is the doubly excited $^1\Sigma_g^+$ state of \ce{H2}. \cite{Yabushita_1985,Sajeev_2009}
This added complexity confers longer lifetimes, often extending to the microsecond range, and gives rise to much narrower spectral features.
In systems such as the para-benzoquinone radical anion, both shape and Feshbach resonances manifest simultaneously. \cite{Schiedt_1999,Kunitsa_2015,Kunitsa_2016,Kunitsa_2017,Loupas_2017,daCosta_2018}

Temporary anions can form either through direct electron attachment to a neutral molecule or via photoexcitation of a bound anion.
Their lifetimes typically span tens of femtoseconds to a few picoseconds, and they manifest spectroscopically as peaks of varying widths.
These transient species play vital roles in a range of chemical and physical processes. 
They are implicated in DNA damage induced by ionizing radiation, \cite{Boudaiffa_2000,Martin_2004,Simons_2006,Alizadeh_2015} modulate the bioactivity of certain radiosensitizing agents, \cite{Sedmidubska_2024} contribute to the chemistry of the interstellar medium, \cite{Wu_2024} and underpin various nanofabrication technologies. \cite{Arumainayagam_2010,Thorman_2015}
Beyond temporary anions, additional types of resonances include multiply charged anions, core-excited and core-ionized states, superexcited states, and Stark resonances formed when molecules are exposed to strong electric fields. \cite{Jagau_2017,Jagau_2022}

Due to their dual character --- exhibiting both bound and continuum-like behavior --- the wavefunctions of temporary anions are not square integrable, unlike those of true bound states. 
This non-normalizability necessitates specialized (non-Hermitian) theoretical treatments. 
One of the most widely used approaches for studying shape resonances is the complex absorbing potential (CAP) method. \cite{Jolicard_1985,Riss_1993,Sommerfeld_1998,Ghosh_2012,Zuev_2014,Jagau_2014a,Jagau_2014b,Sommerfeld_2015,Gyamfi_2024}
In this framework, a one-electron complex-valued potential is added to the bare Hamiltonian to effectively ``absorb'' the oscillatory tail of the resonance wavefunction, rendering it square integrable. 
This transformation allows resonances to be treated similarly to bound states, enabling the calculation of their positions and widths using adapted quantum chemistry techniques.

As with other non-Hermitian approaches, such as complex scaling \cite{Balslev_1971,Moiseyev_1998} or the use of complex basis functions, \cite{McCurdy_1978,White_2015} the CAP-modified Hamiltonian becomes complex-valued and non-Hermitian. \cite{Moiseyev_2011,Jagau_2017,Jagau_2022}
Consequently, resonances emerge directly as eigenstates of this modified Hamiltonian, each characterized by a complex energy whose imaginary part encodes the resonance width.

The CAP formalism must be coupled with a specific electronic structure method. It has been successfully combined with a broad range of wavefunction- and density-based approaches. 
These include selected configuration interaction (SCI), \cite{Damour_2024} equation-of-motion electron-attached coupled-cluster with singles and doubles (EOM-EA-CCSD), \cite{Ghosh_2012,Ghosh_2013,Zuev_2014} Fock-space multireference coupled-cluster theory, \cite{Sajeev_2005a,Sajeev_2005b} coupled-cluster with perturbative triples, \cite{Jagau_2018,Jana_2021} symmetry-adapted-cluster configuration interaction, \cite{Ehara_2012} extended multiconfigurational quasidegenerate second-order perturbation theory (XMCQDPT2), \cite{Kunitsa_2017,Phung_2020} multireference configuration interaction (MRCI), \cite{Sommerfeld_1998,Sommerfeld_2001} and density-functional theory (DFT). \cite{Zhou_2012}

However, studies involving Green's function methods in this context remain rather limited. 
A notable exception is the work of Cederbaum and coworkers, who explored the combination of CAP with the second-order algebraic diagrammatic construction (ADC). \cite{Santra_2002,Feuerbacher_2003,Feuerbacher_2005} 
More recently, the third-order ADC approach was applied to paradigmatic resonances \cite{Belogolova_2021} thanks to the development of the intermediate state representation (ISR) formalism, which allows for the decoupling of the ionization potential (IP) and EA sectors via the so-called non-Dyson framework. \cite{Schirmer_1998,Schirmer_2018} 

Within the broad family of Green's function methods, \cite{MartinBook,CsanakBook,FetterBook} the $GW$ approximation \cite{Hedin_1965,Golze_2019,Marie_2024a} has, to the best of our knowledge, been applied to unbound anions only once. \cite{Falcetta_2014}
In that study, the authors combined $GW$ with the stabilization method by scaling the exponents of the most diffuse basis functions, then analytically continued the computed energies into the complex plane. 
This general framework can be applied to any electronic structure method.

This is somewhat surprising, given that $GW$ is known to perform very well for valence \cite{vanSetten_2015,Caruso_2016,Krause_2017,Lewis_2019,Bruneval_2021,Monino_2023,Marie_2024b} and core \cite{vanSetten_2018,Golze_2018,Golze_2020,Mejia-Rodriguez_2021,Li_2022,Mukatayev_2023,Panades-Barrueta_2023} IPs in molecules. 
In fact, the accuracy of $GW$ is such that designing effective beyond-$GW$ schemes that outperform it at a reasonable computational cost remains a significant challenge. \cite{Baym_1961,Baym_1962,DeDominicis_1964a,DeDominicis_1964b,Bickers_1989a,Bickers_1989b,Bickers_1991,Hedin_1999,Bickers_2004,Shirley_1996,DelSol_1994,Schindlmayr_1998,Morris_2007,Shishkin_2007b,Romaniello_2009a,Romaniello_2012,Gruneis_2014,Hung_2017,Maggio_2017b,Mejuto-Zaera_2022,Wen_2024,Bruneval_2024,Forster_2024,Forster_2025}
Importantly, algorithmic improvements have been designed to lower the computational scaling of $GW$, which can thus be applied to very large systems. \cite{Neuhauser_2013,Neuhauser_2014,Kaltak_2014,Govoni_2015,Vlcek_2017,Wilhelm_2018,Golze_2018,Duchemin_2019,DelBen_2019,Forster_2020,Duchemin_2020,Kaltak_2020,Forster_2021,Duchemin_2021,Wilhelm_2021,Forster_2022,Yu_2022,Panades-Barrueta_2023,Tolle_2024}
However, its performance for anions, particularly unbound ones, is much less well-documented. 
Although some benchmark studies exist for bound anions, \cite{vanSetten_2015,Caruso_2016,Krause_2017,Gallandi_2016,Richard_2016,Knight_2016,Dolgounitcheva_2016} we are not aware of any investigations addressing unbound anions using the standard non-Hermitian frameworks, such as CAP, complex scaling, or complex basis functions.

The goal of the present study is to introduce the CAP-$GW$ approach and evaluate its performance in describing shape resonances.
To this end, we investigate different levels of $GW$ self-consistency:
(i) the widely used one-shot scheme, known as $G_0W_0$;\cite{Strinati_1980,Hybertsen_1985a,Godby_1988,Linden_1988,Northrup_1991,Blase_1994,Rohlfing_1995}
(ii) the eigenvalue self-consistent $GW$ (ev$GW$) method, in which self-consistency is enforced on the quasiparticle energies;\cite{Hybertsen_1986,Shishkin_2007a,Blase_2011a,Faber_2011,Rangel_2016}
and (iii) the quasiparticle self-consistent $GW$ (qs$GW$) scheme, which extends self-consistency to both the energies and orbitals.\cite{Gui_2018,Faleev_2004,vanSchilfgaarde_2006,Kotani_2007,Ke_2011,Kaplan_2016,Forster_2021,Marie_2023}
Calculations are performed for six widely studied shape resonances in small molecules, and the resulting resonance parameters are compared against available experimental data and 
CAP-augmented state-of-the-art theoretical approaches, namely, CAP-SCI and CAP-EOM-EA-CCSD.

%=================================================================%
\section{Theory}
\label{sec:theory}
%=================================================================%
%==============================================
\subsection{Complex Absorbing Potential}
%==============================================
When dealing with resonance states, the wave function exhibits an oscillating, non-decaying tail that prevents it from being square-integrable.  
To address this, a CAP is introduced into the electronic Hamiltonian.  
Specifically, a term $- \ii \eta \hW$, with $\eta > 0$, is added to the original $N$-electron Hamiltonian $\hH$, leading to the modified Hamiltonian
\begin{equation} \label{eq:Heta}
	\hHcap = \hH - \ii \eta \hW
\end{equation}
where $\hW$ is defined as a sum of one-electron contributions:
\begin{equation}
	\hW = \sum_{k=1}^N \hw_k
\end{equation}
The one-body potential $\hw_k$ can take various forms, but the most commonly employed choice is a quadratic form:
\begin{equation}
	\hw_{\alpha_k} =
	\begin{cases}
		\qty(\abs{\alpha_k} - \alpha_0)^2 & \text{if } \abs{\alpha_k} > \alpha_0 \\
		0 & \text{otherwise}
	\end{cases}
\end{equation}
Here, $\hw_k = \hw_{x_k} + \hw_{y_k} + \hw_{z_k}$ sums over the Cartesian components $\alpha_k \in \{x_k, y_k, z_k\}$ of the $k$th electron, and $\alpha_0 \in \{x_0, y_0, z_0\}$ specifies the onset of the absorbing potential along each coordinate axis. The parameter $\eta$ has dimensions of energy per length squared, ensuring that $\eta \hW$ is dimensionally consistent with the Hamiltonian. Throughout this work, $\eta$ is given in atomic units (\si{\hartree\per\bohr\squared}). 

It is important to note that while $\hH$ is Hermitian, the introduction of the CAP makes $\hHcap$ a complex symmetric operator, satisfying $\hHcap = \T{\hHcap}$.  
As a result, the eigenvalues and eigenvectors become complex, and the left and right eigenvectors are related by simple transposition.

Due to the symmetric but non-Hermitian nature of $\hHcap$, the usual stationary principle must be adapted by replacing the standard complex inner product with the so-called \emph{c-product}, as introduced by Moiseyev. \cite{Moiseyev_1978}
This leads to the following stationary expression for the complex energy:
\begin{equation}
	E(\eta) = \cmel{\Psi}{\hHcap}{\Psi}
\label{eq:stationary_principle}
\end{equation}
where the trial wave function $\Psi$ is \emph{c-normalized} such that $\cbraket{\Psi}{\Psi} = 1$, and the c-product is defined as
\begin{equation} \label{eq:c-prod}
	\cbraket{f}{g} = \int f(\br) g(\br) \dd{\br}
\end{equation}
This definition ensures that the energy functional is stationary at eigenfunctions of $\hHcap$, consistent with its complex symmetric structure.
In principle, with a complete basis set, the resonance energy can be obtained by taking the limit $\eta \to 0^+$.  
However, in practical calculations with finite basis sets, one must determine an optimal, small but nonzero value of $\eta$, denoted $\etaopt$.  
This value balances two competing errors: the perturbation introduced by the CAP (which grows with $\eta$) and the incompleteness error of the basis set (which diminishes with $\eta$).

Riss and Meyer\cite{Riss_1993} proposed that $\etaopt$ can be identified by minimizing the so-called \emph{energy velocity}:
\begin{equation} \label{eq:eta_opt}
	\etaopt = \argmin_\eta \abs{\eta \dv{E(\eta)}{\eta}}
\end{equation}
Practically, this involves plotting the trajectory of $E(\eta)$ as a function of $\eta$ in the complex plane and locating the minimum of the energy velocity.

%==============================================
\subsection{Complex Hartree-Fock Theory}
\label{sec:cRHF}
%==============================================
As a first step, we perform a complex restricted Hartree-Fock (cRHF) calculation incorporating the CAP to obtain a mean-field reference wave function and energy for the (closed-shell) neutral system.
Therefore, we add the complex potential to the usual Fock matrix
\begin{equation}
	\bF(\eta) = \bF - \ii \eta \bW
\end{equation}
where $\bF$ is the usual Fock matrix in the (real-valued) atomic orbital basis, which contains the one-electron core Hamiltonian, as well as the two-electron Hartree and exchange contributions.
The elements of $\bW$ are the one-electron CAP contributions in the same basis.

Diagonalizing the $\eta$-dependent Fock matrix, $\bF(\eta)$, yields complex molecular orbitals and energies, represented by the matrix of right eigenvectors $\bC(\eta)$, which are normalized with respect to the c-product, such that
\begin{equation}
	\T{\bC(\eta)} \cdot \bS \cdot \bC(\eta) = \bI
\end{equation}
Here, $\bS$ is the overlap of the atomic basis functions. 
In practice, this step is performed via a modified Cholesky decomposition.\footnotemark[1]
\footnotetext[1]{
Let $\bCt(\eta)$ be the matrix of non-orthonormal eigenvectors, i.e., the results of a diagonalisation method for non-Hermitian complex matrices. 
Then, to ensure the normalization condition, we perform an adapted Cholesky decomposition to get a lower triangular matrix $\bL$ fulfilling $\T{\bCt(\eta)} \cdot \bS \cdot \bCt(\eta) = \bL \cdot \T{\bL}$.
Then, to obtain the orthonormalized eigenvectors, we transform the original coefficients matrix using the inverse transpose of the Cholesky factor $\bL$, as follows: $\bC(\eta) = \bCt(\eta) \bL^{-\intercal}$.
}
The HF equations are iteratively solved until self-consistency is reached. 
To improve the convergence, we rely on Pulay's DIIS algorithm \cite{Pulay_1980,Pulay_1982} adapted for complex algebra with the c-product.
Note that we do not apply any projection to reduce the unphysical perturbation of the occupied orbitals introduced by the CAP. \cite{Santra_1999}

%==============================================
\subsection{Complex $GW$ Approximation}
%==============================================
As a second step, we carry out a complex-valued $GW$ calculation which explicitly incorporates the CAP to access the anionic resonances starting from the neutral reference state.
In this work, we present the essential equations for understanding and implementing the complex $GW$ formalism, and we direct readers interested in further details to several dedicated reviews \cite{Aryasetiawan_1998,Onida_2002,Reining_2017,Golze_2019,Marie_2024a} and textbooks. \cite{MartinBook,CsanakBook,FetterBook,DickhoffBook}

Within the four-point formalism, \cite{Starke_2012,Maggio_2017b,Orlando_2023b,Marie_2024c,Marie_2025} the instantaneous Coulomb potential is expressed as
\begin{equation}
	v(12;1'2') = \delta(11') \frac{\delta(t_1-t_2)}{\abs{\br_{1} - \br_{2}}} \delta(22')
\end{equation}
where $\delta(11')$ represents the Dirac delta function. The indices, such as $1$, denote shorthand notation for both time coordinates ($t_1$) and the spin-space coordinates $\bx{}_1 = (\sigma_1, \br{}_1)$ for each particle.

In practical applications of the $GW$ approximation, the process begins with a reference propagator $G_0$, typically derived from a mean-field model. 
To keep things simple, we illustrate the coupled integro-differential equations for the $GW$ formalism when $G_{0} = G_{\text{HF}}$ is used as the reference propagator. 
Note that, in the present context, the HF orbitals and their corresponding energies are already complex-valued (see Sec.~\ref{sec:cRHF}).

The total self-energy can be decomposed into three components: the Hartree (H), exchange (x), and correlation (c) contributions, such that
\begin{equation}
	\Sigma(11') = \Sigma_{\text{H}}(11') + \Sigma_{\text{x}}(11') + \Sigma_{\text{c}}(11')
\end{equation}
The exchange-correlation part, denoted $\Sigma_{\text{xc}} = \Sigma_{\text{x}} + \Sigma_{\text{c}}$, is calculated as the convolution of the one-body interacting Green's function $G$ with the dynamically screened Coulomb interaction $W$:
\begin{equation}
	\Sigma_{\text{xc}}(11') = \ii \int \dd(22')  G(22')  W(12';21')
\end{equation}
Here, $W$ is determined via the irreducible polarizability $\tilde{L}$ as follows:
\begin{multline}
	W(12;1'2')
	= v(1 2^{-};1' 2') 
	\\
	- \ii \int \dd(343'4')  W(14;1'4')  \tilde{L}(3'4';3^+4)  v(2 3;2' 3')
\end{multline}
A superscript over an index, such as $1^{\pm}$, indicates an infinitesimal time shift, i.e., $t_{1^{\pm}}=t_1 \pm \tau$ ($\tau \to 0^+$).

The irreducible polarizability $\tilde{L}$ in the $GW$ approach is approximated as the product of two Green's functions:
\begin{equation}
	\tilde{L}(12;1'2') = G(1 2')  G(2 1')
\end{equation}
The Green's function itself is computed via the Dyson equation:
\begin{equation} \label{eq:G_Dyson}
	G(11') = G_{\text{HF}}(11') + \int d(22') G_{\text{HF}}(12) \Sigma_{\text{c}}(22') G(2'1')
\end{equation}
To describe quasiparticles, it is often convenient to introduce a set of spin-orbital basis functions $\{ \varphi_p \}_{1 \le p \le K}$ associated with energies $\{ \epsilon_p \}_{1 \le p \le K}$. 
In the absence of a CAP, both quantities are real-valued, reflecting the Hermitian nature of the underlying operators. 
However, when a CAP is present, the energies and the corresponding orbitals generally become complex. 

Within this spin-orbital basis, the Lehmann representation of $G$ is given by:
\begin{equation} \label{eq:G_Lehman_quasipart}
	G(\bx_{1} \bx_{1'};\omega)
	=
	\sum_i \frac{\varphi_i(\bx_{1}) \varphi_i(\bx_{1'})}{\omega - \epsilon_i - \ii \kappa} +
	\sum_a \frac{\varphi_a(\bx_{1}) \varphi_a(\bx_{1'})}{\omega - \epsilon_a + \ii \kappa}
\end{equation}
Here, the indices $a, b, \dots$ represent the $V$ virtual orbitals (states above the Fermi level), $i, j, \dots$ denote the $O$ occupied orbitals (states below the Fermi level), and $\nu$ labels single (de)excitation ($ia$).
The indices $p, q, \dots$ are used for arbitrary orbitals, and we have $K = O + V$. 
The parameter $\kappa > 0$ is a small positive broadening term that ensures the correct causal structure of the Green's function. 
It is commonly denoted as $\eta$ in the literature, but we use $\kappa$ here to avoid confusion with the CAP strength parameter [see Eq.~\eqref{eq:Heta}].

By performing Fourier transforms and projecting onto the spin-orbital basis, we can derive the matrix elements for the correlation part of the self-energy:
\begin{equation}\label{eq:self_energy}
	\left[\Sigma_{\text{c}}(\omega)\right]_{pq} =
	\sum_{i\nu} \frac{M_{pi,\nu} M_{qi,\nu}}{\omega - \epsilon_i + \Omega_\nu - \ii \kappa} +
	\sum_{a\nu} \frac{M_{pa,\nu} M_{qa,\nu}}{\omega - \epsilon_a - \Omega_\nu + \ii \kappa}
\end{equation}
The transition densities are expressed as:
\begin{equation} \label{eq:M}
	M_{pq,\nu} = \sum_{ia} \qty[ \cbraket{pa}{qi} X_{ia,\nu} + \cbraket{pi}{qa} Y_{ia,\nu} ]
\end{equation}
Here, the bracket notation represents the electron repulsion integrals:
\begin{equation}
	\cbraket{pq}{rs} = \iint \frac{\varphi_p(\bx_{1}) \varphi_q(\bx_{2}) \varphi_r(\bx_{1}) \varphi_s(\bx_{2})}{\abs{\br_{1} - \br_{2}}} \dd\bx_{1} \dd\bx_{2}
\end{equation}
The excitation energies $\Omega_\nu$, satisfying $\Re(\Omega_\nu) > 0$, and the corresponding amplitudes $X_{ia,\nu}$ and $Y_{ia,\nu}$ are obtained by solving a Casida-like random-phase approximation (RPA) \cite{Bohm_1951,Pines_1952,Bohm_1953,Nozieres_1958} eigenproblem in the basis of single excitations ($i \to a$) and deexcitations ($a \to i$):
\begin{equation}
	\mqty(\bA & \bB  \\ -\bB & -\bA) \cdot \mqty(\bX & \bY \\ \bY & \bX)
	= \mqty(\bX & \bY \\ \bY & \bX) \cdot \mqty(\bOmega &\bO\\ \bO &-\bOmega)
\label{eq:casidalike}
\end{equation}
Here, $\bX$ and $\bY$ are matrices whose columns are the eigenvectors $\bX_\nu$ and $\bY_\nu$, respectively, and $\bOmega$ is a diagonal matrix with eigenvalues $\Omega_\nu$ on the diagonal, representing the excitation energies.
The matrix elements of $\bA$ and $\bB$ are defined as
\begin{subequations}
\begin{align}
	\label{eq:A}
	A_{ia,jb} &= (\epsilon_a - \epsilon_i) \delta_{ij} \delta_{ab} + \cbraket{aj}{ib} 
	\\
	\label{eq:B}
	B_{ia,jb} &= \cbraket{ab}{ij}
\end{align}
\end{subequations}
Unlike the usually employed complex version of Eq.~\eqref{eq:casidalike}, which involves complex conjugation of the amplitudes, \cite{Holzer_2019} this formulation does not, owing to the use of the complex symmetric c-product in place of the standard Hermitian scalar product, such that $\cbraket{ij}{ab} = \cbraket{ab}{ij}$.

We impose the following normalization conditions on the solutions $\bX$ and $\bY$ similar to the real case: \footnotemark[2] 
\begin{subequations}
	\label{eq:RPA_normalization}
	\begin{align}
		\T{\bX}	\cdot \bX - \T{\bY} \cdot \bY & = \bI
		\\
		\T{\bX}	\cdot \bY - \T{\bY} \cdot \bX & = \bO
	\end{align}
\end{subequations}
\footnotetext[2]{
Practically, we ensure the normalization of the RPA eigenvectors with the same method as introduced in the HF section, i.e., by a Cholesky decomposition of
\begin{equation*}
	 \T{\begin{pmatrix}
		\bX & \bY \\
		\bY & \bX
	\end{pmatrix}}
	\cdot
	\begin{pmatrix}
		\bI & \bO \\
		\bO & -\bI
	\end{pmatrix}
	\cdot
	\begin{pmatrix}
		\bX & \bY \\
		\bY & \bX
	\end{pmatrix}
\end{equation*}
}

%==============================================
\subsection{Self-Consistent $GW$ Schemes}
%==============================================

We now shift our attention to the computation of the Dyson orbitals ${\varphi_p}$ and quasiparticle energies ${\epsilon_p}$. 
The Dyson equation \eqref{eq:G_Dyson} implies that these quantities should satisfy the following dynamical non-Hermitian equation:
\begin{equation} \label{eq:qp_eqt}
	\qty[ f + \Sigma_\text{c}\qty(\omega=\epsilon_p) ] \varphi_p = \epsilon_p \varphi_p
\end{equation}
where $f$ represents the Fock operator in the molecular orbital basis. 
However, the frequency dependence of the correlation self-energy $\Sigma_\text{c}$ introduces non-linearity and complexity to this quasiparticle equation, prompting the use of various approximations.

The $G_0W_0$ scheme, for instance, involves a single-shot iteration of Eq.~\eqref{eq:qp_eqt} by considering only the diagonal elements of the self-energy. \cite{Strinati_1980,Hybertsen_1985a,Godby_1988,Linden_1988,Northrup_1991,Blase_1994,Rohlfing_1995}
In this approach, starting from one-electron HF orbitals $\varphi_p^\HF$ with energies $\epsilon_p^\HF$, the following equation is solved:
\begin{equation} \label{eq:qp_eqt_diag}
	\epsilon_p^{\text{HF}} + \qty[ \Sigma_\text{c}(\omega = \epsilon_p) ]_{pp} = \epsilon_p
\end{equation}
By performing a root search using the Newton-Raphson algorithm starting from the mean-field energies, we aim to locate the quasiparticle solutions, i.e., the roots of the non-linear equations that carry the largest spectral weights:
\begin{equation}
	Z_p = \frac{1}{1 - \eval{\pdv{\Sigma_\text{c}(\omega)}{\omega}}_{\omega = \epsilon_p}}
\end{equation}
In other words, the so-called satellites are discarded.

The ev$GW$ method takes a more iterative approach by updating, at each step, the RPA eigenvectors $\bX_\nu$ and $\bY_\nu$ used to construct the transition densities [see Eq.~\eqref{eq:M}], the RPA neutral excitation energies $\Omega_\nu$ and quasiparticle energies $\epsilon_p$ involved in the expression of the RPA matrix elements [see Eqs.~\eqref{eq:A} and \eqref{eq:B}] and the self-energy [see Eq.~\eqref{eq:self_energy}].
This iterative refinement of the self-energy leads to progressively improved quasiparticle energies until convergence is achieved.  \cite{Hybertsen_1986,Shishkin_2007a,Blase_2011a,Faber_2011,Rangel_2016}

In the qs$GW$ approach, both orbitals and energies are updated self-consistently until convergence is reached. \cite{Gui_2018,Faleev_2004,vanSchilfgaarde_2006,Kotani_2007,Ke_2011,Kaplan_2016,Forster_2021,Marie_2023} 
This is accomplished by iteratively diagonalizing an effective Fock-like operator
\begin{equation} \label{eq:qp_qs}
	\qty( f + \tilde{\Sigma}_\text{c} ) \varphi_p = \epsilon_p \varphi_p
\end{equation}
where the correlation part of the self-energy is approximated by a static, Hermitian operator $\tilde{\Sigma}_\text{c}$ with matrix elements given by
\begin{equation}\label{eq:sigmatilde}
	[\tilde{\Sigma}_\text{c}]_{pq} = \frac{[\Sigma_\text{c}\qty(\omega = \epsilon_p)]_{pq} + [\Sigma_\text{c}\qty(\omega = \epsilon_q)]_{qp}}{2}
\end{equation}
A recent alternative to this approximation, based on a similarity renormalization group (SRG) approach, has been proposed. \cite{Marie_2023} 
We extended this method for complex-valued energies. The expressions for the self-energy can be found in Appendix \ref{app:srg}. The SRG regularizes the self-energy, in contrast to the approach of Eq.~\eqref{eq:sigmatilde}. 
Such regularization is crucial, as achieving convergence in a large basis set, required to treat resonances at the qs$GW$ level, is generally challenging.
The SRG-regularized self-energy and the form given in Eq.~\eqref{eq:sigmatilde} are closely related. The former depends on a flow parameter $s$. In the limit $s \to \infty$, both yield identical diagonal components of the self-energy, leading to very similar final results. Conversely, at $s = 0$, the method reduces to the mean-field approximation, recovering the corresponding orbital energies and orbitals. 
The same renormalization can be used for ev$GW$. 
In the following, unless otherwise stated, we use the renormalized version of the ev$GW$ and qs$GW$ methods to improve convergence.

As discussed in Sec.~\ref{sec:introduction}, resonance states are characterized by complex energies with a small negative imaginary component, reflecting their finite lifetimes. 
To identify such a state, we first perform a calculation on the neutral molecule and then examine the quasiparticle energies $\epsilon_p$ with real part above the Fermi level, corresponding to electron attachment processes.
Among the states within a given real component energy window, the resonance is identified as the one with the smallest imaginary part, indicating the longest lifetime and the most bound-state-like character. 
Its position and width are given by
\begin{align}
	E_R & = \Re(\epsilon_p)
	&
	\Gamma & = -2 \Im(\epsilon_p)
\end{align}
in accordance with Eq.~\eqref{eq:complex_energy}. 
These definitions apply both at the HF and the $GW$ levels of theory. 
At the HF level, this approximation to IPs and EAs is commonly referred to as Koopmans' theorem. \cite{SzaboBook}

As mentioned in Sec.~\ref{sec:introduction}, $GW$ can be implemented with cubic scaling, thanks to numerous innovative algorithmic developments.\cite{Neuhauser_2013,Neuhauser_2014,Kaltak_2014,Govoni_2015,Vlcek_2017,Wilhelm_2018,Golze_2018,Duchemin_2019,DelBen_2019,Forster_2020,Duchemin_2020,Kaltak_2020,Forster_2021,Duchemin_2021,Wilhelm_2021,Forster_2022,Yu_2022,Panades-Barrueta_2023,Tolle_2024} 
In particular, a cubic-scaling implementation of qs$GW$ was recently reported by F\"orster and Visscher. \cite{Forster_2021}
As a result, partially self-consistent $GW$ schemes can now be applied to molecular systems beyond the size limit of wavefunction methods like EOM-CCSD.

%==============================================
\subsection{Spectral Function}
%=============================================

The single-particle Green's function provides direct access to the spectral function 
\begin{equation}
	A_p(\omega) = \frac{1}{\pi} \abs{ \Im G_p(\omega) }
\end{equation}
of each quasiparticle state $p$, as well as to the total spectral density
\begin{equation}
	A(\omega) = \sum_p A_p(\omega)
\end{equation}
In $G_0W_0$ and ev$GW$, where the self-energy is $\omega$-dependent, the spectral function yields
\begin{equation}
		A_p(\omega) = \frac{1}{\pi}\frac{\abs{I(\omega)}}{R(\omega)^2 + I(\omega)^2}
\end{equation}
with
\begin{subequations}
	\begin{align}
		R(\omega) &= \Re{ \omega - \epsilon_p^{\text{HF}}- \qty[ \Sigma_\text{c}(\omega) ]_{pp} }\\
		I(\omega) &= \Im{ \epsilon_p^{\text{HF}} + \qty[ \Sigma_\text{c}(\omega) ]_{pp} - \sgn[\Re(\epsilon_p^{\text{HF}} - \mu)] \kappa }
	\end{align}
\end{subequations}
For the qs$GW$ with static self-energy, the spectral function is given by:
\begin{equation}
	A_p(\omega) =  \frac{1}{\pi}\frac{\abs{ \Im{ \epsilon_p - \sgn[\Re(\epsilon_p - \mu)] \kappa } } }{\Re( \omega - \epsilon_p )^2 + \Im{ \epsilon_p  - \sgn[\Re(\epsilon_p - \mu)] \kappa }^2 }
\end{equation}
In standard $GW$ calculations, a finite broadening parameter $\kappa$ is typically introduced to generate a non-zero width, \ie\ finite lifetime. 
In our CAP-based approach, however, an imaginary part to the energies is inherently introduced, effectively regularizing the Green's function and eliminating the need for artificial broadening.

%=================================================================%
\section{Computational Details}
\label{sec:comp_det}
%=================================================================%

The various flavours of CAP-$GW$ developed here have been implemented in an open-source in-house program, named \textsc{quack}. \cite{quack} 
The one-electron CAP integrals have been computed with \textsc{opencap} 1.2.7. \cite{opencap}  
For the six small molecules investigated here, the CAP onsets, as well as the bond lengths and the bond angles, are given in Table \ref{tab:params}. 
In addition, we provide, in the \SupMat, the geometry files and basis sets. 
For the SRG, we systematically used $s=500$, unless otherwise stated. 
This ensures good convergence to the result obtained without regularization. \cite{Marie_2023} 
Additionally, we employ the DIIS extrapolation using information from up to five previous iterations.
The qs$GW$ calculations are carried out with a convergence threshold of $\SI{5e-4}{\au}$ on the usual commutator norm. \cite{Veril_2018}
Under these conditions, such calculations typically converge within approximately four or five iterations. 
We calculate the derivatives needed to obtain the energy velocity [see Eq.~\eqref{eq:eta_opt}] by the standard second-order finite difference scheme using a step size of $\num{5e-5}$ for $\eta$.

The broadening parameter $\kappa$ is set to zero in all CAP-$GW$ calculations.
Both the $GW$ approximation and the CAP formalism are always applied to all occupied and virtual orbitals.
For $G_0W_0$ and ev$GW$, quasiparticle energies are computed by directly solving the non-linear, frequency-dependent quasiparticle equation, without resorting to linearization.
Our study is limited to closed-shell neutral reference systems, consistently employing a restricted formalism.
Throughout this work, HF orbitals and energies are systematically used as the starting point, a choice justified by the relatively small size of the molecular systems considered.

For all calculations in this work, we employ the aug-cc-pVTZ+3s3p3d basis set.
This corresponds to the standard aug-cc-pVTZ basis set, augmented by three additional diffuse functions of each angular momentum type (s, p, and d), centered at the geometric center of the molecule.
For each angular momentum, the exponent of the first added diffuse function is chosen as half the average of the smallest exponents of the functions centered at the unique non-hydrogen atoms.
The remaining two exponents for each angular momentum are generated by successively halving the previous one, forming a geometric progression.
The same procedure has been used before. \cite{Nestmann_1985b,Zuev_2014,Damour_2024} 
Full details of the basis sets used in this study are provided in the \SupMat.

\begin{table}
	\caption{Parameters employed for the CAP calculations: CAP onset $(x_0,y_0,z_0)$, bond lengths and angles. 
	All lengths are in bohr and all angles in degrees. The molecular orientation with respect to the CAP can be seen in the \SupMat.}
	\label{tab:params}
	\begin{ruledtabular}
		\begin{tabular}{lccc}
			System	&$(x_0,y_0,z_0)$ &Bond length	&Bond angle\\
			\hline
			\ce{N2-}	&(2.76, 2.76, 4.88)	&$R_{\ce{NN}}=2.0740$&--\\
			\ce{CO-}	&(2.76, 2.76, 4.97)	&$R_{\ce{CO}}=2.1316$&--\\
			\ce{C2H2-}	&(3.20, 3.20, 6.35)	&$R_{\ce{CC}}=2.2733$&$\angle_{\ce{CCH}} = 180.0$\\
			&&$R_{\ce{CH}}=2.0088$\\
			\ce{C2H4-}	&(7.1, 4.65, 3.40)	&$R_{\ce{CC}}=2.5303$&$\angle_{\ce{CCH}}=121.2$\\
			&&$R_{\ce{CH}}=2.0522$\\
			\ce{CH2O-}	&(3.85, 2.95, 6.10)	&$R_{\ce{CO}}=2.2771$&$\angle_{\ce{HCO}}=121.9$\\
		    &&$R_{\ce{CH}}=2.0995$\\
			\ce{CO2-}	&(3.33, 3.33, 9.57)	&$R_{\ce{CO}}=2.1978$&--\\
			&&$R_{\ce{CH}}=2.0995$\\
		\end{tabular}
	\end{ruledtabular}
\end{table}

%=================================================================%
\section{Results}
\label{sec:res}

\subsection{\ce{N2-}}
\label{subsec:N2}

\subsubsection{Context and Approach}

The \ce{N2-} temporary anion, corresponding to a $^2\Pi_g$ configuration, is one of the most extensively studied resonances.
It therefore serves as the first benchmark system for our method and is discussed in greater detail to provide the necessary context for interpreting the results of systems examined subsequently. For each system and method, we provide Dyson orbitals that confirm the electronic configuration, as well as energy-velocity plots, in the \SupMat.

We begin with a brief overview of results obtained using various related theoretical approaches. 
While not intended to be comprehensive, this overview serves to place our findings in the appropriate context. 
We then present the results obtained with the methods introduced in this work.

Starting at the lowest level of theory, we present a more qualitative discussion, as these results are inherently less accurate by design. 
As described in Sec.~\ref{sec:cRHF}, we implemented a complex restricted HF method incorporating a CAP.
Koopmans' theorem provides a mean-field estimate of the EA.
This level of theory is highly approximate, even more so than other mean-field approaches that perform self-consistent calculations for both the neutral and anionic species.
The energy difference in the latter yields the EA while accounting for state-specific orbital relaxation, commonly known as the $\Delta$SCF method.

$\Delta$SCF calculations targeting the resonance require a state-specific HF formulation, which is feasible but beyond the scope of this work.
Comparing our Koopmans-level results with literature $\Delta$SCF data,\cite{Zhou_2012} both methods overestimate the resonance position by roughly $\SI{1}{\eV}$, while we overestimate the width by a factor of about two.
In contrast, $\Delta$SCF yields more reasonable widths.
Nevertheless, our method successfully identifies the resonance and provides rough estimates of the position and width.
Thus, it serves as a suitable mean-field starting point for subsequent $GW$ corrections, rather than final predictive results.

At a comparable level of correlation treatment, $GW$ methods have been combined with the stabilization technique, which estimates the position and width by analytic continuation. \cite{Falcetta_2014}
In that study, $G_0W_0$ underestimates the width obtained by EOM-CCSD with analytic continuation by more than a factor of four.
In contrast, a self-consistent $GW$ approach that includes orbital relaxation yields lifetimes of the correct order of magnitude. \cite{Falcetta_2014}

Furthermore, Green's function methods combined with CAP have been applied in previous studies using different basis sets than those employed in our work, making a direct comparison difficult.
Nevertheless, these approaches have been shown to yield reasonable estimates for both the resonance position and width. For instance, reported values including basis set uncertainty are $\SI{2.58\pm0.13}{\eV}$ for the position and $\SI{0.55\pm0.14}{\eV}$ for the width. \cite{Feuerbacher_2003} 
Despite the methodological and basis set differences, our qs$GW$ results for the resonance position largely fall within the reported error margins (see Table \ref{tab:res_opt_N2}). However, the resonance width at the second minimum (see below) deviates by $\SI{0.17}{\eV}$.

We now turn to the highest level of theory combined with a CAP: extrapolated full configuration interaction (exFCI), which treats all electronic correlations. \cite{Damour_2024}
Due to its high computational cost, reference data are limited.
For \ce{N2-} and \ce{CO-}, however, exFCI results\cite{Damour_2024} are available at the optimal CAP strength $\eta$ obtained from EOM-EA-CCSD calculations.\cite{Zuev_2014}
Comparing EOM-EA-CCSD to exFCI reveals smaller deviations than those observed between our $GW$ results and EOM-EA-CCSD. We remark that the deviations are in the expected order of magnitude of CCSD to FCI for neutral excitations (see Table \ref{tab:res_opt_N2}). \cite{Veril_2021}
Since exFCI data is available only for selected systems and lacks $\eta$-optimization, while EOM-EA-CCSD values exist for all systems studied in this work, we adopt CAP-EOM-EA-CCSD as our primary theoretical reference.

For \ce{N2-}, resonance positions and widths derived by CAP-EOM-EA-CCSD have been reported at significantly different CAP strengths: One at $\eta=0.0015$\cite{Zuev_2014} and one at $\eta=0.007$. \cite{Gayvert_2022} Those values correspond to two different local minima in the energy velocity. We observe a qualitatively similar behavior and trajectory using the qs$GW$ method (see Fig.~\ref{fig:traj_N2_1}).  Similar multiple-minimum structures also appear in other systems studied. This phenomenon appears to reflect a broader issue within the CAP framework, which, to our knowledge, remains unresolved. While solving this issue is beyond the scope of the present work, our goal is to assess the performance of CAP-$GW$ approaches relative to higher-level theories. For \ce{N2-}, we analyze both reported minima. For the remaining systems, multiple minima are mentioned when encountered, but we report only the results that correspond to the CAP-EOM-EA-CCSD reference.

\subsubsection{Trajectories, Positions and Lifetimes}

\begin{figure}
	\centering
	\includegraphics{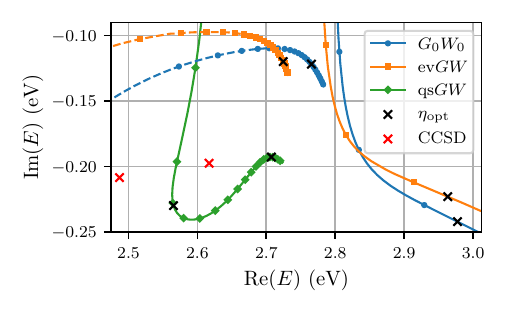}
	\caption{Trajectories of \ce{N2-} computed at different levels of self-consistency. For $G_0W_0$ and ev$GW$, the solid line denotes the trajectory associated with the first energy-velocity minimum, while the dashed line corresponds to the second minimum. Black crosses indicate the positions of the energy-velocity minima, representing the final results for the resonance state. Red crosses show the CAP-EOM-EA-CCSD reference values (see Table \ref{tab:res_opt_N2}). The maximum $\eta$ value shown is $\SI{0.0155}{}$ for $G_0W_0$ and ev$GW$, and $\SI{0.01}{}$ for qs$GW$.
	}
	\label{fig:traj_N2_1}
\end{figure}

Within the qs$GW$ framework, a single state can be followed continuously along the $\eta$-trajectory to identify both minima.
In contrast, for methods based on HF orbitals, such as $G_0W_0$ and ev$GW$, two distinct trajectories must be tracked (see Fig.~\ref{fig:traj_N2_1}).
These apparently describe the same resonance state, but at different values of $\eta$.
Moreover, for these HF-based methods, no local minimum is found in the vicinity of the first reported minimum, except at the Koopmans level of theory, at $\eta = 0.0017$.
We therefore adopt this value to estimate the resonance parameters in $G_0W_0$ and ev$GW$.

Also, in ev$GW$, the $\eta$-trajectory was not smooth enough to reliably identify the second reported minimum of the energy velocity using finite differences.
We thus performed additional calculations with $s = 100$ to identify the optimal $\eta$ value.

Our results are presented in Table \ref{tab:res_opt_N2}. We find that all discussed $GW$ methods overestimate the resonance position compared to reference theoretical methods and experiment. While the $G_0W_0$ and ev$GW$ methods deviate by approximately \SI{0.5}{\eV} from the CAP-EOM-EA-CCSD reference for the first minimum, they are only off by about \SI{0.1}{\eV} for the second minimum. On the other hand, these methods yield a more accurate lifetime for the first minimum than for the second. The qs$GW$ method, by contrast, closely reproduces the reference values, with differences of about \SI{0.1}{\eV} for the resonance position and \SI{0.05}{\eV} for the width $\Gamma$. Compared to experimental data, all $GW$ methods as well as the CAP-EOM-EA-CCSD approach systematically overestimate the resonance position. Furthermore, ev$GW$ and $G_0W_0$ underestimate the resonance width by more than \SI{0.17}{\eV} compared to both theoretical and experimental values.

These observations strongly suggest that orbital relaxation effects, accounted for only in the qs$GW$ method, are essential not only for reliably determining $\etaopt$ but also for accurately describing the underlying resonance state.

%%% FIG 1 %%% 

%%% TAB 1 %%% 
\begin{table}
	\caption{Optimal CAP strengths, resonance positions, and widths of \ce{N2-} obtained using different methods. The $GW$ results are from this work; other data are taken from the cited references. Optimal $\eta$ values marked with $*$ are derived by another method as described in the main text.}
	\label{tab:res_opt_N2}
	\begin{ruledtabular}
		\begin{tabular}{llcc}
            Method & $\etaopt$ (\si{\au}) & $E_\text{R}$ (\si{\eV}) & $\Gamma$ (\si{\eV}) \\
            \hline
            $G_0W_0$ & 0.00170* & 2.977 & 0.484 \\
            $G_0W_0$ & 0.01150 & 2.765 & 0.244 \\
            ev$GW$ & 0.00170* & 2.963 & 0.446 \\
            ev$GW$ & 0.01325 & 2.725 & 0.240 \\
            qs$GW$ & 0.00160 & 2.565 & 0.460 \\
            qs$GW$ & 0.00780 & 2.707 & 0.386 \\	
            \\
			CAP-EOM-EA-CCSD\cite{Zuev_2014} 	&0.0015			&2.487		&0.417\\
			CAP-exFCI\cite{Damour_2024}			&0.0015*		&2.449 		&0.391\\
			CAP-EOM-EA-CCSD\cite{Gayvert_2022}	&0.0070			&2.617		&0.395\\
			%\titou{pCAP}-EOM-EA-CCSD\cite{Gayvert_2022} &0.0070			&2.619		&0.385\\
			Experiment\cite{Berman_1983}		&				&2.316		&0.414	
		\end{tabular}
	\end{ruledtabular}
\end{table}

\subsubsection{Spectral Function}

\begin{figure}
	\includegraphics{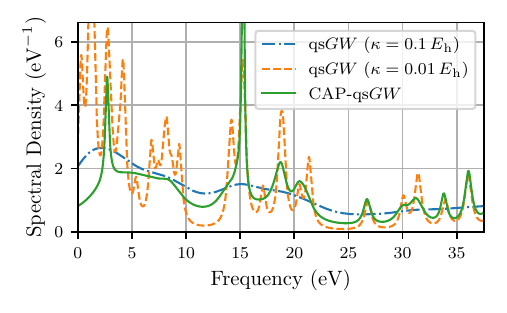}
	\caption{Comparison of the spectral function of \ce{N2} derived by CAP-augmented qs$GW$ (CAP-qs$GW$) with $\etaopt = 0.0078$ (green curve) and qs$GW$ without CAP but with broadening parameter $\kappa = \SI{0.1}{\hartree}$ (blue curve) and $\kappa = \SI{0.01}{\hartree}$ (orange curve).}
	\label{fig:cap_vs_eta}
\end{figure}

Figure \ref{fig:cap_vs_eta} compares spectral functions obtained from the qs$GW$ method with and without the inclusion of a CAP. 
We focus here on the CAP-augmented qs$GW$ (CAP-qs$GW$) result evaluated at the optimal CAP strength, $\etaopt = 0.0078$, which corresponds to the lower of the two energy velocity minima discussed earlier. For comparison, we also include spectra computed using fixed artificial broadening parameters, $\kappa = \SI{0.1}{\hartree}$ and $\kappa = \SI{0.01}{\hartree}$.

While these spectral functions offer valuable qualitative insights, their physical interpretation is limited. First, the chosen Gaussian basis set is clearly too small to accurately describe the continuum, whose states are asymptotically plane-wave-like. As the basis set is enlarged by adding more diffuse functions, one would expect an increasingly dense set of discrete states approximating the continuum over a broad energy range. Second, the inclusion of the CAP discretizes the continuum by turning it into a set of complex-valued eigenvalues, with the goal of isolating resonance features. However, the CAP strength $\eta$ was optimized only for a specific resonance state, so the resulting spectrum captures primarily this state and not the entire unbound spectrum. 

In the absence of a CAP, the broadening parameter $\kappa$ uniformly widens all peaks in the spectral function, regardless of their real part $E_R$. As a result, increasing $\kappa$ leads to overly broad and overlapping peaks, obscuring the identification of individual resonances. Reducing $\kappa$ sharpens the peaks. There is no interpretation of the peak width since $\kappa$ is an artificial parameter and not physically meaningful. All peaks are broadened with the same $\kappa$.

In contrast, the CAP-qs$GW$ approach yields a spectral function where each resonance is broadened according to its actual lifetime. The CAP introduces an energy-dependent, non-Hermitian perturbation that allows for a physically meaningful extraction of both the position $E_R$ and the resonance width $\Gamma$. Importantly, the full width at half maximum of a given peak directly corresponds to $\Gamma$, the inverse lifetime of the resonance state. Hence, the inclusion of the CAP enables an identification of the metastable state and provides access to its intrinsic lifetime, a quantity inaccessible through the use of a constant broadening parameter alone.

Furthermore, it can be observed that the inclusion of the CAP leads to only a slight shift in the peak positions compared to the artificially broadened spectra. This is particularly evident for the resonance state for which the CAP strength was optimized: the first peak in the CAP-qs$GW$ spectral function. Its energy remains nearly unchanged, indicating that the CAP does not significantly alter the resonance position, but rather unveils its physical width in a controlled and interpretable manner.

\subsection{\ce{CO-}}

\begin{figure}
	\centering
	\includegraphics{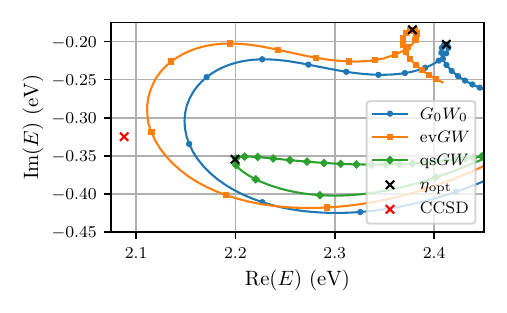}
	\caption{Trajectories of \ce{CO-} computed at different levels of self-consistency. Black crosses indicate the positions of the energy-velocity minima, representing the final results for the resonance state. Red crosses show the CAP-EOM-EA-CCSD reference values (see Table \ref{tab:res_opt_CO}). The maximal $\eta$ value shown is $\SI{0.014}{}$ for all methods.
	}
	\label{fig:traj_CO}
\end{figure}

The resonance state of \ce{CO-} is a $^2\Pi$ shape resonance. The trajectories obtained from the different $GW$ methods are shown in Fig.~\ref{fig:traj_CO}. In contrast to \ce{N2-}, clear minima in the energy velocity are identifiable at all three levels of self-consistency, as well as at the Koopmans level. However, the trajectories of ev$GW$ and $G_0W_0$ differ markedly in shape from that of qs$GW$, and their minima appear at different locations along the trajectories. As a result, the optimal CAP strengths $\eta_\text{opt}$ for ev$GW$ and $G_0W_0$ are more than twice as large as that of qs$GW$.

The ev$GW$ and $G_0W_0$ trajectories exhibit broad turns followed by loops in the upper right corner of the frame, corresponding to the velocity minima. In contrast, the qs$GW$ trajectory features a much sharper turn, closely resembling the shape obtained from CAP-EOM-EA-CCSD. \cite{Gayvert_2022} This qualitative agreement suggests that orbital relaxation plays a significant role in accurately describing the resonance state of \ce{CO-}, and that qs$GW$, through its self-consistency in both eigenvalues and orbitals, is better suited to capture orbital relaxation. We also find another local minimum of the energy velocity at the qs$GW$ level for $\eta=\SI{0.0006}{}$, which is less pronounced and yields a higher energy velocity than the one previously reported. Therefore, we assume it arises from the limitations of the finite basis set.

The numerical resonance parameters are summarized in Table \ref{tab:res_opt_CO}. As shown there, $G_0W_0$ and ev$GW$ yield very similar results, differing by only about \SI{0.04}{\eV} in both the resonance position and width. Both methods, however, underestimate the width predicted by CAP-FCI and CAP-EOM-EA-CCSD by more than \SI{0.2}{\eV}, and deviate even more from experimental data. Additionally, they overestimate the resonance position by more than \SI{0.25}{\eV} compared to the theoretical references.
In contrast, qs$GW$ overestimates the theoretical reference position by less than \SI{0.15}{\eV} and width $\Gamma$ by less than \SI{0.1}{\eV}, representing a noticeable improvement. 

We compare our results to estimates of the \emph{vertical} resonance position, that is, at the bond length of the neutral molecule, 
and not the adiabatic value of \SI{1.50}{\eV} reported in Refs.~\onlinecite{Ehrhardt_1968,Zubek_1977,Zubek_1979}.
From the parameters provided in these works, we estimated an experimental vertical resonance position of \SI{1.93}{\eV},
which aligns well with the cross section maximum of more recent experiments, peaking between \SI{1.9}{\eV} and \SI{2.0}{\eV}. \cite{Buckman_1986,Szmytkowski_1996,Allan_2010}
Thus, both CAP-EOM-EA-CCSD and CAP-exFCI slightly overestimate the resonance position relative to experiment.
The discrepancy between qs$GW$ and experiment is also moderate, on the order of $\SI{0.3}{\eV}$.
Furthermore, for the imaginary part of the energy, qs$GW$ provides better agreement with experimental data than the other theoretical methods.
%%% TAB 3 %%% 
\begin{table}
	\caption{Optimal CAP strengths, resonance positions, and widths of \ce{CO-} obtained using different methods. The $GW$ results are from this work; other data are taken from the cited references.
	The optimal $\eta$ value marked with $*$ is derived by another method as described in the main text.
	}
	\label{tab:res_opt_CO}
	\begin{ruledtabular}
		\begin{tabular}{llll}
            Method & $\etaopt$ (\si{\au}) & $E_\text{R}$ (\si{\eV}) & $\Gamma$ (\si{\eV}) \\
            \hline
            $G_0W_0$ & 0.00870 & 2.412 & 0.407 \\
            ev$GW$ & 0.00875 & 2.378 & 0.369 \\
            qs$GW$ & 0.00295 & 2.200 & 0.709 \\    
            \\
			CAP-EOM-EA-CCSD\cite{Zuev_2014}		&0.0028	&2.088		&0.650\\
			CAP-exFCI\cite{Damour_2024}			&0.0028*	&2.060 		&0.611\\
			Experiment\cite{Zubek_1979,Zubek_1977,Ehrhardt_1968,Buckman_1986,Szmytkowski_1996,Allan_2010}		&&1.9--2.0		&--\\
					&&1.93\cite{Zubek_1979}			&0.75\cite{Zubek_1979}\\
%			Experiment\cite{Zubek_1979,Zubek_1977,Ehrhardt_1968}&		&1.50		&0.75\cite{Zubek_1979}\\
 												   &		&     	&0.80(3)\cite{Zubek_1977}
		\end{tabular}
	\end{ruledtabular}
\end{table}
% Calculation for 1.93 : 14*(1-np.exp(0.23*0.092e-10*2*np.pi/(4.13567e-15)*np.sqrt(1.13866e-26/(2*14*1.6022e-19))))**2 +1.5 = 1.9327472988578478

\subsection{\ce{C2H2-}}

\begin{figure}
	\centering
	\includegraphics{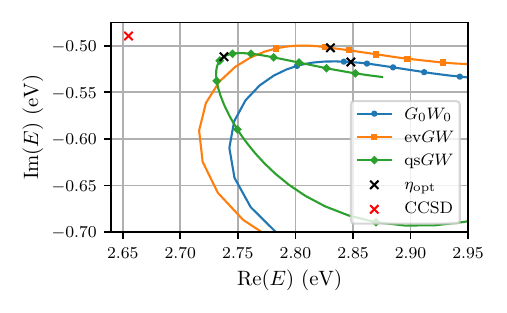}
	\caption{Trajectories of \ce{C2H2-} computed at different levels of self-consistency. Black crosses indicate the positions of the energy-velocity minima, representing the final results for the resonance state. Red crosses show the CAP-EOM-EA-CCSD reference values (see Table \ref{tab:res_opt_C2H2}). The maximal $\eta$ value shown is $\SI{0.007}{}$ for all methods.
	}
	\label{fig:traj_C2H2}
\end{figure}

The resonance state of \ce{C2H2-} is a $^2\Pi_g$ shape resonance. The CAP trajectories obtained from the three $GW$ methods are shown in Fig.~\ref{fig:traj_C2H2}. In contrast to the case of \ce{CO-}, all methods --- $G_0W_0$, ev$GW$, and qs$GW$ --- yield qualitatively similar trajectories, characterized by smoothly curved paths with clearly defined minima in the energy velocity. This similarity in shape indicates that the underlying electronic structure and its response to the CAP are treated consistently across all levels of self-consistency. Even though we find only one minimum for each method, the slope of the energy velocity decreases and then increases again around $\eta=\SI{0.001}{}$. This behavior is consistent with the pattern observed for the other molecules, where shallow local minima appear at small values of $\eta$.

The minor variations in the shape of the trajectories are reflected in the extracted resonance parameters (see Table \ref{tab:res_opt_C2H2}). The optimal CAP strengths $\eta_\text{opt}$ are nearly identical across methods, and the resonance positions differ by less than \SI{0.12}{\eV}. The computed widths $\Gamma$ are also closely aligned, with a spread of only \SI{0.03}{\eV} among the methods. 

All three $GW$ variants overestimate the resonance position relative to the CAP-EOM-EA-CCSD reference (\SI{2.655}{\eV}), with qs$GW$ performing best among them at \SI{2.738}{\eV}. In terms of resonance width, all $GW$ methods predict it slightly above the CAP-EOM-EA-CCSD value of \SI{0.979}{\eV}, yet within deviations of \SI{0.05}{\eV}. Notably, the best agreement for both position and width is again achieved with qs$GW$, reinforcing its overall robustness.

In summary, for the $^2\Pi_g$ resonance of \ce{C2H2-}, all $GW$ methods yield consistent results, both qualitatively and quantitatively. The reduced sensitivity to the level of self-consistency indicates that this system is less impacted by orbital relaxation effects than the systems discussed before.

%%% FIG 4 %%% 
%%% TAB 4 %%% 
\begin{table}
	\caption{Optimal CAP strengths, resonance positions, and widths of \ce{C2H2-} obtained using different methods. The $GW$ results are from this work; other data are taken from the cited references.}
	\label{tab:res_opt_C2H2}
	\begin{ruledtabular}
		\begin{tabular}{llcc}
            Method & $\etaopt$ (\si{\au}) & $E_\text{R}$ (\si{\eV}) & $\Gamma$ (\si{\eV}) \\
            \hline
            $G_0W_0$ & 0.00370 & 2.848 & 1.035 \\
            ev$GW$ & 0.00365 & 2.830 & 1.004 \\
            qs$GW$ & 0.00375 & 2.738 & 1.023 \\              
            \\
			CAP-EOM-EA-CCSD\cite{Zuev_2014}				&0.0036	&2.655		&0.979\\
			Experiment\cite{Kochem_1985,Jordan_1978,Dressler_1987,Andric_1988,Szmytkowski_2014}	&		&2.5--2.6		& --\\
		\end{tabular}
	\end{ruledtabular}
\end{table}

\subsection{\ce{C2H4-}}

\begin{figure}
	\centering
	\includegraphics{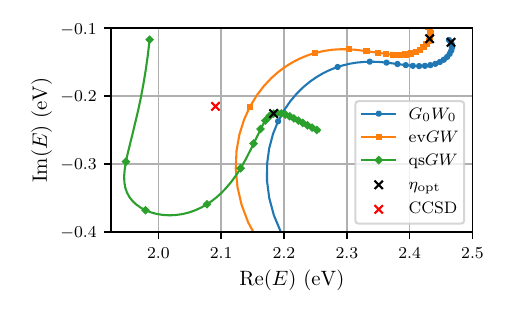}
	\caption{Trajectories of \ce{C2H4-} computed at different levels of self-consistency. Black crosses indicate the positions of the energy-velocity minima, representing the final results for the resonance state. Red crosses show the CAP-EOM-EA-CCSD reference values (see Table \ref{tab:res_opt_C2H4}). The maximal $\eta$ value shown is $\SI{0.013}{}$ for $G_0W_0$, $\SI{0.014}{}$ for ev$GW$ and $\SI{0.01}{}$ for qs$GW$.
	}
	\label{fig:traj_C2H4}
\end{figure}

The resonance state of \ce{C2H4-} corresponds to a $^2B_{2g}$ shape resonance. The CAP trajectories obtained from the three $GW$ variants are shown in Fig.~\ref{fig:traj_C2H4}. As in the case of \ce{CO-}, the $G_0W_0$ and ev$GW$ trajectories exhibit qualitatively similar behavior, featuring broad turns with identifiable minima in the energy velocity located in the upper-right region of the portrayed frame. However, the trajectory produced by qs$GW$ is markedly different in both shape and curvature, showing no apparent similarity to those of $G_0W_0$ and ev$GW$.

Despite these differences in trajectory morphology, all three methods yield well-defined minima in the energy velocity. Notably, the optimal CAP strengths $\etaopt$ differ substantially: $G_0W_0$ and ev$GW$ require relatively large values of $\eta$ ($0.01215$ and $0.01055$, respectively), while qs$GW$ yields a minimum at a significantly smaller value ($0.00475$). This again reflects how the interaction with the CAP is handled differently at different levels of self-consistency, with qs$GW$ requiring a weaker perturbation to stabilize the resonance state. Again, we find another local minimum on the qs$GW$ level at lower $\eta=\SI{0.00095}{}$.

The numerical resonance parameters are summarized in Table \ref{tab:res_opt_C2H4}. The $G_0W_0$ and ev$GW$ variants yield closely aligned results: they differ by only \SI{0.034}{\eV} in the resonance position and \SI{0.01}{\eV} in the width. However, they significantly overestimate the resonance energy and underestimate the width when compared to the high-level CAP-EOM-EA-CCSD calculation.

In contrast, qs$GW$ produces a resonance energy of \SI{2.183}{\eV} and a width of \SI{0.451}{\eV}, in excellent agreement with the CAP-EOM-EA-CCSD reference values of \SI{2.091}{\eV} and \SI{0.430}{\eV}, respectively. All theoretical approaches overestimate the resonance position relative to experiment, likely due to limitations in the basis set and the intrinsic challenges of modeling metastable anions.

%%% FIG 5 %%% 

%%% TAB 5 %%% 
\begin{table}
	\caption{Optimal CAP strengths, resonance positions, and widths of \ce{C2H4-} obtained using different methods. The $GW$ results are from this work; other data are taken from the cited references.
	}
	\label{tab:res_opt_C2H4}
	\begin{ruledtabular}
		\begin{tabular}{llcc}
            Method & $\etaopt$ (\si{\au}) & $E_\text{R}$ (\si{\eV}) & $\Gamma$ (\si{\eV}) \\
            \hline
            $G_0W_0$ & 0.01215 & 2.466 & 0.241 \\
            ev$GW$ & 0.01055 & 2.432 & 0.231 \\
            qs$GW$ & 0.00475 & 2.183 & 0.451 \\   
			\\
			CAP-EOM-EA-CCSD\cite{Zuev_2014}					&0.0046	&2.091		&0.430\\
			Experiment\cite{Sanche_1973,Walker_1978,Lunt_1994,Panajotovic_2003,Szmytkowski_2003,Allan_2008,Khakoo_2016}                                      &		&1.7--1.9	&  $\le$0.70\cite{Sanche_1973} \\
%			Experiment                                      &		&1.76\cite{Sanche_1973}	&  $\le$0.70\cite{Sanche_1973} \\
%			                                             	&		&1.8\cite{Walker_1978,Panajotovic_2003,Allan_2008,Lunt_1994}	&  -- \\
%			                                             	&		&1.9\cite{Szmytkowski_2003}	&  -- 
%			                                             	&		&1.7\cite{Khakoo_2016}	&  -- 
		\end{tabular}
	\end{ruledtabular}
\end{table}

\subsection{\ce{CH2O-}}

\begin{figure}
	\centering
	\includegraphics{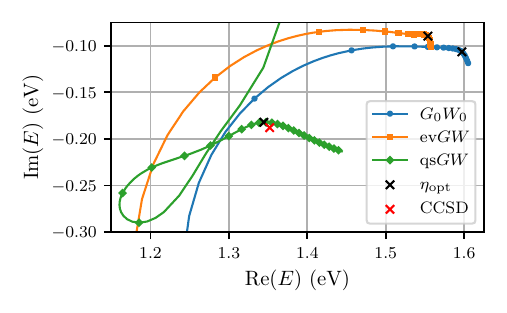}
	\caption{Trajectories of \ce{CH2O-} computed at different levels of self-consistency. Black crosses indicate the positions of the energy-velocity minima, representing the final results for the resonance state. Red crosses show the CAP-EOM-EA-CCSD reference values (see Table \ref{tab:res_opt_CH2O}). The maximal $\eta$ value shown is $\SI{0.012}{}$ for all methods.
	}
	\label{fig:traj_CH2O}
\end{figure}

The resonance state of \ce{CH2O^-} corresponds to a $^2B_1$ shape resonance. The CAP trajectories obtained from the different $GW$ variants are shown in Fig.~\ref{fig:traj_CH2O}. Similar to the case of \ce{C2H4^-}, the trajectories of $G_0W_0$ and ev$GW$ exhibit comparable shapes, both featuring smooth curves with well-defined minima in the energy velocity. In contrast, the qs$GW$ trajectory differs markedly in both geometry and curvature, reflecting the differences in orbital relaxation and self-consistency.

Despite the differences in trajectory shape, all three methods yield identifiable minima in the energy velocity, allowing for the determination of an optimal CAP strength $\etaopt$. The values of $\etaopt$ for $G_0W_0$ and ev$GW$ are nearly identical ($0.00915$ and $0.00900$, respectively), while qs$GW$ again requires a substantially smaller value ($0.00450$). This consistent trend across all systems studied suggests that fully self-consistent $GW$ methods (qs$GW$) can stabilize resonances with weaker perturbations, owing to their more accurate treatment of orbital response in the presence of the CAP. For this molecule, we find an additional local minimum at all levels of self-consistency: at $\eta = \SI{0.00065}{}$ for qs$GW$, and at $\eta = \SI{0.0013}{}$ for the other two methods.

The corresponding resonance parameters are summarized in Table \ref{tab:res_opt_CH2O}. As in previous systems, $G_0W_0$ and ev$GW$ produce similar results, predicting a resonance energy of \SI{1.597}{\eV} and \SI{1.554}{\eV}, respectively, and widths of \SI{0.213}{\eV} and \SI{0.179}{\eV}. However, these methods underestimate the resonance width compared to both high-level theory and experiment.

The qs$GW$ method yields a lower resonance position of \SI{1.345}{\eV} and a broader width of \SI{0.364}{\eV}, in excellent agreement with the CAP-EOM-EA-CCSD values of \SI{1.352}{\eV} and \SI{0.376}{\eV}, respectively. Although all $GW$ methods overestimate the experimental resonance position, the deviation is significantly smaller for qs$GW$, suggesting a more realistic description of the resonance state. To the best of our knowledge, no experimental estimate for the resonance width of this system is available.

%%% FIG 6 %%% 

%%% TAB 6 %%% 
\begin{table}
	\caption{Optimal CAP strengths, resonance positions, and widths of \ce{CH2O-} obtained using different methods. The $GW$ results are from this work; other data are taken from the cited references.}
	\label{tab:res_opt_CH2O}
	\begin{ruledtabular}
		\begin{tabular}{llcc}
            Method & $\etaopt$ (\si{\au}) & $E_\text{R}$ (\si{\eV}) & $\Gamma$ (\si{\eV}) \\
            \hline
            $G_0W_0$ & 0.00915 & 1.597 & 0.213 \\
            ev$GW$ & 0.00900 & 1.554 & 0.179 \\
            qs$GW$ & 0.00450 & 1.345 & 0.364 \\        
			\\
			CAP-EOM-EA-CCSD\cite{Zuev_2014}							&0.0100	&1.352		&0.376\\
			Experiment\cite{Burrow_1976,VanVeen_1976,Benoit_1986}	&		&0.86--0.87		&--
		\end{tabular}
	\end{ruledtabular}
\end{table}

\subsection{\ce{CO2-}}

\begin{figure}
	\centering
	\includegraphics{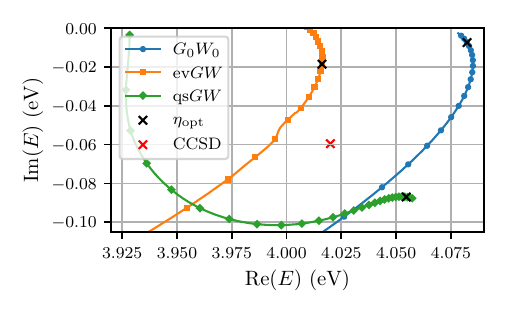}
	\caption{Trajectories of \ce{CO2-} computed at different levels of self-consistency. Black crosses indicate the positions of the energy-velocity minima, representing the final results for the resonance state. Red crosses show the CAP-EOM-EA-CCSD reference values (see Table \ref{tab:res_opt_CO2}). The maximal $\eta$ value shown is $\SI{0.014}{}$ for $G_0W_0$, $\SI{0.0133}{}$ for ev$GW$ and $\SI{0.014}{}$ for qs$GW$.
	}
	\label{fig:traj_CO2}
\end{figure}

The resonance state of \ce{CO2-} corresponds to a $^2\Pi_u$ shape resonance. The trajectories obtained from various $GW$ methods are shown in Fig.~\ref{fig:traj_CO2}. In contrast to \ce{CO-}, the shape and behavior of the trajectories differ markedly across methods. While clear local minima in the energy velocity are observed for qs$GW$ and $G_0W_0$, the ev$GW$ trajectory appears significantly less structured within the examined range. 
To address this, we performed additional calculations with $s = 100$ to determine an optimal value of $\eta$ for ev$GW$. Subsequently, a calculation with $s = 500$ was carried out using this optimal value. 

Another related issue with ev$GW$ is that its self-consistent procedure can converge to different nearby solutions.
As a result, varying $\eta$ does not always lead to the same corresponding $GW$ solution. \cite{Veril_2018,Loos_2020e,Monino_2022}
We hypothesize that this is the cause of the small bump observed in the trajectory at $\Re (E)\approx\SI{4.0}{\eV}$.
This hypothesis is supported by the fact that no such feature appears in the $G_0W_0$ calculation.
The bump corresponds to a local minimum in the energy velocity, which we disregard due to its likely artificial origin.

In addition, as shown in the trajectory plot (see Fig.~\ref{fig:traj_CO2}), the imaginary part of the ev$GW$ energy becomes positive at a certain point.
This behavior is unphysical, so we restrict our search for minima to regions of the trajectory where the imaginary part remains negative.

As observed in other molecules, the qs$GW$ trajectory features a sharp turn that leads to a well-defined minimum, located at a significantly smaller CAP strength, $\eta_\text{opt} = \SI{0.00065}{}$, compared to those of $G_0W_0$ (and ev$GW$), both of which exhibit broader trajectories and minima at $\eta_\text{opt} = \SI{0.01250}{}$ (and $\SI{0.00960}{}$). This qualitative behavior again highlights the stabilizing influence of quasiparticle self-consistency in qs$GW$ and the importance of orbital relaxation for describing resonances.

The resonance parameters are collected in Table \ref{tab:res_opt_CO2}. At the qs$GW$ level, the resonance position is $E_R = \SI{4.055}{\eV}$ with a width of $\Gamma = \SI{0.174}{\eV}$, showing much closer agreement with CAP-EOM-EA-CCSD and experimental values than the other $GW$ variants. Both the $G_0W_0$ and ev$GW$ methods predict higher resonance positions: $E_R = \SI{4.082}{\eV}$ for $G_0W_0$, and $E_R = \SI{4.016}{\eV}$ for ev$GW$. However, they significantly underestimate the resonance width: $\Gamma = \SI{0.037}{\eV}$ for ev$GW$, and $\Gamma = \SI{0.015}{\eV}$ for $G_0W_0$.

qs$GW$ thus provides a much more realistic description of the metastable state in \ce{CO2-}, both in terms of position and lifetime. While it still slightly overestimates the resonance position compared to the experimental value (between \SI{3.6}{\eV} and \SI{3.8}{\eV}), just as CAP-EOM-EA-CCSD, it reproduces the width within the experimental uncertainty (\SI{0.20 \pm 0.07}{\eV}), \cite{Burrow_1972} underlining its strength in treating resonance states where both electron correlation and orbital relaxation are essential.

%%% FIG 7 %%% 

%%% TAB 7 %%% 
\begin{table}
	\caption{Optimal CAP strengths, resonance positions, and widths of \ce{CO2-} obtained using different methods. The $GW$ results are from this work; other data are taken from the cited references.}
	\label{tab:res_opt_CO2}
	\begin{ruledtabular}
		\begin{tabular}{llcc}
            Method & $\etaopt$ (\si{\au}) & $E_\text{R}$ (\si{\eV}) & $\Gamma$ (\si{\eV}) \\
            \hline
            $G_0W_0$ & 0.01250 & 4.082 & 0.015 \\
            ev$GW$ & 0.00960 & 4.016 & 0.037 \\
            qs$GW$ & 0.01235 & 4.055 & 0.174 \\
			\\
			CAP-EOM-EA-CCSD\cite{Zuev_2014}	&0.0074	&4.020		&0.119\\
			Experiment\cite{Sanche_1973,Burrow_1972,Boness_1974,Allan_2001,Lozano_2022}				&		&3.6--3.8 		&0.20$\pm$0.07\cite{Burrow_1972}\\
		\end{tabular}
	\end{ruledtabular}
\end{table}
%%% %%% %%% %%%

%=================================================================%
\section{Conclusion}
\label{sec:conclusion}
%=================================================================%
We have presented an implementation of the CAP formalism within the $GW$ approximation, enabling the computation of metastable electronic resonance states in a Green's function framework. 
Applied to a series of prototypical temporary anions of small systems, this approach yields both resonance positions and lifetimes through a fully complex, non-Hermitian treatment.

Among the $GW$ variants considered, the qs$GW$ scheme offers the most reliable performance. 
Across all studied molecules, qs$GW$ reproduces CAP-EOM-EA-CCSD resonance positions within \SI{0.11}{\eV} and widths within \SI{0.06}{\eV}. 
The mean absolute differences are only \SI{0.07}{\eV} for the position and \SI{0.03}{\eV} for the width.
qs$GW$ systematically overestimates the resonance position relative to CAP-EOM-EA-CCSD, except for \ce{CH2O}, where it slightly underestimates it by only \SI{0.007}{\eV}.
No clear trend emerges for the widths. 

In contrast, $G_0W_0$ and ev$GW$ consistently underestimate resonance widths and predict larger optimal CAP strengths, indicating stronger perturbative effects. 
These methods also struggle to resolve shallow or less pronounced minima in the CAP energy trajectories, limiting their robustness.
Notably, ev$GW$, which includes eigenvalue self-consistency, offers a modest improvement in resonance positions compared to $G_0W_0$, but this trend does not extend to the lifetimes. 

These findings highlight the importance of orbital relaxation, which is fully accounted for only in qs$GW$. 
The close agreement of qs$GW$ with wavefunction benchmarks suggests that a self-consistent treatment of both eigenvalues and orbitals is critical for a robust description of metastable states.
Although all theoretical methods, including CAP-EOM-EA-CCSD, tend to overestimate experimental resonance positions, qs$GW$ provides accuracy on par with high-level wavefunction theory, delivering reliable estimates for both positions and lifetimes.
Thus, qs$GW$  offers a promising and comparatively cheap alternative for describing resonances in larger molecules, where wavefunction-based methods would be too expensive.

This work opens several avenues for further development. 
Incorporating first-order perturbative corrections to the CAP energy may enhance reliability, as demonstrated in wavefunction-based approaches.\cite{Zuev_2014} 
For $G_0W_0$ and ev$GW$, exploring alternative starting points, such as Kohn-Sham orbitals, \cite{Bruneval_2013} could help mitigate current shortcomings.
Furthermore, it would be worthwhile to investigate various flavors of the closely related GF2 (or second Born) methods, \cite{Suhai_1983,Holleboom_1990,Casida_1989,Casida_1991,SzaboBook,Stefanucci_2013,Ortiz_2013,Phillips_2014,Phillips_2015,Rusakov_2014,Rusakov_2016,Hirata_2015,Hirata_2017,Backhouse_2021,Backhouse_2020b,Backhouse_2020a,Pokhilko_2021a,Pokhilko_2021b,Pokhilko_2022} which circumvent solving the RPA eigenvalue problem by employing an excitation energy and amplitude ansatz based on quasiparticle energies and orbitals.
Given their structural similarity to the CAP-$GW$ approach introduced in this work, extending these methods with a CAP should be straightforward and potentially more computationally efficient. 
This would also be closely related to the CAP-ADC methods discussed in the introduction. \cite{Santra_2002,Feuerbacher_2003,Feuerbacher_2005,Monino_2023}

Finally, the methodology presented here could be extended to other metastable phenomena, such as core-ionization and Auger-Meitner decay, where accurate lifetimes are particularly important.
However, the $GW$ formalism is likely not suited for describing Feshbach resonances, which involve two-particle processes, as it is well known that $GW$ struggles to accurately capture satellite transitions. \cite{Marie_2024b,Loos_2024}

%%%%%%%%%%%%%%%%%%%%%%%%%%%%%%%%
\section*{Supporting Information}
%%%%%%%%%%%%%%%%%%%%%%%%%%%%%%%%
The \SupMat contains, for each system considered here, the molecular coordinates, the exponents of the additional diffuse basis functions, the representation of the Dyson orbital associated with the resonance state, the behavior of the energy-velocity as a function of $\eta$, and the spectral functions.

%%%%%%%%%%%%%%%%%%%%%%%%
\section*{Acknowledgements}
%%%%%%%%%%%%%%%%%%%%%%%%

The authors would like to thank Josep Alberola, Antoine Marie, Ivan Duchemin, and Xavier Blase for insightful discussions.
This project has received funding from the European Research Council (ERC) under the European Union's Horizon 2020 research and innovation programme (Grant agreement No.~863481).
This work used the HPC resources from CALMIP (Toulouse) under allocations 2025-18005.

%%%%%%%%%%%%%%%%%%%%%%%%
\appendix
%%%%%%%%%%%%%%%%%%%%%%%%

%%%%%%%%%%%%%%%%%%%%%%%%
\section{SRG renormalization for complex qs$GW$}
\label{app:srg}
%%%%%%%%%%%%%%%%%%%%%%%%

We extended the SRG renormalization method\cite{Marie_2023} for complex-valued energies, which yields the following matrix elements of the self-energy
\begin{equation} \label{eq:SRG-self-energy}
	\begin{split}
		[\tilde{\Sigma}_\text{c}]_{pq}
		& = \sum_{i\nu} \frac{(\tilde{\epsilon}_{pi\nu} + \tilde{\epsilon}_{qi\nu})N_{pqi\nu}}{\tilde{\epsilon}_{pi\nu}^2 + \tilde{\epsilon}_{qi\nu}^2 + \tilde{\kappa}_{pi\nu}^2 + \tilde{\kappa}_{qi\nu}^2} 
		\\
		& + \sum_{a\nu} \frac{(\tilde{\epsilon}_{pa\nu} + \tilde{\epsilon}_{qa\nu})N_{pqa\nu}}{\tilde{\epsilon}_{pa\nu}^2 + \tilde{\epsilon}_{qa\nu}^2 + \tilde{\kappa}_{pa\nu}^2 + \tilde{\kappa}_{qa\nu}^2}
		\\
		&+ \ii \sum_{i\nu} \frac{(\tilde{\kappa}_{pi\nu} + \tilde{\kappa}_{qi\nu})N_{pqi\nu}}{\tilde{\epsilon}_{pi\nu}^2 + \tilde{\epsilon}_{qi\nu}^2 + \tilde{\kappa}_{pi\nu}^2 + \tilde{\kappa}_{qi\nu}^2}
		\\
		&- \ii \sum_{a\nu} \frac{(\tilde{\kappa}_{pa\nu} + \tilde{\kappa}_{qa\nu})N_{pqa\nu}}{\tilde{\epsilon}_{pa\nu}^2 + \tilde{\epsilon}_{qa\nu}^2 + \tilde{\kappa}_{pa\nu}^2 + \tilde{\kappa}_{qa\nu}^2}
	\end{split}
\end{equation}
with the real and imaginary parts of the denominator in Eq.~\eqref{eq:self_energy}
\begin{subequations}
	\begin{align}
		\tilde{\epsilon}_{pq\nu} & = \Re{ \epsilon_p - \epsilon_q - \sgn[\Re(\epsilon_q - \mu)] \Omega_\nu } 
		\\
		\tilde{\kappa}_{pq\nu} & = \kappa + \Im{ \sgn[\Re(\epsilon_q - \mu)] (\epsilon_p - \epsilon_q) - \Omega_\nu }
	\end{align}
\end{subequations}
(where $\mu$ is the chemical potential) and the renormalized elements
\begin{equation}
	N_{pqr\nu} = M_{pr,\nu}M_{qr,\nu}
	\qty[1- e^{-s\qty(\tilde{\epsilon}_{pr\nu}^2 + \tilde{\epsilon}_{qr\nu}^2+\tilde{\kappa}_{pr\nu}^2 + \tilde{\kappa}_{q r\nu}^2)}]
\end{equation}
The exponential term in the renormalized density ensures that the numerator tends faster to zero when both the real and imaginary parts of the denominator tend to zero. 

%%%%%%%%%%%%%%%%%%%%%%%%%%%%%%%%
\section*{References}
%%%%%%%%%%%%%%%%%%%%%%%%%%%%%

%%%%%%%%%%%%%%%%%%%%%%%%
\bibliography{biblio}

%aipnum4-2.bst 2019-01-14 (MD) hand-edited version of apsrev4-1.bst
\begin{thebibliography}{227}%
\makeatletter
\providecommand \@ifxundefined [1]{%
 \@ifx{#1\undefined}
}%
\providecommand \@ifnum [1]{%
 \ifnum #1\expandafter \@firstoftwo
 \else \expandafter \@secondoftwo
 \fi
}%
\providecommand \@ifx [1]{%
 \ifx #1\expandafter \@firstoftwo
 \else \expandafter \@secondoftwo
 \fi
}%
\providecommand \natexlab [1]{#1}%
\providecommand \enquote  [1]{``#1''}%
\providecommand \bibnamefont  [1]{#1}%
\providecommand \bibfnamefont [1]{#1}%
\providecommand \citenamefont [1]{#1}%
\providecommand \href@noop [0]{\@secondoftwo}%
\providecommand \href [0]{\begingroup \@sanitize@url \@href}%
\providecommand \@href[1]{\@@startlink{#1}\@@href}%
\providecommand \@@href[1]{\endgroup#1\@@endlink}%
\providecommand \@sanitize@url [0]{\catcode `\\12\catcode `\$12\catcode
  `\&12\catcode `\#12\catcode `\^12\catcode `\_12\catcode `\%12\relax}%
\providecommand \@@startlink[1]{}%
\providecommand \@@endlink[0]{}%
\providecommand \url  [0]{\begingroup\@sanitize@url \@url }%
\providecommand \@url [1]{\endgroup\@href {#1}{\urlprefix }}%
\providecommand \urlprefix  [0]{URL }%
\providecommand \Eprint [0]{\href }%
\providecommand \doibase [0]{https://doi.org/}%
\providecommand \selectlanguage [0]{\@gobble}%
\providecommand \bibinfo  [0]{\@secondoftwo}%
\providecommand \bibfield  [0]{\@secondoftwo}%
\providecommand \translation [1]{[#1]}%
\providecommand \BibitemOpen [0]{}%
\providecommand \bibitemStop [0]{}%
\providecommand \bibitemNoStop [0]{.\EOS\space}%
\providecommand \EOS [0]{\spacefactor3000\relax}%
\providecommand \BibitemShut  [1]{\csname bibitem#1\endcsname}%
\let\auto@bib@innerbib\@empty
%</preamble>
\bibitem [{\citenamefont {Jordan}\ and\ \citenamefont
  {Burrow}(1987)}]{Jordan_1987}%
  \BibitemOpen
  \bibfield  {author} {\bibinfo {author} {\bibfnamefont {K.~D.}\ \bibnamefont
  {Jordan}}\ and\ \bibinfo {author} {\bibfnamefont {P.~D.}\ \bibnamefont
  {Burrow}},\ }\bibfield  {title} {\enquote {\bibinfo {title} {{Temporary Anion
  States of Polyatomic Hydrocarbons}},}\ }\href
  {https://doi.org/10.1021/cr00079a005} {\bibfield  {journal} {\bibinfo
  {journal} {Chem. Rev.}\ }\textbf {\bibinfo {volume} {87}},\ \bibinfo {pages}
  {557--588} (\bibinfo {year} {1987})}\BibitemShut {NoStop}%
\bibitem [{\citenamefont {Jordan}\ and\ \citenamefont
  {Wang}(2003)}]{Jordan_2003}%
  \BibitemOpen
  \bibfield  {author} {\bibinfo {author} {\bibfnamefont {K.~D.}\ \bibnamefont
  {Jordan}}\ and\ \bibinfo {author} {\bibfnamefont {F.}~\bibnamefont {Wang}},\
  }\bibfield  {title} {\enquote {\bibinfo {title} {{Theory of Dipole-Bound
  Anions}},}\ }\href
  {https://doi.org/https://doi.org/10.1146/annurev.physchem.54.011002.103851}
  {\bibfield  {journal} {\bibinfo  {journal} {Annu. Rev. Phys. Chem.}\ }\textbf
  {\bibinfo {volume} {54}},\ \bibinfo {pages} {367--396} (\bibinfo {year}
  {2003})}\BibitemShut {NoStop}%
\bibitem [{\citenamefont {Simons}(2008)}]{Simons_2008}%
  \BibitemOpen
  \bibfield  {author} {\bibinfo {author} {\bibfnamefont {J.}~\bibnamefont
  {Simons}},\ }\bibfield  {title} {\enquote {\bibinfo {title} {{Molecular
  Anions}},}\ }\href {https://doi.org/10.1021/jp711490b} {\bibfield  {journal}
  {\bibinfo  {journal} {J. Phys. Chem. A}\ }\textbf {\bibinfo {volume} {112}},\
  \bibinfo {pages} {6401--6511} (\bibinfo {year} {2008})}\BibitemShut {NoStop}%
\bibitem [{\citenamefont {Simons}(2011)}]{Simons_2011}%
  \BibitemOpen
  \bibfield  {author} {\bibinfo {author} {\bibfnamefont {J.}~\bibnamefont
  {Simons}},\ }\bibfield  {title} {\enquote {\bibinfo {title} {{Theoretical
  Study of Negative Molecular Ions}},}\ }\href
  {https://doi.org/10.1146/annurev-physchem-032210-103547} {\bibfield
  {journal} {\bibinfo  {journal} {Annu. Rev. Phys. Chem.}\ }\textbf {\bibinfo
  {volume} {62}},\ \bibinfo {pages} {107--128} (\bibinfo {year}
  {2011})}\BibitemShut {NoStop}%
\bibitem [{\citenamefont {Jordan}, \citenamefont {Voora},\ and\ \citenamefont
  {Simons}(2014)}]{Jordan_2014}%
  \BibitemOpen
  \bibfield  {author} {\bibinfo {author} {\bibfnamefont {K.~D.}\ \bibnamefont
  {Jordan}}, \bibinfo {author} {\bibfnamefont {V.~K.}\ \bibnamefont {Voora}},\
  and\ \bibinfo {author} {\bibfnamefont {J.}~\bibnamefont {Simons}},\
  }\bibfield  {title} {\enquote {\bibinfo {title} {{Negative Electron
  Affinities From Conventional Electronic Structure Methods}},}\ }\href
  {https://doi.org/10.1007/s00214-014-1445-1} {\bibfield  {journal} {\bibinfo
  {journal} {Theor. Chem. Acc.}\ }\textbf {\bibinfo {volume} {133}},\ \bibinfo
  {pages} {1445} (\bibinfo {year} {2014})}\BibitemShut {NoStop}%
\bibitem [{\citenamefont {Simons}(2000)}]{Simons_2000}%
  \BibitemOpen
  \bibfield  {author} {\bibinfo {author} {\bibfnamefont {J.}~\bibnamefont
  {Simons}},\ }\enquote {\bibinfo {title} {{Detachment Processes For Molecular
  Anions}},}\ in\ \href {https://doi.org/10.1142/9789812813473_0017} {\emph
  {\bibinfo {booktitle} {Advanced Series in Physical Chemistry: Photoionization
  and Photodetachment}}}\ (\bibinfo  {publisher} {World Scientific Publishing
  Co.: Singapore},\ \bibinfo {year} {2000})\ pp.\ \bibinfo {pages}
  {958--1010}\BibitemShut {NoStop}%
\bibitem [{\citenamefont {Simons}(2023)}]{Simons_2023}%
  \BibitemOpen
  \bibfield  {author} {\bibinfo {author} {\bibfnamefont {J.}~\bibnamefont
  {Simons}},\ }\bibfield  {title} {\enquote {\bibinfo {title} {Molecular anions
  perspective},}\ }\href {https://doi.org/10.1021/acs.jpca.3c01564} {\bibfield
  {journal} {\bibinfo  {journal} {J. Phys. Chem. A}\ }\textbf {\bibinfo
  {volume} {127}},\ \bibinfo {pages} {3940--3957} (\bibinfo {year}
  {2023})}\BibitemShut {NoStop}%
\bibitem [{\citenamefont {Boudaiffa}\ \emph {et~al.}(2000)\citenamefont
  {Boudaiffa}, \citenamefont {Cloutier}, \citenamefont {Hunting}, \citenamefont
  {Huels},\ and\ \citenamefont {Sanche}}]{Boudaiffa_2000}%
  \BibitemOpen
  \bibfield  {author} {\bibinfo {author} {\bibfnamefont {B.}~\bibnamefont
  {Boudaiffa}}, \bibinfo {author} {\bibfnamefont {P.}~\bibnamefont {Cloutier}},
  \bibinfo {author} {\bibfnamefont {D.}~\bibnamefont {Hunting}}, \bibinfo
  {author} {\bibfnamefont {M.~A.}\ \bibnamefont {Huels}},\ and\ \bibinfo
  {author} {\bibfnamefont {L.}~\bibnamefont {Sanche}},\ }\bibfield  {title}
  {\enquote {\bibinfo {title} {{Resonant Formation of DNA Strand Breaks by
  Low-Energy (3 to 20 eV) Electrons}},}\ }\href
  {https://doi.org/10.1126/science.287.5458.1658} {\bibfield  {journal}
  {\bibinfo  {journal} {Science}\ }\textbf {\bibinfo {volume} {287}},\ \bibinfo
  {pages} {1658--1660} (\bibinfo {year} {2000})}\BibitemShut {NoStop}%
\bibitem [{\citenamefont {Martin}\ \emph {et~al.}(2004)\citenamefont {Martin},
  \citenamefont {Burrow}, \citenamefont {Cai}, \citenamefont {Cloutier},
  \citenamefont {Hunting},\ and\ \citenamefont {Sanche}}]{Martin_2004}%
  \BibitemOpen
  \bibfield  {author} {\bibinfo {author} {\bibfnamefont {F.}~\bibnamefont
  {Martin}}, \bibinfo {author} {\bibfnamefont {P.~D.}\ \bibnamefont {Burrow}},
  \bibinfo {author} {\bibfnamefont {Z.}~\bibnamefont {Cai}}, \bibinfo {author}
  {\bibfnamefont {P.}~\bibnamefont {Cloutier}}, \bibinfo {author}
  {\bibfnamefont {D.}~\bibnamefont {Hunting}},\ and\ \bibinfo {author}
  {\bibfnamefont {L.}~\bibnamefont {Sanche}},\ }\bibfield  {title} {\enquote
  {\bibinfo {title} {{DNA Strand Breaks Induced by 0--4 eV Electrons: The Role
  of Shape Resonances}},}\ }\href
  {https://doi.org/10.1103/PhysRevLett.93.068101} {\bibfield  {journal}
  {\bibinfo  {journal} {Phys. Rev. Lett.}\ }\textbf {\bibinfo {volume} {93}},\
  \bibinfo {pages} {068101} (\bibinfo {year} {2004})}\BibitemShut {NoStop}%
\bibitem [{\citenamefont {Simons}(2006)}]{Simons_2006}%
  \BibitemOpen
  \bibfield  {author} {\bibinfo {author} {\bibfnamefont {J.}~\bibnamefont
  {Simons}},\ }\bibfield  {title} {\enquote {\bibinfo {title} {{How Do
  Low-Energy (0.1--2 eV) Electrons Cause DNA-Strand Breaks?}}}\ }\href
  {https://doi.org/10.1021/ar0680769} {\bibfield  {journal} {\bibinfo
  {journal} {Acc. Chem. Res.}\ }\textbf {\bibinfo {volume} {39}},\ \bibinfo
  {pages} {772--779} (\bibinfo {year} {2006})}\BibitemShut {NoStop}%
\bibitem [{\citenamefont {Alizadeh}, \citenamefont {Orlando},\ and\
  \citenamefont {Sanche}(2015)}]{Alizadeh_2015}%
  \BibitemOpen
  \bibfield  {author} {\bibinfo {author} {\bibfnamefont {E.}~\bibnamefont
  {Alizadeh}}, \bibinfo {author} {\bibfnamefont {T.~M.}\ \bibnamefont
  {Orlando}},\ and\ \bibinfo {author} {\bibfnamefont {L.}~\bibnamefont
  {Sanche}},\ }\bibfield  {title} {\enquote {\bibinfo {title} {{Biomolecular
  Damage Induced by Ionizing Radiation: The Direct and Indirect Effects of
  Low-Energy Electrons on DNA}},}\ }\href
  {https://doi.org/10.1146/annurev-physchem-040513-103605} {\bibfield
  {journal} {\bibinfo  {journal} {Annu. Rev. Phys. Chem.}\ ,\ \bibinfo {pages}
  {379--398}} (\bibinfo {year} {2015})}\BibitemShut {NoStop}%
\bibitem [{\citenamefont {Hammer}\ \emph {et~al.}(2004)\citenamefont {Hammer},
  \citenamefont {Shin}, \citenamefont {Headrick}, \citenamefont {Diken},
  \citenamefont {Roscioli}, \citenamefont {Weddle},\ and\ \citenamefont
  {Johnson}}]{Hammer_2004}%
  \BibitemOpen
  \bibfield  {author} {\bibinfo {author} {\bibfnamefont {N.~I.}\ \bibnamefont
  {Hammer}}, \bibinfo {author} {\bibfnamefont {J.-W.}\ \bibnamefont {Shin}},
  \bibinfo {author} {\bibfnamefont {J.~M.}\ \bibnamefont {Headrick}}, \bibinfo
  {author} {\bibfnamefont {E.~G.}\ \bibnamefont {Diken}}, \bibinfo {author}
  {\bibfnamefont {J.~R.}\ \bibnamefont {Roscioli}}, \bibinfo {author}
  {\bibfnamefont {G.~H.}\ \bibnamefont {Weddle}},\ and\ \bibinfo {author}
  {\bibfnamefont {M.~A.}\ \bibnamefont {Johnson}},\ }\bibfield  {title}
  {\enquote {\bibinfo {title} {{How Do Small Water Clusters Bind an Excess
  Electron?}}}\ }\href {https://doi.org/10.1126/science.1102792} {\bibfield
  {journal} {\bibinfo  {journal} {Science}\ }\textbf {\bibinfo {volume}
  {306}},\ \bibinfo {pages} {675--679} (\bibinfo {year} {2004})}\BibitemShut
  {NoStop}%
\bibitem [{\citenamefont {Clarke}\ and\ \citenamefont
  {Verlet}(2024)}]{Clarke_2024}%
  \BibitemOpen
  \bibfield  {author} {\bibinfo {author} {\bibfnamefont {C.~J.}\ \bibnamefont
  {Clarke}}\ and\ \bibinfo {author} {\bibfnamefont {J.~R.}\ \bibnamefont
  {Verlet}},\ }\bibfield  {title} {\enquote {\bibinfo {title} {Dynamics of
  anions: From bound to unbound states and everything in between},}\ }\href
  {https://doi.org/https://doi.org/10.1146/annurev-physchem-090722-125031}
  {\bibfield  {journal} {\bibinfo  {journal} {Annu. Rev. Phys. Chem.}\ }\textbf
  {\bibinfo {volume} {75}},\ \bibinfo {pages} {89--110} (\bibinfo {year}
  {2024})}\BibitemShut {NoStop}%
\bibitem [{\citenamefont {Gamow}(1928)}]{Gamow_1928}%
  \BibitemOpen
  \bibfield  {author} {\bibinfo {author} {\bibfnamefont {G.}~\bibnamefont
  {Gamow}},\ }\bibfield  {title} {\enquote {\bibinfo {title} {{Zur
  Quantentheorie des Atomkernes}},}\ }\href
  {https://doi.org/10.1007/BF01343196} {\bibfield  {journal} {\bibinfo
  {journal} {Z. Phys.}\ }\textbf {\bibinfo {volume} {51}},\ \bibinfo {pages}
  {204--212} (\bibinfo {year} {1928})}\BibitemShut {NoStop}%
\bibitem [{\citenamefont {Siegert}(1939)}]{Siegert_1939}%
  \BibitemOpen
  \bibfield  {author} {\bibinfo {author} {\bibfnamefont {A.~J.~F.}\
  \bibnamefont {Siegert}},\ }\bibfield  {title} {\enquote {\bibinfo {title}
  {{On the Derivation of the Dispersion Formula for Nuclear Reactions}},}\
  }\href {https://doi.org/10.1103/PhysRev.56.750} {\bibfield  {journal}
  {\bibinfo  {journal} {Phys. Rev.}\ }\textbf {\bibinfo {volume} {56}},\
  \bibinfo {pages} {750--752} (\bibinfo {year} {1939})}\BibitemShut {NoStop}%
\bibitem [{\citenamefont {Klaiman}\ and\ \citenamefont
  {Gilary}(2012)}]{Klaiman_2012}%
  \BibitemOpen
  \bibfield  {author} {\bibinfo {author} {\bibfnamefont {S.}~\bibnamefont
  {Klaiman}}\ and\ \bibinfo {author} {\bibfnamefont {I.}~\bibnamefont
  {Gilary}},\ }\bibfield  {title} {\enquote {\bibinfo {title} {{On Resonance: A
  First Glance into the Behavior of Unstable States}},}\ }\href
  {https://doi.org/10.1016/B978-0-12-397009-1.00001-1} {\bibfield  {journal}
  {\bibinfo  {journal} {Adv. Quantum Chem.}\ }\textbf {\bibinfo {volume}
  {63}},\ \bibinfo {pages} {1--31} (\bibinfo {year} {2012})}\BibitemShut
  {NoStop}%
\bibitem [{\citenamefont {Zuev}\ \emph {et~al.}(2011)\citenamefont {Zuev},
  \citenamefont {Bravaya}, \citenamefont {Crawford}, \citenamefont {Lindh},\
  and\ \citenamefont {Krylov}}]{Zuev_2011}%
  \BibitemOpen
  \bibfield  {author} {\bibinfo {author} {\bibfnamefont {D.}~\bibnamefont
  {Zuev}}, \bibinfo {author} {\bibfnamefont {K.~B.}\ \bibnamefont {Bravaya}},
  \bibinfo {author} {\bibfnamefont {T.~D.}\ \bibnamefont {Crawford}}, \bibinfo
  {author} {\bibfnamefont {R.}~\bibnamefont {Lindh}},\ and\ \bibinfo {author}
  {\bibfnamefont {A.~I.}\ \bibnamefont {Krylov}},\ }\bibfield  {title}
  {\enquote {\bibinfo {title} {{Electronic structure of the two isomers of the
  anionic form of p-coumaric acid chromophore}},}\ }\href
  {https://doi.org/10.1063/1.3516211} {\bibfield  {journal} {\bibinfo
  {journal} {J. Chem. Phys.}\ }\textbf {\bibinfo {volume} {134}},\ \bibinfo
  {pages} {034310} (\bibinfo {year} {2011})}\BibitemShut {NoStop}%
\bibitem [{\citenamefont {Bravaya}\ \emph {et~al.}(2013)\citenamefont
  {Bravaya}, \citenamefont {Zuev}, \citenamefont {Epifanovsky},\ and\
  \citenamefont {Krylov}}]{Bravaya_2013}%
  \BibitemOpen
  \bibfield  {author} {\bibinfo {author} {\bibfnamefont {K.~B.}\ \bibnamefont
  {Bravaya}}, \bibinfo {author} {\bibfnamefont {D.}~\bibnamefont {Zuev}},
  \bibinfo {author} {\bibfnamefont {E.}~\bibnamefont {Epifanovsky}},\ and\
  \bibinfo {author} {\bibfnamefont {A.~I.}\ \bibnamefont {Krylov}},\ }\bibfield
   {title} {\enquote {\bibinfo {title} {{Complex-Scaled Equation-Of-Motion
  Coupled-Cluster Method with Single and Double Substitutions for Autoionizing
  Excited States: Theory, Implementation, and Examples}},}\ }\href
  {https://doi.org/10.1063/1.4795750} {\bibfield  {journal} {\bibinfo
  {journal} {J. Chem. Phys.}\ }\textbf {\bibinfo {volume} {138}},\ \bibinfo
  {pages} {124106} (\bibinfo {year} {2013})}\BibitemShut {NoStop}%
\bibitem [{\citenamefont {Kunitsa}\ and\ \citenamefont
  {Bravaya}(2015)}]{Kunitsa_2015}%
  \BibitemOpen
  \bibfield  {author} {\bibinfo {author} {\bibfnamefont {A.~A.}\ \bibnamefont
  {Kunitsa}}\ and\ \bibinfo {author} {\bibfnamefont {K.~B.}\ \bibnamefont
  {Bravaya}},\ }\bibfield  {title} {\enquote {\bibinfo {title}
  {{First-Principles Calculations of the Energy and Width of the $^2A_u$ Shape
  Resonance in p-Benzoquinone: A Gateway State for Electron Transfer}},}\
  }\href {https://doi.org/10.1021/acs.jpclett.5b00207} {\bibfield  {journal}
  {\bibinfo  {journal} {J. Phys. Chem. Lett.}\ }\textbf {\bibinfo {volume}
  {6}},\ \bibinfo {pages} {1053--1058} (\bibinfo {year} {2015})}\BibitemShut
  {NoStop}%
\bibitem [{\citenamefont {Feshbach}(1958)}]{Feshbach_1958}%
  \BibitemOpen
  \bibfield  {author} {\bibinfo {author} {\bibfnamefont {H.}~\bibnamefont
  {Feshbach}},\ }\bibfield  {title} {\enquote {\bibinfo {title} {{Unified
  Theory of Nuclear Reactions}},}\ }\href
  {https://doi.org/10.1016/0003-4916(58)90007-1} {\bibfield  {journal}
  {\bibinfo  {journal} {Ann. Phys.}\ }\textbf {\bibinfo {volume} {5}},\
  \bibinfo {pages} {357--390} (\bibinfo {year} {1958})}\BibitemShut {NoStop}%
\bibitem [{\citenamefont {Feshbach}(1962)}]{Feshbach_1962}%
  \BibitemOpen
  \bibfield  {author} {\bibinfo {author} {\bibfnamefont {H.}~\bibnamefont
  {Feshbach}},\ }\bibfield  {title} {\enquote {\bibinfo {title} {{A Unified
  Theory of Nuclear Reactions. II}},}\ }\href
  {https://doi.org/10.1016/0003-4916(62)90221-X} {\bibfield  {journal}
  {\bibinfo  {journal} {Ann. Phys.}\ }\textbf {\bibinfo {volume} {19}},\
  \bibinfo {pages} {287--313} (\bibinfo {year} {1962})}\BibitemShut {NoStop}%
\bibitem [{\citenamefont {Domcke}(1991)}]{Domcke_1991}%
  \BibitemOpen
  \bibfield  {author} {\bibinfo {author} {\bibfnamefont {W.}~\bibnamefont
  {Domcke}},\ }\bibfield  {title} {\enquote {\bibinfo {title} {Theory of
  resonance and threshold effects in electron-molecule collisions: The
  projection-operator approach},}\ }\href
  {https://doi.org/https://doi.org/10.1016/0370-1573(91)90125-6} {\bibfield
  {journal} {\bibinfo  {journal} {Phys. Rep.}\ }\textbf {\bibinfo {volume}
  {208}},\ \bibinfo {pages} {97--188} (\bibinfo {year} {1991})}\BibitemShut
  {NoStop}%
\bibitem [{\citenamefont {Averbukh}\ and\ \citenamefont
  {Cederbaum}(2005)}]{Averbukh_2005}%
  \BibitemOpen
  \bibfield  {author} {\bibinfo {author} {\bibfnamefont {V.}~\bibnamefont
  {Averbukh}}\ and\ \bibinfo {author} {\bibfnamefont {L.~S.}\ \bibnamefont
  {Cederbaum}},\ }\bibfield  {title} {\enquote {\bibinfo {title} {Ab initio
  calculation of interatomic decay rates by a combination of the fano ansatz,
  green's-function methods, and the stieltjes imaging technique},}\ }\href
  {https://doi.org/10.1063/1.2126976} {\bibfield  {journal} {\bibinfo
  {journal} {J. Chem. Phys.}\ }\textbf {\bibinfo {volume} {123}},\ \bibinfo
  {pages} {204107} (\bibinfo {year} {2005})}\BibitemShut {NoStop}%
\bibitem [{\citenamefont {Koloren{\v c}}\ \emph {et~al.}(2008)\citenamefont
  {Koloren{\v c}}, \citenamefont {Averbukh}, \citenamefont {Gokhberg},\ and\
  \citenamefont {Cederbaum}}]{Kolorenc_2008}%
  \BibitemOpen
  \bibfield  {author} {\bibinfo {author} {\bibfnamefont {P.}~\bibnamefont
  {Koloren{\v c}}}, \bibinfo {author} {\bibfnamefont {V.}~\bibnamefont
  {Averbukh}}, \bibinfo {author} {\bibfnamefont {K.}~\bibnamefont {Gokhberg}},\
  and\ \bibinfo {author} {\bibfnamefont {L.~S.}\ \bibnamefont {Cederbaum}},\
  }\bibfield  {title} {\enquote {\bibinfo {title} {Ab initio calculation of
  interatomic decay rates of excited doubly ionized states in clusters},}\
  }\href {https://doi.org/10.1063/1.3043437} {\bibfield  {journal} {\bibinfo
  {journal} {J. Chem. Phys.}\ }\textbf {\bibinfo {volume} {129}},\ \bibinfo
  {pages} {244102} (\bibinfo {year} {2008})}\BibitemShut {NoStop}%
\bibitem [{\citenamefont {Kolorenc}\ and\ \citenamefont
  {Mitas}(2011)}]{Kolorenc_2011}%
  \BibitemOpen
  \bibfield  {author} {\bibinfo {author} {\bibfnamefont {J.}~\bibnamefont
  {Kolorenc}}\ and\ \bibinfo {author} {\bibfnamefont {L.}~\bibnamefont
  {Mitas}},\ }\bibfield  {title} {\enquote {\bibinfo {title} {{Applications of
  quantum Monte Carlo methods in condensed systems}},}\ }\href
  {https://doi.org/10.1088/0034-4885/74/2/026502} {\bibfield  {journal}
  {\bibinfo  {journal} {Rep. Prog. Phys.}\ }\textbf {\bibinfo {volume} {74}},\
  \bibinfo {pages} {026502} (\bibinfo {year} {2011})}\BibitemShut {NoStop}%
\bibitem [{\citenamefont {Yabushita}\ and\ \citenamefont
  {McCurdy}(1985)}]{Yabushita_1985}%
  \BibitemOpen
  \bibfield  {author} {\bibinfo {author} {\bibfnamefont {S.}~\bibnamefont
  {Yabushita}}\ and\ \bibinfo {author} {\bibfnamefont {C.~W.}\ \bibnamefont
  {McCurdy}},\ }\bibfield  {title} {\enquote {\bibinfo {title} {Feshbach
  resonances in electron--molecule scattering by the complex multiconfiguration
  scf and configuration interaction procedures: The $^1 \sigma_g^+$
  autoionizing states of h2},}\ }\href {https://doi.org/10.1063/1.449160}
  {\bibfield  {journal} {\bibinfo  {journal} {J. Chem. Phys.}\ }\textbf
  {\bibinfo {volume} {83}},\ \bibinfo {pages} {3547--3559} (\bibinfo {year}
  {1985})}\BibitemShut {NoStop}%
\bibitem [{\citenamefont {Sajeev}\ \emph {et~al.}(2009)\citenamefont {Sajeev},
  \citenamefont {Vysotskiy}, \citenamefont {Cederbaum},\ and\ \citenamefont
  {Moiseyev}}]{Sajeev_2009}%
  \BibitemOpen
  \bibfield  {author} {\bibinfo {author} {\bibfnamefont {Y.}~\bibnamefont
  {Sajeev}}, \bibinfo {author} {\bibfnamefont {V.}~\bibnamefont {Vysotskiy}},
  \bibinfo {author} {\bibfnamefont {L.~S.}\ \bibnamefont {Cederbaum}},\ and\
  \bibinfo {author} {\bibfnamefont {N.}~\bibnamefont {Moiseyev}},\ }\bibfield
  {title} {\enquote {\bibinfo {title} {{Continuum remover-complex absorbing
  potential: Efficient removal of the nonphysical stabilization points}},}\
  }\href {https://doi.org/10.1063/1.3271350} {\bibfield  {journal} {\bibinfo
  {journal} {J. Chem. Phys.}\ }\textbf {\bibinfo {volume} {131}},\ \bibinfo
  {pages} {211102} (\bibinfo {year} {2009})}\BibitemShut {NoStop}%
\bibitem [{\citenamefont {Schiedt}\ and\ \citenamefont
  {Weinkauf}(1999)}]{Schiedt_1999}%
  \BibitemOpen
  \bibfield  {author} {\bibinfo {author} {\bibfnamefont {J.}~\bibnamefont
  {Schiedt}}\ and\ \bibinfo {author} {\bibfnamefont {R.}~\bibnamefont
  {Weinkauf}},\ }\bibfield  {title} {\enquote {\bibinfo {title} {Resonant
  photodetachment via shape and feshbach resonances: p-benzoquinone anions as a
  model system},}\ }\href {https://doi.org/10.1063/1.478066} {\bibfield
  {journal} {\bibinfo  {journal} {J. Chem. Phys.}\ }\textbf {\bibinfo {volume}
  {110}},\ \bibinfo {pages} {304--314} (\bibinfo {year} {1999})}\BibitemShut
  {NoStop}%
\bibitem [{\citenamefont {Kunitsa}\ and\ \citenamefont
  {Bravaya}(2016)}]{Kunitsa_2016}%
  \BibitemOpen
  \bibfield  {author} {\bibinfo {author} {\bibfnamefont {A.~A.}\ \bibnamefont
  {Kunitsa}}\ and\ \bibinfo {author} {\bibfnamefont {K.~B.}\ \bibnamefont
  {Bravaya}},\ }\bibfield  {title} {\enquote {\bibinfo {title} {{Electronic
  structure of the para-benzoquinone radical anion revisited}},}\ }\href
  {https://doi.org/10.1039/C5CP06476G} {\bibfield  {journal} {\bibinfo
  {journal} {Phys. Chem. Chem. Phys.}\ }\textbf {\bibinfo {volume} {18}},\
  \bibinfo {pages} {3454--3462} (\bibinfo {year} {2016})}\BibitemShut {NoStop}%
\bibitem [{\citenamefont {Kunitsa}, \citenamefont {Granovsky},\ and\
  \citenamefont {Bravaya}(2017)}]{Kunitsa_2017}%
  \BibitemOpen
  \bibfield  {author} {\bibinfo {author} {\bibfnamefont {A.~A.}\ \bibnamefont
  {Kunitsa}}, \bibinfo {author} {\bibfnamefont {A.~A.}\ \bibnamefont
  {Granovsky}},\ and\ \bibinfo {author} {\bibfnamefont {K.~B.}\ \bibnamefont
  {Bravaya}},\ }\bibfield  {title} {\enquote {\bibinfo {title} {{CAP-XMCQDPT2
  Method for Molecular Electronic Resonances}},}\ }\href
  {https://doi.org/10.1063/1.4982950} {\bibfield  {journal} {\bibinfo
  {journal} {J. Chem. Phys.}\ }\textbf {\bibinfo {volume} {146}},\ \bibinfo
  {pages} {184107} (\bibinfo {year} {2017})}\BibitemShut {NoStop}%
\bibitem [{\citenamefont {Loupas}\ and\ \citenamefont
  {Gorfinkiel}(2017)}]{Loupas_2017}%
  \BibitemOpen
  \bibfield  {author} {\bibinfo {author} {\bibfnamefont {A.}~\bibnamefont
  {Loupas}}\ and\ \bibinfo {author} {\bibfnamefont {J.~D.}\ \bibnamefont
  {Gorfinkiel}},\ }\bibfield  {title} {\enquote {\bibinfo {title} {Resonances
  in low-energy electron scattering from para-benzoquinone},}\ }\href
  {https://doi.org/10.1039/C7CP02916K} {\bibfield  {journal} {\bibinfo
  {journal} {Phys. Chem. Chem. Phys.}\ }\textbf {\bibinfo {volume} {19}},\
  \bibinfo {pages} {18252--18261} (\bibinfo {year} {2017})}\BibitemShut
  {NoStop}%
\bibitem [{\citenamefont {da~Costa}\ \emph {et~al.}(2018)\citenamefont
  {da~Costa}, \citenamefont {Ruivo}, \citenamefont {Kossoski}, \citenamefont
  {Varella}, \citenamefont {Bettega}, \citenamefont {Jones}, \citenamefont
  {Brunger},\ and\ \citenamefont {Lima}}]{daCosta_2018}%
  \BibitemOpen
  \bibfield  {author} {\bibinfo {author} {\bibfnamefont {R.~F.}\ \bibnamefont
  {da~Costa}}, \bibinfo {author} {\bibfnamefont {J.~C.}\ \bibnamefont {Ruivo}},
  \bibinfo {author} {\bibfnamefont {F.}~\bibnamefont {Kossoski}}, \bibinfo
  {author} {\bibfnamefont {M.~T. d.~N.}\ \bibnamefont {Varella}}, \bibinfo
  {author} {\bibfnamefont {M.~H.~F.}\ \bibnamefont {Bettega}}, \bibinfo
  {author} {\bibfnamefont {D.~B.}\ \bibnamefont {Jones}}, \bibinfo {author}
  {\bibfnamefont {M.~J.}\ \bibnamefont {Brunger}},\ and\ \bibinfo {author}
  {\bibfnamefont {M.~A.~P.}\ \bibnamefont {Lima}},\ }\bibfield  {title}
  {\enquote {\bibinfo {title} {An ab initio investigation for elastic and
  electronically inelastic electron scattering from para-benzoquinone},}\
  }\href {https://doi.org/10.1063/1.5050622} {\bibfield  {journal} {\bibinfo
  {journal} {J. Chem. Phys.}\ }\textbf {\bibinfo {volume} {149}},\ \bibinfo
  {pages} {174308} (\bibinfo {year} {2018})}\BibitemShut {NoStop}%
\bibitem [{\citenamefont {Sedmidubsk{\ifmmode\acute{a}\else\'{a}\fi}}\ and\
  \citenamefont
  {Ko{\ifmmode\check{c}\else\v{c}\fi}i{\ifmmode\check{s}\else\v{s}\fi}ek}(2024)}]{Sedmidubska_2024}%
  \BibitemOpen
  \bibfield  {author} {\bibinfo {author} {\bibfnamefont {B.}~\bibnamefont
  {Sedmidubsk{\ifmmode\acute{a}\else\'{a}\fi}}}\ and\ \bibinfo {author}
  {\bibfnamefont {J.}~\bibnamefont
  {Ko{\ifmmode\check{c}\else\v{c}\fi}i{\ifmmode\check{s}\else\v{s}\fi}ek}},\
  }\bibfield  {title} {\enquote {\bibinfo {title} {{Interaction of low-energy
  electrons with radiosensitizers}},}\ }\href
  {https://doi.org/10.1039/D3CP06003A} {\bibfield  {journal} {\bibinfo
  {journal} {Phys. Chem. Chem. Phys.}\ }\textbf {\bibinfo {volume} {26}},\
  \bibinfo {pages} {9112--9136} (\bibinfo {year} {2024})}\BibitemShut {NoStop}%
\bibitem [{\citenamefont {Wu}\ \emph {et~al.}(2024)\citenamefont {Wu},
  \citenamefont {Anderson}, \citenamefont {Watkins}, \citenamefont {Arora},
  \citenamefont {Barnes}, \citenamefont {Padovani}, \citenamefont
  {Shingledecker}, \citenamefont {Arumainayagam},\ and\ \citenamefont
  {Battat}}]{Wu_2024}%
  \BibitemOpen
  \bibfield  {author} {\bibinfo {author} {\bibfnamefont {Q.~T.}\ \bibnamefont
  {Wu}}, \bibinfo {author} {\bibfnamefont {H.}~\bibnamefont {Anderson}},
  \bibinfo {author} {\bibfnamefont {A.~K.}\ \bibnamefont {Watkins}}, \bibinfo
  {author} {\bibfnamefont {D.}~\bibnamefont {Arora}}, \bibinfo {author}
  {\bibfnamefont {K.}~\bibnamefont {Barnes}}, \bibinfo {author} {\bibfnamefont
  {M.}~\bibnamefont {Padovani}}, \bibinfo {author} {\bibfnamefont {C.~N.}\
  \bibnamefont {Shingledecker}}, \bibinfo {author} {\bibfnamefont {C.~R.}\
  \bibnamefont {Arumainayagam}},\ and\ \bibinfo {author} {\bibfnamefont
  {J.~B.~R.}\ \bibnamefont {Battat}},\ }\bibfield  {title} {\enquote {\bibinfo
  {title} {{Role of Low-Energy ({$<$}20 eV) Secondary Electrons in the
  Extraterrestrial Synthesis of Prebiotic Molecules}},}\ }\href
  {https://doi.org/10.1021/acsearthspacechem.3c00259} {\bibfield  {journal}
  {\bibinfo  {journal} {ACS Earth Space Chem.}\ }\textbf {\bibinfo {volume}
  {8}},\ \bibinfo {pages} {79--88} (\bibinfo {year} {2024})}\BibitemShut
  {NoStop}%
\bibitem [{\citenamefont {Arumainayagam}\ \emph {et~al.}(2010)\citenamefont
  {Arumainayagam}, \citenamefont {Lee}, \citenamefont {Nelson}, \citenamefont
  {Haines},\ and\ \citenamefont {Gunawardane}}]{Arumainayagam_2010}%
  \BibitemOpen
  \bibfield  {author} {\bibinfo {author} {\bibfnamefont {C.~R.}\ \bibnamefont
  {Arumainayagam}}, \bibinfo {author} {\bibfnamefont {H.-L.}\ \bibnamefont
  {Lee}}, \bibinfo {author} {\bibfnamefont {R.~B.}\ \bibnamefont {Nelson}},
  \bibinfo {author} {\bibfnamefont {D.~R.}\ \bibnamefont {Haines}},\ and\
  \bibinfo {author} {\bibfnamefont {R.~P.}\ \bibnamefont {Gunawardane}},\
  }\bibfield  {title} {\enquote {\bibinfo {title} {Low-energy electron-induced
  reactions in condensed matter},}\ }\href
  {https://doi.org/https://doi.org/10.1016/j.surfrep.2009.09.001} {\bibfield
  {journal} {\bibinfo  {journal} {Surf. Sci. Rep.}\ }\textbf {\bibinfo {volume}
  {65}},\ \bibinfo {pages} {1--44} (\bibinfo {year} {2010})}\BibitemShut
  {NoStop}%
\bibitem [{\citenamefont {Thorman}\ \emph {et~al.}(2015)\citenamefont
  {Thorman}, \citenamefont {P.}, \citenamefont {Fairbrother},\ and\
  \citenamefont {Ing{\'o}lfsson}}]{Thorman_2015}%
  \BibitemOpen
  \bibfield  {author} {\bibinfo {author} {\bibfnamefont {R.~M.}\ \bibnamefont
  {Thorman}}, \bibinfo {author} {\bibfnamefont {R.~K.~T.}\ \bibnamefont {P.}},
  \bibinfo {author} {\bibfnamefont {D.~H.}\ \bibnamefont {Fairbrother}},\ and\
  \bibinfo {author} {\bibfnamefont {O.}~\bibnamefont {Ing{\'o}lfsson}},\
  }\bibfield  {title} {\enquote {\bibinfo {title} {The role of low-energy
  electrons in focused electron beam induced deposition: four case studies of
  representative precursors},}\ }\href {https://doi.org/10.3762/bjnano.6.194}
  {\bibfield  {journal} {\bibinfo  {journal} {Beilstein J. Nanotechnol.}\
  }\textbf {\bibinfo {volume} {6}},\ \bibinfo {pages} {1904--1926} (\bibinfo
  {year} {2015})}\BibitemShut {NoStop}%
\bibitem [{\citenamefont {Jagau}, \citenamefont {Bravaya},\ and\ \citenamefont
  {Krylov}(2017)}]{Jagau_2017}%
  \BibitemOpen
  \bibfield  {author} {\bibinfo {author} {\bibfnamefont {T.-C.}\ \bibnamefont
  {Jagau}}, \bibinfo {author} {\bibfnamefont {K.~B.}\ \bibnamefont {Bravaya}},\
  and\ \bibinfo {author} {\bibfnamefont {A.~I.}\ \bibnamefont {Krylov}},\
  }\bibfield  {title} {\enquote {\bibinfo {title} {{Extending Quantum Chemistry
  of Bound States to Electronic Resonances}},}\ }\href
  {https://doi.org/10.1146/annurev-physchem-052516-050622} {\bibfield
  {journal} {\bibinfo  {journal} {Annu. Rev. Phys. Chem.}\ ,\ \bibinfo {pages}
  {525--553}} (\bibinfo {year} {2017})}\BibitemShut {NoStop}%
\bibitem [{\citenamefont {Jagau}(2022)}]{Jagau_2022}%
  \BibitemOpen
  \bibfield  {author} {\bibinfo {author} {\bibfnamefont {T.-C.}\ \bibnamefont
  {Jagau}},\ }\bibfield  {title} {\enquote {\bibinfo {title} {{Theory of
  electronic resonances: fundamental aspects and recent advances}},}\ }\href
  {https://doi.org/10.1039/D1CC07090H} {\bibfield  {journal} {\bibinfo
  {journal} {Chem. Commun.}\ }\textbf {\bibinfo {volume} {58}},\ \bibinfo
  {pages} {5205--5224} (\bibinfo {year} {2022})}\BibitemShut {NoStop}%
\bibitem [{\citenamefont {Jolicard}\ and\ \citenamefont
  {Austin}(1985)}]{Jolicard_1985}%
  \BibitemOpen
  \bibfield  {author} {\bibinfo {author} {\bibfnamefont {G.}~\bibnamefont
  {Jolicard}}\ and\ \bibinfo {author} {\bibfnamefont {E.~J.}\ \bibnamefont
  {Austin}},\ }\bibfield  {title} {\enquote {\bibinfo {title} {{Optical
  potential stabilisation method for predicting resonance levels}},}\ }\href
  {https://doi.org/10.1016/0009-2614(85)87164-5} {\bibfield  {journal}
  {\bibinfo  {journal} {Chem. Phys. Lett.}\ }\textbf {\bibinfo {volume}
  {121}},\ \bibinfo {pages} {106--110} (\bibinfo {year} {1985})}\BibitemShut
  {NoStop}%
\bibitem [{\citenamefont {Riss}\ and\ \citenamefont {Meyer}(1993)}]{Riss_1993}%
  \BibitemOpen
  \bibfield  {author} {\bibinfo {author} {\bibfnamefont {U.~V.}\ \bibnamefont
  {Riss}}\ and\ \bibinfo {author} {\bibfnamefont {H.-D.}\ \bibnamefont
  {Meyer}},\ }\bibfield  {title} {\enquote {\bibinfo {title} {{Calculation of
  resonance energies and widths using the complex absorbing potential
  method}},}\ }\href {https://doi.org/10.1088/0953-4075/26/23/021} {\bibfield
  {journal} {\bibinfo  {journal} {J. Phys. B: At. Mol. Opt. Phys.}\ }\textbf
  {\bibinfo {volume} {26}},\ \bibinfo {pages} {4503} (\bibinfo {year}
  {1993})}\BibitemShut {NoStop}%
\bibitem [{\citenamefont {Sommerfeld}\ \emph {et~al.}(1998)\citenamefont
  {Sommerfeld}, \citenamefont {Riss}, \citenamefont {Meyer}, \citenamefont
  {Cederbaum}, \citenamefont {Engels},\ and\ \citenamefont
  {Suter}}]{Sommerfeld_1998}%
  \BibitemOpen
  \bibfield  {author} {\bibinfo {author} {\bibfnamefont {T.}~\bibnamefont
  {Sommerfeld}}, \bibinfo {author} {\bibfnamefont {U.~V.}\ \bibnamefont
  {Riss}}, \bibinfo {author} {\bibfnamefont {H.-D.}\ \bibnamefont {Meyer}},
  \bibinfo {author} {\bibfnamefont {L.~S.}\ \bibnamefont {Cederbaum}}, \bibinfo
  {author} {\bibfnamefont {B.}~\bibnamefont {Engels}},\ and\ \bibinfo {author}
  {\bibfnamefont {H.~U.}\ \bibnamefont {Suter}},\ }\bibfield  {title} {\enquote
  {\bibinfo {title} {{Temporary anions - calculation of energy and lifetime by
  absorbing potentials: the resonance}},}\ }\href
  {https://doi.org/10.1088/0953-4075/31/18/009} {\bibfield  {journal} {\bibinfo
   {journal} {J. Phys. B: At. Mol. Opt. Phys.}\ }\textbf {\bibinfo {volume}
  {31}},\ \bibinfo {pages} {4107} (\bibinfo {year} {1998})}\BibitemShut
  {NoStop}%
\bibitem [{\citenamefont {Ghosh}, \citenamefont {Vaval},\ and\ \citenamefont
  {Pal}(2012)}]{Ghosh_2012}%
  \BibitemOpen
  \bibfield  {author} {\bibinfo {author} {\bibfnamefont {A.}~\bibnamefont
  {Ghosh}}, \bibinfo {author} {\bibfnamefont {N.}~\bibnamefont {Vaval}},\ and\
  \bibinfo {author} {\bibfnamefont {S.}~\bibnamefont {Pal}},\ }\bibfield
  {title} {\enquote {\bibinfo {title} {{Equation-of-motion coupled-cluster
  method for the study of shape resonance}},}\ }\href
  {https://doi.org/10.1063/1.4729464} {\bibfield  {journal} {\bibinfo
  {journal} {J. Chem. Phys.}\ }\textbf {\bibinfo {volume} {136}},\ \bibinfo
  {pages} {234110} (\bibinfo {year} {2012})}\BibitemShut {NoStop}%
\bibitem [{\citenamefont {Zuev}\ \emph {et~al.}(2014)\citenamefont {Zuev},
  \citenamefont {Jagau}, \citenamefont {Bravaya}, \citenamefont {Epifanovsky},
  \citenamefont {Shao}, \citenamefont {Sundstrom}, \citenamefont
  {Head-Gordon},\ and\ \citenamefont {Krylov}}]{Zuev_2014}%
  \BibitemOpen
  \bibfield  {author} {\bibinfo {author} {\bibfnamefont {D.}~\bibnamefont
  {Zuev}}, \bibinfo {author} {\bibfnamefont {T.-C.}\ \bibnamefont {Jagau}},
  \bibinfo {author} {\bibfnamefont {K.~B.}\ \bibnamefont {Bravaya}}, \bibinfo
  {author} {\bibfnamefont {E.}~\bibnamefont {Epifanovsky}}, \bibinfo {author}
  {\bibfnamefont {Y.}~\bibnamefont {Shao}}, \bibinfo {author} {\bibfnamefont
  {E.}~\bibnamefont {Sundstrom}}, \bibinfo {author} {\bibfnamefont
  {M.}~\bibnamefont {Head-Gordon}},\ and\ \bibinfo {author} {\bibfnamefont
  {A.~I.}\ \bibnamefont {Krylov}},\ }\bibfield  {title} {\enquote {\bibinfo
  {title} {{Complex Absorbing Potentials Within EOM-CC Family of Methods:
  Theory, Implementation, and Benchmarks}},}\ }\href
  {https://doi.org/10.1063/1.4885056} {\bibfield  {journal} {\bibinfo
  {journal} {J. Chem. Phys.}\ }\textbf {\bibinfo {volume} {141}},\ \bibinfo
  {pages} {024102} (\bibinfo {year} {2014})}\BibitemShut {NoStop}%
\bibitem [{\citenamefont {Jagau}\ \emph {et~al.}(2014)\citenamefont {Jagau},
  \citenamefont {Zuev}, \citenamefont {Bravaya}, \citenamefont {Epifanovsky},\
  and\ \citenamefont {Krylov}}]{Jagau_2014a}%
  \BibitemOpen
  \bibfield  {author} {\bibinfo {author} {\bibfnamefont {T.-C.}\ \bibnamefont
  {Jagau}}, \bibinfo {author} {\bibfnamefont {D.}~\bibnamefont {Zuev}},
  \bibinfo {author} {\bibfnamefont {K.~B.}\ \bibnamefont {Bravaya}}, \bibinfo
  {author} {\bibfnamefont {E.}~\bibnamefont {Epifanovsky}},\ and\ \bibinfo
  {author} {\bibfnamefont {A.~I.}\ \bibnamefont {Krylov}},\ }\bibfield  {title}
  {\enquote {\bibinfo {title} {{A Fresh Look at Resonances and Complex
  Absorbing Potentials: Density Matrix-Based Approach}},}\ }\href
  {https://doi.org/10.1021/jz402482a} {\bibfield  {journal} {\bibinfo
  {journal} {J. Phys. Chem. Lett.}\ }\textbf {\bibinfo {volume} {5}},\ \bibinfo
  {pages} {310--315} (\bibinfo {year} {2014})}\BibitemShut {NoStop}%
\bibitem [{\citenamefont {Jagau}\ and\ \citenamefont
  {Krylov}(2014)}]{Jagau_2014b}%
  \BibitemOpen
  \bibfield  {author} {\bibinfo {author} {\bibfnamefont {T.-C.}\ \bibnamefont
  {Jagau}}\ and\ \bibinfo {author} {\bibfnamefont {A.~I.}\ \bibnamefont
  {Krylov}},\ }\bibfield  {title} {\enquote {\bibinfo {title} {{Complex
  Absorbing Potential Equation-of-Motion Coupled-Cluster Method Yields Smooth
  and Internally Consistent Potential Energy Surfaces and Lifetimes for
  Molecular Resonances}},}\ }\href {https://doi.org/10.1021/jz501515j}
  {\bibfield  {journal} {\bibinfo  {journal} {J. Phys. Chem. Lett.}\ }\textbf
  {\bibinfo {volume} {5}},\ \bibinfo {pages} {3078--3085} (\bibinfo {year}
  {2014})}\BibitemShut {NoStop}%
\bibitem [{\citenamefont {Sommerfeld}\ and\ \citenamefont
  {Ehara}(2015)}]{Sommerfeld_2015}%
  \BibitemOpen
  \bibfield  {author} {\bibinfo {author} {\bibfnamefont {T.}~\bibnamefont
  {Sommerfeld}}\ and\ \bibinfo {author} {\bibfnamefont {M.}~\bibnamefont
  {Ehara}},\ }\bibfield  {title} {\enquote {\bibinfo {title} {Complex absorbing
  potentials with voronoi isosurfaces wrapping perfectly around molecules},}\
  }\href {https://doi.org/10.1021/acs.jctc.5b00465} {\bibfield  {journal}
  {\bibinfo  {journal} {J. Chem. Theory Comput.}\ }\textbf {\bibinfo {volume}
  {11}},\ \bibinfo {pages} {4627--4633} (\bibinfo {year} {2015})}\BibitemShut
  {NoStop}%
\bibitem [{\citenamefont {Gyamfi}\ and\ \citenamefont
  {Jagau}(2024)}]{Gyamfi_2024}%
  \BibitemOpen
  \bibfield  {author} {\bibinfo {author} {\bibfnamefont {J.~A.}\ \bibnamefont
  {Gyamfi}}\ and\ \bibinfo {author} {\bibfnamefont {T.-C.}\ \bibnamefont
  {Jagau}},\ }\bibfield  {title} {\enquote {\bibinfo {title} {{A New Strategy
  to Optimize Complex Absorbing Potentials for the Computation of Resonance
  Energies and Widths}},}\ }\href {https://doi.org/10.1021/acs.jctc.3c01039}
  {\bibfield  {journal} {\bibinfo  {journal} {J. Chem. Theory Comput.}\
  }\textbf {\bibinfo {volume} {20}},\ \bibinfo {pages} {1096--1107} (\bibinfo
  {year} {2024})}\BibitemShut {NoStop}%
\bibitem [{\citenamefont {Balslev}\ and\ \citenamefont
  {Combes}(1971)}]{Balslev_1971}%
  \BibitemOpen
  \bibfield  {author} {\bibinfo {author} {\bibfnamefont {E.}~\bibnamefont
  {Balslev}}\ and\ \bibinfo {author} {\bibfnamefont {J.~M.}\ \bibnamefont
  {Combes}},\ }\bibfield  {title} {\enquote {\bibinfo {title} {{Spectral
  properties of many-body Schr{\ifmmode\ddot{o}\else\"{o}\fi}dinger operators
  with dilatation-analytic interactions}},}\ }\href
  {https://doi.org/10.1007/BF01877511} {\bibfield  {journal} {\bibinfo
  {journal} {Commun. Math. Phys.}\ }\textbf {\bibinfo {volume} {22}},\ \bibinfo
  {pages} {280--294} (\bibinfo {year} {1971})}\BibitemShut {NoStop}%
\bibitem [{\citenamefont {Moiseyev}(1998)}]{Moiseyev_1998}%
  \BibitemOpen
  \bibfield  {author} {\bibinfo {author} {\bibfnamefont {N.}~\bibnamefont
  {Moiseyev}},\ }\bibfield  {title} {\enquote {\bibinfo {title} {{Quantum
  theory of resonances: calculating energies, widths and cross-sections by
  complex scaling}},}\ }\href {https://doi.org/10.1016/S0370-1573(98)00002-7}
  {\bibfield  {journal} {\bibinfo  {journal} {Phys. Rep.}\ }\textbf {\bibinfo
  {volume} {302}},\ \bibinfo {pages} {212--293} (\bibinfo {year}
  {1998})}\BibitemShut {NoStop}%
\bibitem [{\citenamefont {McCurdy}\ and\ \citenamefont
  {Rescigno}(1978)}]{McCurdy_1978}%
  \BibitemOpen
  \bibfield  {author} {\bibinfo {author} {\bibfnamefont {C.~W.}\ \bibnamefont
  {McCurdy}}\ and\ \bibinfo {author} {\bibfnamefont {T.~N.}\ \bibnamefont
  {Rescigno}},\ }\bibfield  {title} {\enquote {\bibinfo {title} {Extension of
  the method of complex basis functions to molecular resonances},}\ }\href
  {https://doi.org/10.1103/PhysRevLett.41.1364} {\bibfield  {journal} {\bibinfo
   {journal} {Phys. Rev. Lett.}\ }\textbf {\bibinfo {volume} {41}},\ \bibinfo
  {pages} {1364--1368} (\bibinfo {year} {1978})}\BibitemShut {NoStop}%
\bibitem [{\citenamefont {White}, \citenamefont {Head-Gordon},\ and\
  \citenamefont {McCurdy}(2015)}]{White_2015}%
  \BibitemOpen
  \bibfield  {author} {\bibinfo {author} {\bibfnamefont {A.~F.}\ \bibnamefont
  {White}}, \bibinfo {author} {\bibfnamefont {M.}~\bibnamefont {Head-Gordon}},\
  and\ \bibinfo {author} {\bibfnamefont {C.~W.}\ \bibnamefont {McCurdy}},\
  }\bibfield  {title} {\enquote {\bibinfo {title} {{Complex basis functions
  revisited: Implementation with applications to carbon tetrafluoride and
  aromatic N-containing heterocycles within the static-exchange
  approximation}},}\ }\href {https://doi.org/10.1063/1.4906940} {\bibfield
  {journal} {\bibinfo  {journal} {J. Chem. Phys.}\ }\textbf {\bibinfo {volume}
  {142}},\ \bibinfo {pages} {054103} (\bibinfo {year} {2015})}\BibitemShut
  {NoStop}%
\bibitem [{\citenamefont {Moiseyev}(2011)}]{Moiseyev_2011}%
  \BibitemOpen
  \bibfield  {author} {\bibinfo {author} {\bibfnamefont {N.}~\bibnamefont
  {Moiseyev}},\ }\href {https://doi.org/10.1017/CBO9780511976186} {\emph
  {\bibinfo {title} {{Non-Hermitian Quantum Mechanics}}}}\ (\bibinfo
  {publisher} {Cambridge University Press},\ \bibinfo {address} {Cambridge,
  England, UK},\ \bibinfo {year} {2011})\BibitemShut {NoStop}%
\bibitem [{\citenamefont {Damour}\ \emph {et~al.}(2024)\citenamefont {Damour},
  \citenamefont {Scemama}, \citenamefont {Kossoski},\ and\ \citenamefont
  {Loos}}]{Damour_2024}%
  \BibitemOpen
  \bibfield  {author} {\bibinfo {author} {\bibfnamefont {Y.}~\bibnamefont
  {Damour}}, \bibinfo {author} {\bibfnamefont {A.}~\bibnamefont {Scemama}},
  \bibinfo {author} {\bibfnamefont {F.}~\bibnamefont {Kossoski}},\ and\
  \bibinfo {author} {\bibfnamefont {P.-F.}\ \bibnamefont {Loos}},\ }\bibfield
  {title} {\enquote {\bibinfo {title} {{Selected Configuration Interaction for
  Resonances}},}\ }\href {https://doi.org/10.1021/acs.jpclett.4c02060}
  {\bibfield  {journal} {\bibinfo  {journal} {J. Phys. Chem. Lett.}\ }\textbf
  {\bibinfo {volume} {15}},\ \bibinfo {pages} {8296--8305} (\bibinfo {year}
  {2024})}\BibitemShut {NoStop}%
\bibitem [{\citenamefont {Ghosh}\ \emph {et~al.}(2013)\citenamefont {Ghosh},
  \citenamefont {Karne}, \citenamefont {Pal},\ and\ \citenamefont
  {Vaval}}]{Ghosh_2013}%
  \BibitemOpen
  \bibfield  {author} {\bibinfo {author} {\bibfnamefont {A.}~\bibnamefont
  {Ghosh}}, \bibinfo {author} {\bibfnamefont {A.}~\bibnamefont {Karne}},
  \bibinfo {author} {\bibfnamefont {S.}~\bibnamefont {Pal}},\ and\ \bibinfo
  {author} {\bibfnamefont {N.}~\bibnamefont {Vaval}},\ }\bibfield  {title}
  {\enquote {\bibinfo {title} {{CAP/EOM-CCSD method for the study of potential
  curves of resonant states}},}\ }\href {https://doi.org/10.1039/C3CP52552J}
  {\bibfield  {journal} {\bibinfo  {journal} {Phys. Chem. Chem. Phys.}\
  }\textbf {\bibinfo {volume} {15}},\ \bibinfo {pages} {17915--17921} (\bibinfo
  {year} {2013})}\BibitemShut {NoStop}%
\bibitem [{\citenamefont {Sajeev}, \citenamefont {Santra},\ and\ \citenamefont
  {Pal}(2005)}]{Sajeev_2005a}%
  \BibitemOpen
  \bibfield  {author} {\bibinfo {author} {\bibfnamefont {Y.}~\bibnamefont
  {Sajeev}}, \bibinfo {author} {\bibfnamefont {R.}~\bibnamefont {Santra}},\
  and\ \bibinfo {author} {\bibfnamefont {S.}~\bibnamefont {Pal}},\ }\bibfield
  {title} {\enquote {\bibinfo {title} {{Analytically continued Fock space
  multireference coupled-cluster theory: Application to the {$\Pi$}g2 shape
  resonance in e-N2 scattering}},}\ }\href {https://doi.org/10.1063/1.1938887}
  {\bibfield  {journal} {\bibinfo  {journal} {J. Chem. Phys.}\ }\textbf
  {\bibinfo {volume} {122}},\ \bibinfo {pages} {234320} (\bibinfo {year}
  {2005})}\BibitemShut {NoStop}%
\bibitem [{\citenamefont {Sajeev}\ and\ \citenamefont
  {Pal}(2005)}]{Sajeev_2005b}%
  \BibitemOpen
  \bibfield  {author} {\bibinfo {author} {\bibfnamefont {Y.}~\bibnamefont
  {Sajeev}}\ and\ \bibinfo {author} {\bibfnamefont {S.}~\bibnamefont {Pal}},\
  }\bibfield  {title} {\enquote {\bibinfo {title} {{A general formalism of the
  Fock space multireference coupled cluster method for investigating molecular
  electronic resonances}},}\ }\href
  {https://www.tandfonline.com/doi/abs/10.1080/00268970500084158} {\bibfield
  {journal} {\bibinfo  {journal} {Mol. Phys.}\ }\textbf {\bibinfo {volume}
  {103}},\ \bibinfo {pages} {2267--2275} (\bibinfo {year} {2005})}\BibitemShut
  {NoStop}%
\bibitem [{\citenamefont {Jagau}(2018)}]{Jagau_2018}%
  \BibitemOpen
  \bibfield  {author} {\bibinfo {author} {\bibfnamefont {T.-C.}\ \bibnamefont
  {Jagau}},\ }\bibfield  {title} {\enquote {\bibinfo {title} {{Non-iterative
  triple excitations in equation-of-motion coupled-cluster theory for electron
  attachment with applications to bound and temporary anions}},}\ }\href
  {https://doi.org/10.1063/1.5006374} {\bibfield  {journal} {\bibinfo
  {journal} {J. Chem. Phys.}\ }\textbf {\bibinfo {volume} {148}},\ \bibinfo
  {pages} {024104} (\bibinfo {year} {2018})}\BibitemShut {NoStop}%
\bibitem [{\citenamefont {Jana}\ \emph {et~al.}(2021)\citenamefont {Jana},
  \citenamefont {Basumallick}, \citenamefont {Pal},\ and\ \citenamefont
  {Vaval}}]{Jana_2021}%
  \BibitemOpen
  \bibfield  {author} {\bibinfo {author} {\bibfnamefont {I.}~\bibnamefont
  {Jana}}, \bibinfo {author} {\bibfnamefont {S.}~\bibnamefont {Basumallick}},
  \bibinfo {author} {\bibfnamefont {S.}~\bibnamefont {Pal}},\ and\ \bibinfo
  {author} {\bibfnamefont {N.}~\bibnamefont {Vaval}},\ }\bibfield  {title}
  {\enquote {\bibinfo {title} {{Resonance study: Effect of partial triples
  excitation using complex absorbing potential-based Fock-space multi-reference
  coupled cluster}},}\ }\href
  {https://doi.org/https://doi.org/10.1002/qua.26738} {\bibfield  {journal}
  {\bibinfo  {journal} {Int. J. Quantum Chem.}\ }\textbf {\bibinfo {volume}
  {121}},\ \bibinfo {pages} {e26738} (\bibinfo {year} {2021})}\BibitemShut
  {NoStop}%
\bibitem [{\citenamefont {Ehara}\ and\ \citenamefont
  {Sommerfeld}(2012)}]{Ehara_2012}%
  \BibitemOpen
  \bibfield  {author} {\bibinfo {author} {\bibfnamefont {M.}~\bibnamefont
  {Ehara}}\ and\ \bibinfo {author} {\bibfnamefont {T.}~\bibnamefont
  {Sommerfeld}},\ }\bibfield  {title} {\enquote {\bibinfo {title} {{CAP/SAC-CI
  method for calculating resonance states of metastable anions}},}\ }\href
  {https://doi.org/10.1016/j.cplett.2012.03.104} {\bibfield  {journal}
  {\bibinfo  {journal} {Chem. Phys. Lett.}\ }\textbf {\bibinfo {volume}
  {537}},\ \bibinfo {pages} {107--112} (\bibinfo {year} {2012})}\BibitemShut
  {NoStop}%
\bibitem [{\citenamefont {Phung}\ \emph {et~al.}(2020)\citenamefont {Phung},
  \citenamefont {Komori}, \citenamefont {Yanai}, \citenamefont {Sommerfeld},\
  and\ \citenamefont {Ehara}}]{Phung_2020}%
  \BibitemOpen
  \bibfield  {author} {\bibinfo {author} {\bibfnamefont {Q.~M.}\ \bibnamefont
  {Phung}}, \bibinfo {author} {\bibfnamefont {Y.}~\bibnamefont {Komori}},
  \bibinfo {author} {\bibfnamefont {T.}~\bibnamefont {Yanai}}, \bibinfo
  {author} {\bibfnamefont {T.}~\bibnamefont {Sommerfeld}},\ and\ \bibinfo
  {author} {\bibfnamefont {M.}~\bibnamefont {Ehara}},\ }\bibfield  {title}
  {\enquote {\bibinfo {title} {{Combination of a Voronoi-Type Complex Absorbing
  Potential with the XMS-CASPT2 Method and Pilot Applications}},}\ }\href
  {https://doi.org/10.1021/acs.jctc.9b01032} {\bibfield  {journal} {\bibinfo
  {journal} {J. Chem. Theory Comput.}\ }\textbf {\bibinfo {volume} {16}},\
  \bibinfo {pages} {2606--2616} (\bibinfo {year} {2020})}\BibitemShut {NoStop}%
\bibitem [{\citenamefont {Sommerfeld}\ and\ \citenamefont
  {Santra}(2001)}]{Sommerfeld_2001}%
  \BibitemOpen
  \bibfield  {author} {\bibinfo {author} {\bibfnamefont {T.}~\bibnamefont
  {Sommerfeld}}\ and\ \bibinfo {author} {\bibfnamefont {R.}~\bibnamefont
  {Santra}},\ }\bibfield  {title} {\enquote {\bibinfo {title} {{Efficient
  method to perform CAP/CI calculations for temporary anions}},}\ }\href
  {https://doi.org/10.1002/qua.1042} {\bibfield  {journal} {\bibinfo  {journal}
  {Int. J. Quantum Chem.}\ }\textbf {\bibinfo {volume} {82}},\ \bibinfo {pages}
  {218--226} (\bibinfo {year} {2001})}\BibitemShut {NoStop}%
\bibitem [{\citenamefont {Zhou}\ and\ \citenamefont
  {Ernzerhof}(2012)}]{Zhou_2012}%
  \BibitemOpen
  \bibfield  {author} {\bibinfo {author} {\bibfnamefont {Y.}~\bibnamefont
  {Zhou}}\ and\ \bibinfo {author} {\bibfnamefont {M.}~\bibnamefont
  {Ernzerhof}},\ }\bibfield  {title} {\enquote {\bibinfo {title} {Calculating
  the lifetimes of metastable states with complex density functional theory},}\
  }\href {https://doi.org/10.1021/jz3006805} {\bibfield  {journal} {\bibinfo
  {journal} {J. Phys. Chem. Lett.}\ }\textbf {\bibinfo {volume} {3}},\ \bibinfo
  {pages} {1916--1920} (\bibinfo {year} {2012})}\BibitemShut {NoStop}%
\bibitem [{\citenamefont {Santra}\ and\ \citenamefont
  {Cederbaum}(2002)}]{Santra_2002}%
  \BibitemOpen
  \bibfield  {author} {\bibinfo {author} {\bibfnamefont {R.}~\bibnamefont
  {Santra}}\ and\ \bibinfo {author} {\bibfnamefont {L.~S.}\ \bibnamefont
  {Cederbaum}},\ }\bibfield  {title} {\enquote {\bibinfo {title} {{Complex
  absorbing potentials in the framework of electron propagator theory. I.
  General formalism}},}\ }\href {https://doi.org/10.1063/1.1501903} {\bibfield
  {journal} {\bibinfo  {journal} {J. Chem. Phys.}\ }\textbf {\bibinfo {volume}
  {117}},\ \bibinfo {pages} {5511--5521} (\bibinfo {year} {2002})}\BibitemShut
  {NoStop}%
\bibitem [{\citenamefont {Feuerbacher}\ \emph {et~al.}(2003)\citenamefont
  {Feuerbacher}, \citenamefont {Sommerfeld}, \citenamefont {Santra},\ and\
  \citenamefont {Cederbaum}}]{Feuerbacher_2003}%
  \BibitemOpen
  \bibfield  {author} {\bibinfo {author} {\bibfnamefont {S.}~\bibnamefont
  {Feuerbacher}}, \bibinfo {author} {\bibfnamefont {T.}~\bibnamefont
  {Sommerfeld}}, \bibinfo {author} {\bibfnamefont {R.}~\bibnamefont {Santra}},\
  and\ \bibinfo {author} {\bibfnamefont {L.~S.}\ \bibnamefont {Cederbaum}},\
  }\bibfield  {title} {\enquote {\bibinfo {title} {{Complex absorbing
  potentials in the framework of electron propagator theory. II. Application to
  temporary anions}},}\ }\href {https://doi.org/10.1063/1.1557452} {\bibfield
  {journal} {\bibinfo  {journal} {J. Chem. Phys.}\ }\textbf {\bibinfo {volume}
  {118}},\ \bibinfo {pages} {6188--6199} (\bibinfo {year} {2003})}\BibitemShut
  {NoStop}%
\bibitem [{\citenamefont {Feuerbacher}\ and\ \citenamefont
  {Santra}(2005)}]{Feuerbacher_2005}%
  \BibitemOpen
  \bibfield  {author} {\bibinfo {author} {\bibfnamefont {S.}~\bibnamefont
  {Feuerbacher}}\ and\ \bibinfo {author} {\bibfnamefont {R.}~\bibnamefont
  {Santra}},\ }\bibfield  {title} {\enquote {\bibinfo {title} {{Calculating
  molecular Rydberg states using the one-particle Green's function: Application
  to HCO and C(NH2)3}},}\ }\href {https://doi.org/10.1063/1.2122687} {\bibfield
   {journal} {\bibinfo  {journal} {J. Chem. Phys.}\ }\textbf {\bibinfo {volume}
  {123}},\ \bibinfo {pages} {194310} (\bibinfo {year} {2005})}\BibitemShut
  {NoStop}%
\bibitem [{\citenamefont {Belogolova}\ \emph {et~al.}(2021)\citenamefont
  {Belogolova}, \citenamefont {Dempwolff}, \citenamefont {Dreuw},\ and\
  \citenamefont {Trofimov}}]{Belogolova_2021}%
  \BibitemOpen
  \bibfield  {author} {\bibinfo {author} {\bibfnamefont {A.~M.}\ \bibnamefont
  {Belogolova}}, \bibinfo {author} {\bibfnamefont {A.~L.}\ \bibnamefont
  {Dempwolff}}, \bibinfo {author} {\bibfnamefont {A.}~\bibnamefont {Dreuw}},\
  and\ \bibinfo {author} {\bibfnamefont {A.~B.}\ \bibnamefont {Trofimov}},\
  }\bibfield  {title} {\enquote {\bibinfo {title} {{A complex absorbing
  potential electron propagator approach to resonance states of metastable
  anions}},}\ }\href {https://doi.org/10.1088/1742-6596/1847/1/012050}
  {\bibfield  {journal} {\bibinfo  {journal} {J. Phys. Conf. Ser.}\ }\textbf
  {\bibinfo {volume} {1847}},\ \bibinfo {pages} {012050} (\bibinfo {year}
  {2021})}\BibitemShut {NoStop}%
\bibitem [{\citenamefont {Schirmer}, \citenamefont {Trofimov},\ and\
  \citenamefont {Stelter}(1998)}]{Schirmer_1998}%
  \BibitemOpen
  \bibfield  {author} {\bibinfo {author} {\bibfnamefont {J.}~\bibnamefont
  {Schirmer}}, \bibinfo {author} {\bibfnamefont {A.~B.}\ \bibnamefont
  {Trofimov}},\ and\ \bibinfo {author} {\bibfnamefont {G.}~\bibnamefont
  {Stelter}},\ }\bibfield  {title} {\enquote {\bibinfo {title} {{A non-Dyson
  third-order approximation scheme for the electron propagator}},}\ }\href
  {https://doi.org/10.1063/1.477085} {\bibfield  {journal} {\bibinfo  {journal}
  {J. Chem. Phys.}\ }\textbf {\bibinfo {volume} {109}},\ \bibinfo {pages}
  {4734--4744} (\bibinfo {year} {1998})}\BibitemShut {NoStop}%
\bibitem [{\citenamefont {Schirmer}(2018)}]{Schirmer_2018}%
  \BibitemOpen
  \bibfield  {author} {\bibinfo {author} {\bibfnamefont {J.}~\bibnamefont
  {Schirmer}},\ }\href@noop {} {\emph {\bibinfo {title} {Many-Body Methods for
  Atoms, Molecules and Clusters}}}\ (\bibinfo  {publisher} {Springer},\
  \bibinfo {year} {2018})\BibitemShut {NoStop}%
\bibitem [{\citenamefont {Martin}, \citenamefont {Reining},\ and\ \citenamefont
  {Ceperley}(2016)}]{MartinBook}%
  \BibitemOpen
  \bibfield  {author} {\bibinfo {author} {\bibfnamefont {R.~M.}\ \bibnamefont
  {Martin}}, \bibinfo {author} {\bibfnamefont {L.}~\bibnamefont {Reining}},\
  and\ \bibinfo {author} {\bibfnamefont {D.~M.}\ \bibnamefont {Ceperley}},\
  }\href@noop {} {\emph {\bibinfo {title} {Interacting Electrons: Theory and
  Computational Approaches}}}\ (\bibinfo  {publisher} {Cambridge University
  Press},\ \bibinfo {year} {2016})\BibitemShut {NoStop}%
\bibitem [{\citenamefont {Csanak}, \citenamefont {Taylor},\ and\ \citenamefont
  {Yaris}(1971)}]{CsanakBook}%
  \BibitemOpen
  \bibfield  {author} {\bibinfo {author} {\bibfnamefont {G.}~\bibnamefont
  {Csanak}}, \bibinfo {author} {\bibfnamefont {H.}~\bibnamefont {Taylor}},\
  and\ \bibinfo {author} {\bibfnamefont {R.}~\bibnamefont {Yaris}},\ }\bibfield
   {title} {\enquote {\bibinfo {title} {Green's function technique in atomic
  and molecular physics},}\ }in\ \href@noop {} {\emph {\bibinfo {booktitle}
  {Advances in atomic and molecular physics}}},\ Vol.~\bibinfo {volume} {7}\
  (\bibinfo  {publisher} {Elsevier},\ \bibinfo {year} {1971})\ pp.\ \bibinfo
  {pages} {287--361}\BibitemShut {NoStop}%
\bibitem [{\citenamefont {Fetter}\ and\ \citenamefont
  {Waleck}(1971)}]{FetterBook}%
  \BibitemOpen
  \bibfield  {author} {\bibinfo {author} {\bibfnamefont {A.~L.}\ \bibnamefont
  {Fetter}}\ and\ \bibinfo {author} {\bibfnamefont {J.~D.}\ \bibnamefont
  {Waleck}},\ }\href@noop {} {\emph {\bibinfo {title} {Quantum Theory of Many
  Particle Systems}}}\ (\bibinfo  {publisher} {McGraw Hill, San Francisco},\
  \bibinfo {year} {1971})\BibitemShut {NoStop}%
\bibitem [{\citenamefont {Hedin}(1965)}]{Hedin_1965}%
  \BibitemOpen
  \bibfield  {author} {\bibinfo {author} {\bibfnamefont {L.}~\bibnamefont
  {Hedin}},\ }\bibfield  {title} {\enquote {\bibinfo {title} {New method for
  calculating the one-particle {{Green}}'s function with application to the
  electron-gas problem},}\ }\href {https://doi.org/10.1103/PhysRev.139.A796}
  {\bibfield  {journal} {\bibinfo  {journal} {Phys. Rev.}\ }\textbf {\bibinfo
  {volume} {139}},\ \bibinfo {pages} {A796} (\bibinfo {year}
  {1965})}\BibitemShut {NoStop}%
\bibitem [{\citenamefont {Golze}, \citenamefont {Dvorak},\ and\ \citenamefont
  {Rinke}(2019)}]{Golze_2019}%
  \BibitemOpen
  \bibfield  {author} {\bibinfo {author} {\bibfnamefont {D.}~\bibnamefont
  {Golze}}, \bibinfo {author} {\bibfnamefont {M.}~\bibnamefont {Dvorak}},\ and\
  \bibinfo {author} {\bibfnamefont {P.}~\bibnamefont {Rinke}},\ }\bibfield
  {title} {\enquote {\bibinfo {title} {{The $GW$ Compendium: A Practical Guide
  to Theoretical Photoemission Spectroscopy}},}\ }\href
  {https://doi.org/10.3389/fchem.2019.00377} {\bibfield  {journal} {\bibinfo
  {journal} {Front. Chem.}\ }\textbf {\bibinfo {volume} {7}},\ \bibinfo {pages}
  {377} (\bibinfo {year} {2019})}\BibitemShut {NoStop}%
\bibitem [{\citenamefont {Marie}, \citenamefont {Ammar},\ and\ \citenamefont
  {Loos}(2024)}]{Marie_2024a}%
  \BibitemOpen
  \bibfield  {author} {\bibinfo {author} {\bibfnamefont {A.}~\bibnamefont
  {Marie}}, \bibinfo {author} {\bibfnamefont {A.}~\bibnamefont {Ammar}},\ and\
  \bibinfo {author} {\bibfnamefont {P.-F.}\ \bibnamefont {Loos}},\ }\bibfield
  {title} {\enquote {\bibinfo {title} {The {{$GW$}} approximation: {{A}}
  quantum chemistry perspective},}\ }in\ \href
  {https://doi.org/10.1016/bs.aiq.2024.04.001} {\emph {\bibinfo {booktitle}
  {Advances in {{Quantum Chemistry}}}}},\ \bibinfo {series} {Novel
  {{Treatments}} of {{Strong Correlations}}}, Vol.~\bibinfo {volume} {90}\
  (\bibinfo {year} {2024})\ pp.\ \bibinfo {pages} {157--184}\BibitemShut
  {NoStop}%
\bibitem [{\citenamefont {Falcetta}\ \emph {et~al.}(2014)\citenamefont
  {Falcetta}, \citenamefont {DiFalco}, \citenamefont {Ackerman}, \citenamefont
  {Barlow},\ and\ \citenamefont {Jordan}}]{Falcetta_2014}%
  \BibitemOpen
  \bibfield  {author} {\bibinfo {author} {\bibfnamefont {M.~F.}\ \bibnamefont
  {Falcetta}}, \bibinfo {author} {\bibfnamefont {L.~A.}\ \bibnamefont
  {DiFalco}}, \bibinfo {author} {\bibfnamefont {D.~S.}\ \bibnamefont
  {Ackerman}}, \bibinfo {author} {\bibfnamefont {J.~C.}\ \bibnamefont
  {Barlow}},\ and\ \bibinfo {author} {\bibfnamefont {K.~D.}\ \bibnamefont
  {Jordan}},\ }\bibfield  {title} {\enquote {\bibinfo {title} {Assessment of
  various electronic structure methods for characterizing temporary anion
  states: Application to the ground state anions of n2, c2h2, c2h4, and
  c6h6},}\ }\href {https://doi.org/10.1021/jp5003287} {\bibfield  {journal}
  {\bibinfo  {journal} {J. Phys. Chem. A}\ }\textbf {\bibinfo {volume} {118}},\
  \bibinfo {pages} {7489--7497} (\bibinfo {year} {2014})}\BibitemShut {NoStop}%
\bibitem [{\citenamefont {{van Setten}}\ \emph {et~al.}(2015)\citenamefont
  {{van Setten}}, \citenamefont {Caruso}, \citenamefont {Sharifzadeh},
  \citenamefont {Ren}, \citenamefont {Scheffler}, \citenamefont {Liu},
  \citenamefont {Lischner}, \citenamefont {Lin}, \citenamefont {Deslippe},
  \citenamefont {Louie}, \citenamefont {Yang}, \citenamefont {Weigend},
  \citenamefont {Neaton}, \citenamefont {Evers},\ and\ \citenamefont
  {Rinke}}]{vanSetten_2015}%
  \BibitemOpen
  \bibfield  {author} {\bibinfo {author} {\bibfnamefont {M.~J.}\ \bibnamefont
  {{van Setten}}}, \bibinfo {author} {\bibfnamefont {F.}~\bibnamefont
  {Caruso}}, \bibinfo {author} {\bibfnamefont {S.}~\bibnamefont {Sharifzadeh}},
  \bibinfo {author} {\bibfnamefont {X.}~\bibnamefont {Ren}}, \bibinfo {author}
  {\bibfnamefont {M.}~\bibnamefont {Scheffler}}, \bibinfo {author}
  {\bibfnamefont {F.}~\bibnamefont {Liu}}, \bibinfo {author} {\bibfnamefont
  {J.}~\bibnamefont {Lischner}}, \bibinfo {author} {\bibfnamefont
  {L.}~\bibnamefont {Lin}}, \bibinfo {author} {\bibfnamefont {J.~R.}\
  \bibnamefont {Deslippe}}, \bibinfo {author} {\bibfnamefont {S.~G.}\
  \bibnamefont {Louie}}, \bibinfo {author} {\bibfnamefont {C.}~\bibnamefont
  {Yang}}, \bibinfo {author} {\bibfnamefont {F.}~\bibnamefont {Weigend}},
  \bibinfo {author} {\bibfnamefont {J.~B.}\ \bibnamefont {Neaton}}, \bibinfo
  {author} {\bibfnamefont {F.}~\bibnamefont {Evers}},\ and\ \bibinfo {author}
  {\bibfnamefont {P.}~\bibnamefont {Rinke}},\ }\bibfield  {title} {\enquote
  {\bibinfo {title} {{{{\emph{GW}}}} 100: {{Benchmarking}}
  {{{\emph{G}}}}{\textsubscript{0}}{{{\emph{W}}}}{\textsubscript{0}} for
  {{Molecular Systems}}},}\ }\href {https://doi.org/10.1021/acs.jctc.5b00453}
  {\bibfield  {journal} {\bibinfo  {journal} {J. Chem. Theory Comput.}\
  }\textbf {\bibinfo {volume} {11}},\ \bibinfo {pages} {5665--5687} (\bibinfo
  {year} {2015})}\BibitemShut {NoStop}%
\bibitem [{\citenamefont {Caruso}\ \emph {et~al.}(2016)\citenamefont {Caruso},
  \citenamefont {Dauth}, \citenamefont {{van Setten}},\ and\ \citenamefont
  {Rinke}}]{Caruso_2016}%
  \BibitemOpen
  \bibfield  {author} {\bibinfo {author} {\bibfnamefont {F.}~\bibnamefont
  {Caruso}}, \bibinfo {author} {\bibfnamefont {M.}~\bibnamefont {Dauth}},
  \bibinfo {author} {\bibfnamefont {M.~J.}\ \bibnamefont {{van Setten}}},\ and\
  \bibinfo {author} {\bibfnamefont {P.}~\bibnamefont {Rinke}},\ }\bibfield
  {title} {\enquote {\bibinfo {title} {{Benchmark of $GW$ Approaches for the
  GW100 Test Set}},}\ }\href {https://doi.org/10.1021/acs.jctc.6b00774}
  {\bibfield  {journal} {\bibinfo  {journal} {J. Chem. Theory Comput.}\
  }\textbf {\bibinfo {volume} {12}},\ \bibinfo {pages} {5076} (\bibinfo {year}
  {2016})}\BibitemShut {NoStop}%
\bibitem [{\citenamefont {Krause}\ and\ \citenamefont
  {Klopper}(2017)}]{Krause_2017}%
  \BibitemOpen
  \bibfield  {author} {\bibinfo {author} {\bibfnamefont {K.}~\bibnamefont
  {Krause}}\ and\ \bibinfo {author} {\bibfnamefont {W.}~\bibnamefont
  {Klopper}},\ }\bibfield  {title} {\enquote {\bibinfo {title} {Implementation
  of the bethe-salpeter equation in the turbomole program},}\ }\href
  {https://doi.org/10.1002/jcc.24688} {\bibfield  {journal} {\bibinfo
  {journal} {J. Comput. Chem.}\ }\textbf {\bibinfo {volume} {38}},\ \bibinfo
  {pages} {383--388} (\bibinfo {year} {2017})}\BibitemShut {NoStop}%
\bibitem [{\citenamefont {Lewis}\ and\ \citenamefont
  {Berkelbach}(2019)}]{Lewis_2019}%
  \BibitemOpen
  \bibfield  {author} {\bibinfo {author} {\bibfnamefont {A.~M.}\ \bibnamefont
  {Lewis}}\ and\ \bibinfo {author} {\bibfnamefont {T.~C.}\ \bibnamefont
  {Berkelbach}},\ }\bibfield  {title} {\enquote {\bibinfo {title} {Vertex
  corrections to the polarizability do not improve the gw approximation for the
  ionization potential of molecules},}\ }\href
  {https://doi.org/10.1021/acs.jctc.8b00995} {\bibfield  {journal} {\bibinfo
  {journal} {J. Chem. Theory Comput.}\ }\textbf {\bibinfo {volume} {15}},\
  \bibinfo {pages} {2925} (\bibinfo {year} {2019})}\BibitemShut {NoStop}%
\bibitem [{\citenamefont {Bruneval}, \citenamefont {Dattani},\ and\
  \citenamefont {van Setten}(2021)}]{Bruneval_2021}%
  \BibitemOpen
  \bibfield  {author} {\bibinfo {author} {\bibfnamefont {F.}~\bibnamefont
  {Bruneval}}, \bibinfo {author} {\bibfnamefont {N.}~\bibnamefont {Dattani}},\
  and\ \bibinfo {author} {\bibfnamefont {M.~J.}\ \bibnamefont {van Setten}},\
  }\bibfield  {title} {\enquote {\bibinfo {title} {{The $GW$ Miracle in
  Many-Body Perturbation Theory for the Ionization Potential of Molecules}},}\
  }\href {https://doi.org/10.3389/fchem.2021.749779} {\bibfield  {journal}
  {\bibinfo  {journal} {Front. Chem.}\ }\textbf {\bibinfo {volume} {9}},\
  \bibinfo {pages} {749779} (\bibinfo {year} {2021})}\BibitemShut {NoStop}%
\bibitem [{\citenamefont {Monino}\ and\ \citenamefont
  {Loos}(2023)}]{Monino_2023}%
  \BibitemOpen
  \bibfield  {author} {\bibinfo {author} {\bibfnamefont {E.}~\bibnamefont
  {Monino}}\ and\ \bibinfo {author} {\bibfnamefont {P.-F.}\ \bibnamefont
  {Loos}},\ }\bibfield  {title} {\enquote {\bibinfo {title} {Connections and
  performances of green's function methods for charged and neutral
  excitations},}\ }\href {https://doi.org/10.1063/5.0159853} {\bibfield
  {journal} {\bibinfo  {journal} {J. Chem. Phys.}\ }\textbf {\bibinfo {volume}
  {159}},\ \bibinfo {pages} {034105} (\bibinfo {year} {2023})}\BibitemShut
  {NoStop}%
\bibitem [{\citenamefont {Marie}\ and\ \citenamefont
  {Loos}(2024)}]{Marie_2024b}%
  \BibitemOpen
  \bibfield  {author} {\bibinfo {author} {\bibfnamefont {A.}~\bibnamefont
  {Marie}}\ and\ \bibinfo {author} {\bibfnamefont {P.-F.}\ \bibnamefont
  {Loos}},\ }\bibfield  {title} {\enquote {\bibinfo {title} {Reference
  {{Energies}} for {{Valence Ionizations}} and {{Satellite Transitions}}},}\
  }\href {https://doi.org/10.1021/acs.jctc.4c00216} {\bibfield  {journal}
  {\bibinfo  {journal} {J. Chem. Theory Comput.}\ }\textbf {\bibinfo {volume}
  {20}},\ \bibinfo {pages} {4751--4777} (\bibinfo {year} {2024})}\BibitemShut
  {NoStop}%
\bibitem [{\citenamefont {{van Setten}}\ \emph {et~al.}(2018)\citenamefont
  {{van Setten}}, \citenamefont {Costa}, \citenamefont {Vi{\~n}es},\ and\
  \citenamefont {Illas}}]{vanSetten_2018}%
  \BibitemOpen
  \bibfield  {author} {\bibinfo {author} {\bibfnamefont {M.~J.}\ \bibnamefont
  {{van Setten}}}, \bibinfo {author} {\bibfnamefont {R.}~\bibnamefont {Costa}},
  \bibinfo {author} {\bibfnamefont {F.}~\bibnamefont {Vi{\~n}es}},\ and\
  \bibinfo {author} {\bibfnamefont {F.}~\bibnamefont {Illas}},\ }\bibfield
  {title} {\enquote {\bibinfo {title} {Assessing {{{\emph{GW}}}} {{Approaches}}
  for {{Predicting Core Level Binding Energies}}},}\ }\href
  {https://doi.org/10.1021/acs.jctc.7b01192} {\bibfield  {journal} {\bibinfo
  {journal} {J. Chem. Theory Comput.}\ }\textbf {\bibinfo {volume} {14}},\
  \bibinfo {pages} {877--883} (\bibinfo {year} {2018})}\BibitemShut {NoStop}%
\bibitem [{\citenamefont {Golze}\ \emph {et~al.}(2018)\citenamefont {Golze},
  \citenamefont {Wilhelm}, \citenamefont {van Setten},\ and\ \citenamefont
  {Rinke}}]{Golze_2018}%
  \BibitemOpen
  \bibfield  {author} {\bibinfo {author} {\bibfnamefont {D.}~\bibnamefont
  {Golze}}, \bibinfo {author} {\bibfnamefont {J.}~\bibnamefont {Wilhelm}},
  \bibinfo {author} {\bibfnamefont {M.~J.}\ \bibnamefont {van Setten}},\ and\
  \bibinfo {author} {\bibfnamefont {P.}~\bibnamefont {Rinke}},\ }\bibfield
  {title} {\enquote {\bibinfo {title} {{Core-Level Binding Energies from $GW$:
  An Efficient Full-Frequency Approach within a Localized Basis}},}\ }\href
  {https://doi.org/10.1021/acs.jctc.8b00458} {\bibfield  {journal} {\bibinfo
  {journal} {J. Chem. Theory Comput.}\ }\textbf {\bibinfo {volume} {14}},\
  \bibinfo {pages} {4856--4869} (\bibinfo {year} {2018})}\BibitemShut {NoStop}%
\bibitem [{\citenamefont {Golze}, \citenamefont {Keller},\ and\ \citenamefont
  {Rinke}(2020)}]{Golze_2020}%
  \BibitemOpen
  \bibfield  {author} {\bibinfo {author} {\bibfnamefont {D.}~\bibnamefont
  {Golze}}, \bibinfo {author} {\bibfnamefont {L.}~\bibnamefont {Keller}},\ and\
  \bibinfo {author} {\bibfnamefont {P.}~\bibnamefont {Rinke}},\ }\bibfield
  {title} {\enquote {\bibinfo {title} {{Accurate Absolute and Relative
  Core-Level Binding Energies from $GW$}},}\ }\href
  {https://doi.org/10.1021/acs.jpclett.9b03423} {\bibfield  {journal} {\bibinfo
   {journal} {J. Phys. Chem. Lett.}\ }\textbf {\bibinfo {volume} {11}},\
  \bibinfo {pages} {1840--1847} (\bibinfo {year} {2020})}\BibitemShut {NoStop}%
\bibitem [{\citenamefont {{Mejia-Rodriguez}}\ \emph {et~al.}(2021)\citenamefont
  {{Mejia-Rodriguez}}, \citenamefont {Kunitsa}, \citenamefont {Apr{\`a}},\ and\
  \citenamefont {Govind}}]{Mejia-Rodriguez_2021}%
  \BibitemOpen
  \bibfield  {author} {\bibinfo {author} {\bibfnamefont {D.}~\bibnamefont
  {{Mejia-Rodriguez}}}, \bibinfo {author} {\bibfnamefont {A.}~\bibnamefont
  {Kunitsa}}, \bibinfo {author} {\bibfnamefont {E.}~\bibnamefont {Apr{\`a}}},\
  and\ \bibinfo {author} {\bibfnamefont {N.}~\bibnamefont {Govind}},\
  }\bibfield  {title} {\enquote {\bibinfo {title} {Scalable {{Molecular GW
  Calculations}}: {{Valence}} and {{Core Spectra}}},}\ }\href
  {https://doi.org/10.1021/acs.jctc.1c00738} {\bibfield  {journal} {\bibinfo
  {journal} {J. Chem. Theory Comput.}\ }\textbf {\bibinfo {volume} {17}},\
  \bibinfo {pages} {7504--7517} (\bibinfo {year} {2021})}\BibitemShut {NoStop}%
\bibitem [{\citenamefont {Li}\ \emph {et~al.}(2022)\citenamefont {Li},
  \citenamefont {Jin}, \citenamefont {Rinke}, \citenamefont {Yang},\ and\
  \citenamefont {Golze}}]{Li_2022}%
  \BibitemOpen
  \bibfield  {author} {\bibinfo {author} {\bibfnamefont {J.}~\bibnamefont
  {Li}}, \bibinfo {author} {\bibfnamefont {Y.}~\bibnamefont {Jin}}, \bibinfo
  {author} {\bibfnamefont {P.}~\bibnamefont {Rinke}}, \bibinfo {author}
  {\bibfnamefont {W.}~\bibnamefont {Yang}},\ and\ \bibinfo {author}
  {\bibfnamefont {D.}~\bibnamefont {Golze}},\ }\bibfield  {title} {\enquote
  {\bibinfo {title} {Benchmark of gw methods for core-level binding
  energies},}\ }\href {https://doi.org/10.1021/acs.jctc.2c00617} {\bibfield
  {journal} {\bibinfo  {journal} {J. Chem. Theory Comput.}\ }\textbf {\bibinfo
  {volume} {18}},\ \bibinfo {pages} {7570--7585} (\bibinfo {year}
  {2022})}\BibitemShut {NoStop}%
\bibitem [{\citenamefont {Mukatayev}\ \emph {et~al.}(2023)\citenamefont
  {Mukatayev}, \citenamefont {Moevus}, \citenamefont {Skl{\'e}nard},
  \citenamefont {Olevano},\ and\ \citenamefont {Li}}]{Mukatayev_2023}%
  \BibitemOpen
  \bibfield  {author} {\bibinfo {author} {\bibfnamefont {I.}~\bibnamefont
  {Mukatayev}}, \bibinfo {author} {\bibfnamefont {F.}~\bibnamefont {Moevus}},
  \bibinfo {author} {\bibfnamefont {B.}~\bibnamefont {Skl{\'e}nard}}, \bibinfo
  {author} {\bibfnamefont {V.}~\bibnamefont {Olevano}},\ and\ \bibinfo {author}
  {\bibfnamefont {J.}~\bibnamefont {Li}},\ }\bibfield  {title} {\enquote
  {\bibinfo {title} {{{XPS Core-Level Chemical Shift}} by {{Ab Initio Many-Body
  Theory}}},}\ }\href {https://doi.org/10.1021/acs.jpca.3c00173} {\bibfield
  {journal} {\bibinfo  {journal} {J. Phys. Chem. A}\ }\textbf {\bibinfo
  {volume} {127}},\ \bibinfo {pages} {1642--1648} (\bibinfo {year}
  {2023})}\BibitemShut {NoStop}%
\bibitem [{\citenamefont {Panad{\'e}s-Barrueta}\ and\ \citenamefont
  {Golze}(2023)}]{Panades-Barrueta_2023}%
  \BibitemOpen
  \bibfield  {author} {\bibinfo {author} {\bibfnamefont {R.~L.}\ \bibnamefont
  {Panad{\'e}s-Barrueta}}\ and\ \bibinfo {author} {\bibfnamefont
  {D.}~\bibnamefont {Golze}},\ }\bibfield  {title} {\enquote {\bibinfo {title}
  {{Accelerating Core-Level $GW$ Calculations by Combining the Contour
  Deformation Approach with the Analytic Continuation of W}},}\ }\href
  {https://doi.org/10.1021/acs.jctc.3c00555} {\bibfield  {journal} {\bibinfo
  {journal} {J. Chem. Theory Comput.}\ }\textbf {\bibinfo {volume} {19}},\
  \bibinfo {pages} {5450--5464} (\bibinfo {year} {2023})}\BibitemShut {NoStop}%
\bibitem [{\citenamefont {Baym}\ and\ \citenamefont
  {Kadanoff}(1961)}]{Baym_1961}%
  \BibitemOpen
  \bibfield  {author} {\bibinfo {author} {\bibfnamefont {G.}~\bibnamefont
  {Baym}}\ and\ \bibinfo {author} {\bibfnamefont {L.~P.}\ \bibnamefont
  {Kadanoff}},\ }\bibfield  {title} {\enquote {\bibinfo {title} {Conservation
  laws and correlation functions},}\ }\href
  {https://doi.org/10.1103/PhysRev.124.287} {\bibfield  {journal} {\bibinfo
  {journal} {Phys. Rev.}\ }\textbf {\bibinfo {volume} {124}},\ \bibinfo {pages}
  {287--299} (\bibinfo {year} {1961})}\BibitemShut {NoStop}%
\bibitem [{\citenamefont {Baym}(1962)}]{Baym_1962}%
  \BibitemOpen
  \bibfield  {author} {\bibinfo {author} {\bibfnamefont {G.}~\bibnamefont
  {Baym}},\ }\bibfield  {title} {\enquote {\bibinfo {title} {Self-consistent
  approximations in many-body systems},}\ }\href
  {https://doi.org/10.1103/PhysRev.127.1391} {\bibfield  {journal} {\bibinfo
  {journal} {Phys. Rev.}\ }\textbf {\bibinfo {volume} {127}},\ \bibinfo {pages}
  {1391--1401} (\bibinfo {year} {1962})}\BibitemShut {NoStop}%
\bibitem [{\citenamefont {De~Dominicis}\ and\ \citenamefont
  {Martin}(1964{\natexlab{a}})}]{DeDominicis_1964a}%
  \BibitemOpen
  \bibfield  {author} {\bibinfo {author} {\bibfnamefont {C.}~\bibnamefont
  {De~Dominicis}}\ and\ \bibinfo {author} {\bibfnamefont {P.~C.}\ \bibnamefont
  {Martin}},\ }\bibfield  {title} {\enquote {\bibinfo {title} {Stationary
  entropy principle and renormalization in normal and superfluid systems. i.
  algebraic formulation},}\ }\href {https://doi.org/10.1063/1.1704062}
  {\bibfield  {journal} {\bibinfo  {journal} {J. Math. Phys.}\ }\textbf
  {\bibinfo {volume} {5}},\ \bibinfo {pages} {14--30} (\bibinfo {year}
  {1964}{\natexlab{a}})}\BibitemShut {NoStop}%
\bibitem [{\citenamefont {De~Dominicis}\ and\ \citenamefont
  {Martin}(1964{\natexlab{b}})}]{DeDominicis_1964b}%
  \BibitemOpen
  \bibfield  {author} {\bibinfo {author} {\bibfnamefont {C.}~\bibnamefont
  {De~Dominicis}}\ and\ \bibinfo {author} {\bibfnamefont {P.~C.}\ \bibnamefont
  {Martin}},\ }\bibfield  {title} {\enquote {\bibinfo {title} {Stationary
  entropy principle and renormalization in normal and superfluid systems. ii.
  diagrammatic formulation},}\ }\href {https://doi.org/10.1063/1.1704064}
  {\bibfield  {journal} {\bibinfo  {journal} {J. Math. Phys.}\ }\textbf
  {\bibinfo {volume} {5}},\ \bibinfo {pages} {31--59} (\bibinfo {year}
  {1964}{\natexlab{b}})}\BibitemShut {NoStop}%
\bibitem [{\citenamefont {Bickers}, \citenamefont {Scalapino},\ and\
  \citenamefont {White}(1989)}]{Bickers_1989a}%
  \BibitemOpen
  \bibfield  {author} {\bibinfo {author} {\bibfnamefont {N.~E.}\ \bibnamefont
  {Bickers}}, \bibinfo {author} {\bibfnamefont {D.~J.}\ \bibnamefont
  {Scalapino}},\ and\ \bibinfo {author} {\bibfnamefont {S.~R.}\ \bibnamefont
  {White}},\ }\bibfield  {title} {\enquote {\bibinfo {title} {{Conserving
  Approximations for Strongly Correlated Electron Systems: Bethe-Salpeter
  Equation and Dynamics for the Two-Dimensional Hubbard Model}},}\ }\href
  {https://doi.org/10.1103/PhysRevLett.62.961} {\bibfield  {journal} {\bibinfo
  {journal} {Phys. Rev. Lett.}\ }\textbf {\bibinfo {volume} {62}},\ \bibinfo
  {pages} {961--964} (\bibinfo {year} {1989})}\BibitemShut {NoStop}%
\bibitem [{\citenamefont {Bickers}\ and\ \citenamefont
  {Scalapino}(1989)}]{Bickers_1989b}%
  \BibitemOpen
  \bibfield  {author} {\bibinfo {author} {\bibfnamefont {N.}~\bibnamefont
  {Bickers}}\ and\ \bibinfo {author} {\bibfnamefont {D.}~\bibnamefont
  {Scalapino}},\ }\bibfield  {title} {\enquote {\bibinfo {title} {{Conserving
  approximations for strongly fluctuating electron systems. I. Formalism and
  calculational approach}},}\ }\href
  {https://doi.org/https://doi.org/10.1016/0003-4916(89)90359-X} {\bibfield
  {journal} {\bibinfo  {journal} {Ann. Phys.}\ }\textbf {\bibinfo {volume}
  {193}},\ \bibinfo {pages} {206--251} (\bibinfo {year} {1989})}\BibitemShut
  {NoStop}%
\bibitem [{\citenamefont {Bickers}\ and\ \citenamefont
  {White}(1991)}]{Bickers_1991}%
  \BibitemOpen
  \bibfield  {author} {\bibinfo {author} {\bibfnamefont {N.~E.}\ \bibnamefont
  {Bickers}}\ and\ \bibinfo {author} {\bibfnamefont {S.~R.}\ \bibnamefont
  {White}},\ }\bibfield  {title} {\enquote {\bibinfo {title} {{Conserving
  approximations for strongly fluctuating electron systems. II. Numerical
  results and parquet extension}},}\ }\href
  {https://doi.org/10.1103/PhysRevB.43.8044} {\bibfield  {journal} {\bibinfo
  {journal} {Phys. Rev. B}\ }\textbf {\bibinfo {volume} {43}},\ \bibinfo
  {pages} {8044--8064} (\bibinfo {year} {1991})}\BibitemShut {NoStop}%
\bibitem [{\citenamefont {Hedin}(1999)}]{Hedin_1999}%
  \BibitemOpen
  \bibfield  {author} {\bibinfo {author} {\bibfnamefont {L.}~\bibnamefont
  {Hedin}},\ }\bibfield  {title} {\enquote {\bibinfo {title} {On correlation
  effects in electron spectroscopies and the {{$GW$}} approximation},}\ }\href
  {https://doi.org/10.1088/0953-8984/11/42/201} {\bibfield  {journal} {\bibinfo
   {journal} {J. Phys. Condens. Matter}\ }\textbf {\bibinfo {volume} {11}},\
  \bibinfo {pages} {R489--R528} (\bibinfo {year} {1999})}\BibitemShut {NoStop}%
\bibitem [{\citenamefont {Bickers}(2004)}]{Bickers_2004}%
  \BibitemOpen
  \bibfield  {author} {\bibinfo {author} {\bibfnamefont {N.~E.}\ \bibnamefont
  {Bickers}},\ }\enquote {\bibinfo {title} {{Self-Consistent Many-Body Theory
  for Condensed Matter Systems}},}\ in\ \href
  {https://doi.org/10.1007/0-387-21717-7_6} {\emph {\bibinfo {booktitle}
  {Theoretical Methods for Strongly Correlated Electrons}}},\ \bibinfo {editor}
  {edited by\ \bibinfo {editor} {\bibfnamefont {D.}~\bibnamefont
  {S{\'e}n{\'e}chal}}, \bibinfo {editor} {\bibfnamefont {A.-M.}\ \bibnamefont
  {Tremblay}},\ and\ \bibinfo {editor} {\bibfnamefont {C.}~\bibnamefont
  {Bourbonnais}}}\ (\bibinfo  {publisher} {Springer New York},\ \bibinfo
  {address} {New York, NY},\ \bibinfo {year} {2004})\ pp.\ \bibinfo {pages}
  {237--296}\BibitemShut {NoStop}%
\bibitem [{\citenamefont {Shirley}(1996)}]{Shirley_1996}%
  \BibitemOpen
  \bibfield  {author} {\bibinfo {author} {\bibfnamefont {E.~L.}\ \bibnamefont
  {Shirley}},\ }\bibfield  {title} {\enquote {\bibinfo {title} {Self-consistent
  gw and higher-order calculations of electron states in metals},}\ }\href
  {https://doi.org/10.1103/PhysRevB.54.7758} {\bibfield  {journal} {\bibinfo
  {journal} {Phys. Rev. B}\ }\textbf {\bibinfo {volume} {54}},\ \bibinfo
  {pages} {7758--7764} (\bibinfo {year} {1996})}\BibitemShut {NoStop}%
\bibitem [{\citenamefont {Del~Sole}, \citenamefont {Reining},\ and\
  \citenamefont {Godby}(1994)}]{DelSol_1994}%
  \BibitemOpen
  \bibfield  {author} {\bibinfo {author} {\bibfnamefont {R.}~\bibnamefont
  {Del~Sole}}, \bibinfo {author} {\bibfnamefont {L.}~\bibnamefont {Reining}},\
  and\ \bibinfo {author} {\bibfnamefont {R.~W.}\ \bibnamefont {Godby}},\
  }\bibfield  {title} {\enquote {\bibinfo {title} {{$GW\Gamma$ Approximation
  for Electron Self-Energies in Semiconductors and Insulators}},}\ }\href
  {https://doi.org/10.1103/PhysRevB.49.8024} {\bibfield  {journal} {\bibinfo
  {journal} {Phys. Rev. B}\ }\textbf {\bibinfo {volume} {49}},\ \bibinfo
  {pages} {8024--8028} (\bibinfo {year} {1994})}\BibitemShut {NoStop}%
\bibitem [{\citenamefont {Schindlmayr}\ and\ \citenamefont
  {Godby}(1998)}]{Schindlmayr_1998}%
  \BibitemOpen
  \bibfield  {author} {\bibinfo {author} {\bibfnamefont {A.}~\bibnamefont
  {Schindlmayr}}\ and\ \bibinfo {author} {\bibfnamefont {R.~W.}\ \bibnamefont
  {Godby}},\ }\bibfield  {title} {\enquote {\bibinfo {title} {Systematic
  {{Vertex Corrections}} through {{Iterative Solution}} of {{Hedin}}'s
  {{Equations Beyond}} the \$\textbackslash
  mathit\{\vphantom\}{{GW}}\vphantom\{\}\$ {{Approximation}}},}\ }\href
  {https://doi.org/10.1103/PhysRevLett.80.1702} {\bibfield  {journal} {\bibinfo
   {journal} {Phys. Rev. Lett.}\ }\textbf {\bibinfo {volume} {80}},\ \bibinfo
  {pages} {1702--1705} (\bibinfo {year} {1998})}\BibitemShut {NoStop}%
\bibitem [{\citenamefont {Morris}\ \emph {et~al.}(2007)\citenamefont {Morris},
  \citenamefont {Stankovski}, \citenamefont {Delaney}, \citenamefont {Rinke},
  \citenamefont {Garc\'{\i}a-Gonz\'alez},\ and\ \citenamefont
  {Godby}}]{Morris_2007}%
  \BibitemOpen
  \bibfield  {author} {\bibinfo {author} {\bibfnamefont {A.~J.}\ \bibnamefont
  {Morris}}, \bibinfo {author} {\bibfnamefont {M.}~\bibnamefont {Stankovski}},
  \bibinfo {author} {\bibfnamefont {K.~T.}\ \bibnamefont {Delaney}}, \bibinfo
  {author} {\bibfnamefont {P.}~\bibnamefont {Rinke}}, \bibinfo {author}
  {\bibfnamefont {P.}~\bibnamefont {Garc\'{\i}a-Gonz\'alez}},\ and\ \bibinfo
  {author} {\bibfnamefont {R.~W.}\ \bibnamefont {Godby}},\ }\bibfield  {title}
  {\enquote {\bibinfo {title} {Vertex corrections in localized and extended
  systems},}\ }\href {https://doi.org/10.1103/PhysRevB.76.155106} {\bibfield
  {journal} {\bibinfo  {journal} {Phys. Rev. B}\ }\textbf {\bibinfo {volume}
  {76}},\ \bibinfo {pages} {155106} (\bibinfo {year} {2007})}\BibitemShut
  {NoStop}%
\bibitem [{\citenamefont {Shishkin}, \citenamefont {Marsman},\ and\
  \citenamefont {Kresse}(2007)}]{Shishkin_2007b}%
  \BibitemOpen
  \bibfield  {author} {\bibinfo {author} {\bibfnamefont {M.}~\bibnamefont
  {Shishkin}}, \bibinfo {author} {\bibfnamefont {M.}~\bibnamefont {Marsman}},\
  and\ \bibinfo {author} {\bibfnamefont {G.}~\bibnamefont {Kresse}},\
  }\bibfield  {title} {\enquote {\bibinfo {title} {{Accurate Quasiparticle
  Spectra from Self-Consistent $GW$ Calculations with Vertex Corrections}},}\
  }\href {https://doi.org/10.1103/PhysRevLett.99.246403} {\bibfield  {journal}
  {\bibinfo  {journal} {Phys. Rev. Lett.}\ }\textbf {\bibinfo {volume} {99}},\
  \bibinfo {pages} {246403} (\bibinfo {year} {2007})}\BibitemShut {NoStop}%
\bibitem [{\citenamefont {Romaniello}, \citenamefont {Guyot},\ and\
  \citenamefont {Reining}(2009)}]{Romaniello_2009a}%
  \BibitemOpen
  \bibfield  {author} {\bibinfo {author} {\bibfnamefont {P.}~\bibnamefont
  {Romaniello}}, \bibinfo {author} {\bibfnamefont {S.}~\bibnamefont {Guyot}},\
  and\ \bibinfo {author} {\bibfnamefont {L.}~\bibnamefont {Reining}},\
  }\bibfield  {title} {\enquote {\bibinfo {title} {The self-energy beyond
  {{GW}}: {{Local}} and nonlocal vertex corrections},}\ }\href
  {https://doi.org/10.1063/1.3249965} {\bibfield  {journal} {\bibinfo
  {journal} {J. Chem. Phys.}\ }\textbf {\bibinfo {volume} {131}},\ \bibinfo
  {pages} {154111} (\bibinfo {year} {2009})}\BibitemShut {NoStop}%
\bibitem [{\citenamefont {Romaniello}, \citenamefont {Bechstedt},\ and\
  \citenamefont {Reining}(2012)}]{Romaniello_2012}%
  \BibitemOpen
  \bibfield  {author} {\bibinfo {author} {\bibfnamefont {P.}~\bibnamefont
  {Romaniello}}, \bibinfo {author} {\bibfnamefont {F.}~\bibnamefont
  {Bechstedt}},\ and\ \bibinfo {author} {\bibfnamefont {L.}~\bibnamefont
  {Reining}},\ }\bibfield  {title} {\enquote {\bibinfo {title} {{Beyond the
  $GW$ Approximation: Combining Correlation Channels}},}\ }\href
  {https://doi.org/10.1103/PhysRevB.85.155131} {\bibfield  {journal} {\bibinfo
  {journal} {Phys. Rev. B}\ }\textbf {\bibinfo {volume} {85}},\ \bibinfo
  {pages} {155131} (\bibinfo {year} {2012})}\BibitemShut {NoStop}%
\bibitem [{\citenamefont {Gr\"uneis}\ \emph {et~al.}(2014)\citenamefont
  {Gr\"uneis}, \citenamefont {Kresse}, \citenamefont {Hinuma},\ and\
  \citenamefont {Oba}}]{Gruneis_2014}%
  \BibitemOpen
  \bibfield  {author} {\bibinfo {author} {\bibfnamefont {A.}~\bibnamefont
  {Gr\"uneis}}, \bibinfo {author} {\bibfnamefont {G.}~\bibnamefont {Kresse}},
  \bibinfo {author} {\bibfnamefont {Y.}~\bibnamefont {Hinuma}},\ and\ \bibinfo
  {author} {\bibfnamefont {F.}~\bibnamefont {Oba}},\ }\bibfield  {title}
  {\enquote {\bibinfo {title} {Ionization potentials of solids: The importance
  of vertex corrections},}\ }\href
  {https://doi.org/10.1103/PhysRevLett.112.096401} {\bibfield  {journal}
  {\bibinfo  {journal} {Phys. Rev. Lett.}\ }\textbf {\bibinfo {volume} {112}},\
  \bibinfo {pages} {096401} (\bibinfo {year} {2014})}\BibitemShut {NoStop}%
\bibitem [{\citenamefont {Hung}\ \emph {et~al.}(2017)\citenamefont {Hung},
  \citenamefont {Bruneval}, \citenamefont {Baishya},\ and\ \citenamefont
  {{\"O}{\u g}{\"u}t}}]{Hung_2017}%
  \BibitemOpen
  \bibfield  {author} {\bibinfo {author} {\bibfnamefont {L.}~\bibnamefont
  {Hung}}, \bibinfo {author} {\bibfnamefont {F.}~\bibnamefont {Bruneval}},
  \bibinfo {author} {\bibfnamefont {K.}~\bibnamefont {Baishya}},\ and\ \bibinfo
  {author} {\bibfnamefont {S.}~\bibnamefont {{\"O}{\u g}{\"u}t}},\ }\bibfield
  {title} {\enquote {\bibinfo {title} {Benchmarking the {{{\emph{GW}}}}
  {{Approximation}} and {{Bethe}}\textendash{}{{Salpeter Equation}} for
  {{Groups IB}} and {{IIB Atoms}} and {{Monoxides}}},}\ }\href
  {https://doi.org/10.1021/acs.jctc.7b00123} {\bibfield  {journal} {\bibinfo
  {journal} {J. Chem. Theory Comput.}\ }\textbf {\bibinfo {volume} {13}},\
  \bibinfo {pages} {2135--2146} (\bibinfo {year} {2017})}\BibitemShut {NoStop}%
\bibitem [{\citenamefont {Maggio}\ and\ \citenamefont
  {Kresse}(2017)}]{Maggio_2017b}%
  \BibitemOpen
  \bibfield  {author} {\bibinfo {author} {\bibfnamefont {E.}~\bibnamefont
  {Maggio}}\ and\ \bibinfo {author} {\bibfnamefont {G.}~\bibnamefont
  {Kresse}},\ }\bibfield  {title} {\enquote {\bibinfo {title} {{$GW$ Vertex
  Corrected Calculations for Molecular Systems}},}\ }\href
  {https://doi.org/10.1021/acs.jctc.7b00586} {\bibfield  {journal} {\bibinfo
  {journal} {J. Chem. Theory Comput.}\ }\textbf {\bibinfo {volume} {13}},\
  \bibinfo {pages} {4765--4778} (\bibinfo {year} {2017})}\BibitemShut {NoStop}%
\bibitem [{\citenamefont {Mejuto-Zaera}\ and\ \citenamefont
  {Vlcek}(2022)}]{Mejuto-Zaera_2022}%
  \BibitemOpen
  \bibfield  {author} {\bibinfo {author} {\bibfnamefont {C.}~\bibnamefont
  {Mejuto-Zaera}}\ and\ \bibinfo {author} {\bibfnamefont {V.~c.~v.}\
  \bibnamefont {Vlcek}},\ }\bibfield  {title} {\enquote {\bibinfo {title}
  {{Self-consistency in $GW\Gamma$ formalism leading to
  quasiparticle-quasiparticle couplings}},}\ }\href
  {https://doi.org/10.1103/PhysRevB.106.165129} {\bibfield  {journal} {\bibinfo
   {journal} {Phys. Rev. B}\ }\textbf {\bibinfo {volume} {106}},\ \bibinfo
  {pages} {165129} (\bibinfo {year} {2022})}\BibitemShut {NoStop}%
\bibitem [{\citenamefont {Wen}\ \emph {et~al.}(2024)\citenamefont {Wen},
  \citenamefont {Abraham}, \citenamefont {Harsha}, \citenamefont {Shee},
  \citenamefont {Whaley},\ and\ \citenamefont {Zgid}}]{Wen_2024}%
  \BibitemOpen
  \bibfield  {author} {\bibinfo {author} {\bibfnamefont {M.}~\bibnamefont
  {Wen}}, \bibinfo {author} {\bibfnamefont {V.}~\bibnamefont {Abraham}},
  \bibinfo {author} {\bibfnamefont {G.}~\bibnamefont {Harsha}}, \bibinfo
  {author} {\bibfnamefont {A.}~\bibnamefont {Shee}}, \bibinfo {author}
  {\bibfnamefont {B.}~\bibnamefont {Whaley}},\ and\ \bibinfo {author}
  {\bibfnamefont {D.}~\bibnamefont {Zgid}},\ }\bibfield  {title} {\enquote
  {\bibinfo {title} {{Comparing Self-Consistent $GW$ and Vertex Corrected
  $G_0W_0$ Accuracy for Molecular Ionization Potentials}},}\ }\href
  {https://doi.org/10.1021/acs.jctc.3c01279} {\bibfield  {journal} {\bibinfo
  {journal} {J. Chem. Theory Comput.}\ ,\ \bibinfo {pages} {in press}}
  (\bibinfo {year} {2024})}\BibitemShut {NoStop}%
\bibitem [{\citenamefont {Bruneval}\ and\ \citenamefont
  {F{\"o}rster}(2024)}]{Bruneval_2024}%
  \BibitemOpen
  \bibfield  {author} {\bibinfo {author} {\bibfnamefont {F.}~\bibnamefont
  {Bruneval}}\ and\ \bibinfo {author} {\bibfnamefont {A.}~\bibnamefont
  {F{\"o}rster}},\ }\bibfield  {title} {\enquote {\bibinfo {title} {{Fully
  Dynamic $G3W2$ Self-Energy for Finite Systems: Formulas and Benchmark}},}\
  }\href {https://doi.org/10.1021/acs.jctc.4c00090} {\bibfield  {journal}
  {\bibinfo  {journal} {J. Chem. Theory Comput.}\ }\textbf {\bibinfo {volume}
  {20}},\ \bibinfo {pages} {3218--3230} (\bibinfo {year} {2024})}\BibitemShut
  {NoStop}%
\bibitem [{\citenamefont {F{\"o}rster}\ and\ \citenamefont
  {Bruneval}(2024)}]{Forster_2024}%
  \BibitemOpen
  \bibfield  {author} {\bibinfo {author} {\bibfnamefont {A.}~\bibnamefont
  {F{\"o}rster}}\ and\ \bibinfo {author} {\bibfnamefont {F.}~\bibnamefont
  {Bruneval}},\ }\bibfield  {title} {\enquote {\bibinfo {title} {{Why Does the
  $GW$ Approximation Give Accurate Quasiparticle Energies? The Cancellation of
  Vertex Corrections Quantified}},}\ }\href
  {https://doi.org/10.1021/acs.jpclett.4c03126} {\bibfield  {journal} {\bibinfo
   {journal} {J. Phys. Chem. Lett.}\ }\textbf {\bibinfo {volume} {15}},\
  \bibinfo {pages} {12526--12534} (\bibinfo {year} {2024})}\BibitemShut
  {NoStop}%
\bibitem [{\citenamefont {F{\"o}rster}(2025)}]{Forster_2025}%
  \BibitemOpen
  \bibfield  {author} {\bibinfo {author} {\bibfnamefont {A.}~\bibnamefont
  {F{\"o}rster}},\ }\bibfield  {title} {\enquote {\bibinfo {title} {{Beyond
  Quasi-Particle Self-Consistent $GW$ for Molecules with Vertex
  Corrections}},}\ }\href {https://doi.org/10.1021/acs.jctc.4c01639} {\bibfield
   {journal} {\bibinfo  {journal} {J. Chem. Theory Comput.}\ }\textbf {\bibinfo
  {volume} {21}},\ \bibinfo {pages} {1709--1721} (\bibinfo {year}
  {2025})}\BibitemShut {NoStop}%
\bibitem [{\citenamefont {Neuhauser}, \citenamefont {Rabani},\ and\
  \citenamefont {Baer}(2013)}]{Neuhauser_2013}%
  \BibitemOpen
  \bibfield  {author} {\bibinfo {author} {\bibfnamefont {D.}~\bibnamefont
  {Neuhauser}}, \bibinfo {author} {\bibfnamefont {E.}~\bibnamefont {Rabani}},\
  and\ \bibinfo {author} {\bibfnamefont {R.}~\bibnamefont {Baer}},\ }\bibfield
  {title} {\enquote {\bibinfo {title} {Expeditious stochastic calculation of
  random-phase approximation energies for thousands of electrons in three
  dimensions},}\ }\href {https://doi.org/10.1021/jz3021606} {\bibfield
  {journal} {\bibinfo  {journal} {J. Phys. Chem. Lett.}\ }\textbf {\bibinfo
  {volume} {4}},\ \bibinfo {pages} {1172--1176} (\bibinfo {year}
  {2013})}\BibitemShut {NoStop}%
\bibitem [{\citenamefont {Neuhauser}\ \emph {et~al.}(2014)\citenamefont
  {Neuhauser}, \citenamefont {Gao}, \citenamefont {Arntsen}, \citenamefont
  {Karshenas}, \citenamefont {Rabani},\ and\ \citenamefont
  {Baer}}]{Neuhauser_2014}%
  \BibitemOpen
  \bibfield  {author} {\bibinfo {author} {\bibfnamefont {D.}~\bibnamefont
  {Neuhauser}}, \bibinfo {author} {\bibfnamefont {Y.}~\bibnamefont {Gao}},
  \bibinfo {author} {\bibfnamefont {C.}~\bibnamefont {Arntsen}}, \bibinfo
  {author} {\bibfnamefont {C.}~\bibnamefont {Karshenas}}, \bibinfo {author}
  {\bibfnamefont {E.}~\bibnamefont {Rabani}},\ and\ \bibinfo {author}
  {\bibfnamefont {R.}~\bibnamefont {Baer}},\ }\bibfield  {title} {\enquote
  {\bibinfo {title} {{Breaking the {{Theoretical Scaling Limit}} for
  {{Predicting Quasiparticle Energies}}: {{The Stochastic $GW$ Approach}}}},}\
  }\href {https://doi.org/10.1103/PhysRevLett.113.076402} {\bibfield  {journal}
  {\bibinfo  {journal} {Phys. Rev. Lett.}\ }\textbf {\bibinfo {volume} {113}},\
  \bibinfo {pages} {076402} (\bibinfo {year} {2014})}\BibitemShut {NoStop}%
\bibitem [{\citenamefont {Kaltak}, \citenamefont {Klime\v{s}},\ and\
  \citenamefont {Kresse}(2014)}]{Kaltak_2014}%
  \BibitemOpen
  \bibfield  {author} {\bibinfo {author} {\bibfnamefont {M.}~\bibnamefont
  {Kaltak}}, \bibinfo {author} {\bibfnamefont {J.}~\bibnamefont {Klime\v{s}}},\
  and\ \bibinfo {author} {\bibfnamefont {G.}~\bibnamefont {Kresse}},\
  }\bibfield  {title} {\enquote {\bibinfo {title} {Low scaling algorithms for
  the random phase approximation: Imaginary time and laplace
  transformations},}\ }\href {https://doi.org/10.1021/ct5001268} {\bibfield
  {journal} {\bibinfo  {journal} {J. Chem. Theory Comput.}\ }\textbf {\bibinfo
  {volume} {10}},\ \bibinfo {pages} {2498--2507} (\bibinfo {year}
  {2014})}\BibitemShut {NoStop}%
\bibitem [{\citenamefont {Govoni}\ and\ \citenamefont
  {Galli}(2015)}]{Govoni_2015}%
  \BibitemOpen
  \bibfield  {author} {\bibinfo {author} {\bibfnamefont {M.}~\bibnamefont
  {Govoni}}\ and\ \bibinfo {author} {\bibfnamefont {G.}~\bibnamefont {Galli}},\
  }\bibfield  {title} {\enquote {\bibinfo {title} {{Large Scale $GW$
  Calculations}},}\ }\href {https://doi.org/10.1021/ct500958p} {\bibfield
  {journal} {\bibinfo  {journal} {J. Chem. Theory Comput.}\ }\textbf {\bibinfo
  {volume} {11}},\ \bibinfo {pages} {2680--2696} (\bibinfo {year}
  {2015})}\BibitemShut {NoStop}%
\bibitem [{\citenamefont {Vl{\v c}ek}\ \emph {et~al.}(2017)\citenamefont {Vl{\v
  c}ek}, \citenamefont {Rabani}, \citenamefont {Neuhauser},\ and\ \citenamefont
  {Baer}}]{Vlcek_2017}%
  \BibitemOpen
  \bibfield  {author} {\bibinfo {author} {\bibfnamefont {V.}~\bibnamefont
  {Vl{\v c}ek}}, \bibinfo {author} {\bibfnamefont {E.}~\bibnamefont {Rabani}},
  \bibinfo {author} {\bibfnamefont {D.}~\bibnamefont {Neuhauser}},\ and\
  \bibinfo {author} {\bibfnamefont {R.}~\bibnamefont {Baer}},\ }\bibfield
  {title} {\enquote {\bibinfo {title} {Stochastic {{GW Calculations}} for
  {{Molecules}}},}\ }\href {https://doi.org/10.1021/acs.jctc.7b00770}
  {\bibfield  {journal} {\bibinfo  {journal} {J. Chem. Theory Comput.}\
  }\textbf {\bibinfo {volume} {13}},\ \bibinfo {pages} {4997--5003} (\bibinfo
  {year} {2017})}\BibitemShut {NoStop}%
\bibitem [{\citenamefont {Wilhelm}\ \emph {et~al.}(2018)\citenamefont
  {Wilhelm}, \citenamefont {Golze}, \citenamefont {Talirz}, \citenamefont
  {Hutter},\ and\ \citenamefont {Pignedoli}}]{Wilhelm_2018}%
  \BibitemOpen
  \bibfield  {author} {\bibinfo {author} {\bibfnamefont {J.}~\bibnamefont
  {Wilhelm}}, \bibinfo {author} {\bibfnamefont {D.}~\bibnamefont {Golze}},
  \bibinfo {author} {\bibfnamefont {L.}~\bibnamefont {Talirz}}, \bibinfo
  {author} {\bibfnamefont {J.}~\bibnamefont {Hutter}},\ and\ \bibinfo {author}
  {\bibfnamefont {C.~A.}\ \bibnamefont {Pignedoli}},\ }\bibfield  {title}
  {\enquote {\bibinfo {title} {{Toward $GW$ Calculations on Thousands of
  Atoms}},}\ }\href {https://doi.org/10.1021/acs.jpclett.7b02740} {\bibfield
  {journal} {\bibinfo  {journal} {J. Phys. Chem. Lett.}\ }\textbf {\bibinfo
  {volume} {9}},\ \bibinfo {pages} {306--312} (\bibinfo {year}
  {2018})}\BibitemShut {NoStop}%
\bibitem [{\citenamefont {Duchemin}\ and\ \citenamefont
  {Blase}(2019)}]{Duchemin_2019}%
  \BibitemOpen
  \bibfield  {author} {\bibinfo {author} {\bibfnamefont {I.}~\bibnamefont
  {Duchemin}}\ and\ \bibinfo {author} {\bibfnamefont {X.}~\bibnamefont
  {Blase}},\ }\bibfield  {title} {\enquote {\bibinfo {title} {Separable
  resolution-of-the-identity with all-electron gaussian bases: Application to
  cubic-scaling rpa},}\ }\href {https://doi.org/10.1063/1.5090605} {\bibfield
  {journal} {\bibinfo  {journal} {J. Chem. Phys.}\ }\textbf {\bibinfo {volume}
  {150}},\ \bibinfo {pages} {174120} (\bibinfo {year} {2019})}\BibitemShut
  {NoStop}%
\bibitem [{\citenamefont {Ben}\ \emph {et~al.}(2019)\citenamefont {Ben},
  \citenamefont {da~Jornada}, \citenamefont {Canning}, \citenamefont
  {Wichmann}, \citenamefont {Raman}, \citenamefont {Sasanka}, \citenamefont
  {Yang}, \citenamefont {Louie},\ and\ \citenamefont {Deslippe}}]{DelBen_2019}%
  \BibitemOpen
  \bibfield  {author} {\bibinfo {author} {\bibfnamefont {M.~D.}\ \bibnamefont
  {Ben}}, \bibinfo {author} {\bibfnamefont {F.~H.}\ \bibnamefont {da~Jornada}},
  \bibinfo {author} {\bibfnamefont {A.}~\bibnamefont {Canning}}, \bibinfo
  {author} {\bibfnamefont {N.}~\bibnamefont {Wichmann}}, \bibinfo {author}
  {\bibfnamefont {K.}~\bibnamefont {Raman}}, \bibinfo {author} {\bibfnamefont
  {R.}~\bibnamefont {Sasanka}}, \bibinfo {author} {\bibfnamefont
  {C.}~\bibnamefont {Yang}}, \bibinfo {author} {\bibfnamefont {S.~G.}\
  \bibnamefont {Louie}},\ and\ \bibinfo {author} {\bibfnamefont
  {J.}~\bibnamefont {Deslippe}},\ }\bibfield  {title} {\enquote {\bibinfo
  {title} {Large-scale {GW} calculations on pre-exascale {HPC} systems},}\
  }\href {https://doi.org/10.1016/j.cpc.2018.09.003} {\bibfield  {journal}
  {\bibinfo  {journal} {Comp. Phys. Comm.}\ }\textbf {\bibinfo {volume}
  {235}},\ \bibinfo {pages} {187--195} (\bibinfo {year} {2019})}\BibitemShut
  {NoStop}%
\bibitem [{\citenamefont {F{\"o}rster}\ and\ \citenamefont
  {Visscher}(2020)}]{Forster_2020}%
  \BibitemOpen
  \bibfield  {author} {\bibinfo {author} {\bibfnamefont {A.}~\bibnamefont
  {F{\"o}rster}}\ and\ \bibinfo {author} {\bibfnamefont {L.}~\bibnamefont
  {Visscher}},\ }\bibfield  {title} {\enquote {\bibinfo {title} {{Low-Order
  Scaling G0W0 by Pair Atomic Density Fitting}},}\ }\href
  {https://doi.org/10.1021/acs.jctc.0c00693} {\bibfield  {journal} {\bibinfo
  {journal} {J. Chem. Theory Comput.}\ }\textbf {\bibinfo {volume} {16}},\
  \bibinfo {pages} {7381--7399} (\bibinfo {year} {2020})}\BibitemShut {NoStop}%
\bibitem [{\citenamefont {Duchemin}\ and\ \citenamefont
  {Blase}(2020)}]{Duchemin_2020}%
  \BibitemOpen
  \bibfield  {author} {\bibinfo {author} {\bibfnamefont {I.}~\bibnamefont
  {Duchemin}}\ and\ \bibinfo {author} {\bibfnamefont {X.}~\bibnamefont
  {Blase}},\ }\bibfield  {title} {\enquote {\bibinfo {title} {Robust
  analytic-continuation approach to many-body {{GW}} calculations},}\ }\href
  {https://doi.org/10.1021/acs.jctc.9b01235} {\bibfield  {journal} {\bibinfo
  {journal} {J. Chem. Theory Comput.}\ }\textbf {\bibinfo {volume} {16}},\
  \bibinfo {pages} {1742--1756} (\bibinfo {year} {2020})}\BibitemShut {NoStop}%
\bibitem [{\citenamefont {Kaltak}\ and\ \citenamefont
  {Kresse}(2020)}]{Kaltak_2020}%
  \BibitemOpen
  \bibfield  {author} {\bibinfo {author} {\bibfnamefont {M.}~\bibnamefont
  {Kaltak}}\ and\ \bibinfo {author} {\bibfnamefont {G.}~\bibnamefont
  {Kresse}},\ }\bibfield  {title} {\enquote {\bibinfo {title} {Minimax isometry
  method: A compressive sensing approach for matsubara summation in many-body
  perturbation theory},}\ }\href {https://doi.org/10.1103/PhysRevB.101.205145}
  {\bibfield  {journal} {\bibinfo  {journal} {Phys. Rev. B}\ }\textbf {\bibinfo
  {volume} {101}},\ \bibinfo {pages} {205145} (\bibinfo {year}
  {2020})}\BibitemShut {NoStop}%
\bibitem [{\citenamefont {F{\"o}rster}\ and\ \citenamefont
  {Visscher}(2021)}]{Forster_2021}%
  \BibitemOpen
  \bibfield  {author} {\bibinfo {author} {\bibfnamefont {A.}~\bibnamefont
  {F{\"o}rster}}\ and\ \bibinfo {author} {\bibfnamefont {L.}~\bibnamefont
  {Visscher}},\ }\bibfield  {title} {\enquote {\bibinfo {title} {{Low-Order
  Scaling Quasiparticle Self-Consistent GW for Molecules}},}\ }\href
  {https://doi.org/10.3389/fchem.2021.736591} {\bibfield  {journal} {\bibinfo
  {journal} {Front. Chem.}\ }\textbf {\bibinfo {volume} {9}},\ \bibinfo {pages}
  {736591} (\bibinfo {year} {2021})}\BibitemShut {NoStop}%
\bibitem [{\citenamefont {Duchemin}\ and\ \citenamefont
  {Blase}(2021)}]{Duchemin_2021}%
  \BibitemOpen
  \bibfield  {author} {\bibinfo {author} {\bibfnamefont {I.}~\bibnamefont
  {Duchemin}}\ and\ \bibinfo {author} {\bibfnamefont {X.}~\bibnamefont
  {Blase}},\ }\bibfield  {title} {\enquote {\bibinfo {title} {Cubic-scaling
  all-electron gw calculations with a separable density-fitting space--time
  approach},}\ }\href {https://doi.org/10.1021/acs.jctc.1c00101} {\bibfield
  {journal} {\bibinfo  {journal} {J. Chem. Theory Comput.}\ }\textbf {\bibinfo
  {volume} {17}},\ \bibinfo {pages} {2383--2393} (\bibinfo {year}
  {2021})}\BibitemShut {NoStop}%
\bibitem [{\citenamefont {Wilhelm}, \citenamefont {Seewald},\ and\
  \citenamefont {Golze}(2021)}]{Wilhelm_2021}%
  \BibitemOpen
  \bibfield  {author} {\bibinfo {author} {\bibfnamefont {J.}~\bibnamefont
  {Wilhelm}}, \bibinfo {author} {\bibfnamefont {P.}~\bibnamefont {Seewald}},\
  and\ \bibinfo {author} {\bibfnamefont {D.}~\bibnamefont {Golze}},\ }\bibfield
   {title} {\enquote {\bibinfo {title} {{Low-Scaling $GW$ with Benchmark
  Accuracy and Application to Phosphorene Nanosheets}},}\ }\href
  {https://doi.org/10.1021/acs.jctc.0c01282} {\bibfield  {journal} {\bibinfo
  {journal} {J. Chem. Theory Comput.}\ }\textbf {\bibinfo {volume} {17}},\
  \bibinfo {pages} {1662--1677} (\bibinfo {year} {2021})}\BibitemShut {NoStop}%
\bibitem [{\citenamefont {F{\"o}rster}\ and\ \citenamefont
  {Visscher}(2022)}]{Forster_2022}%
  \BibitemOpen
  \bibfield  {author} {\bibinfo {author} {\bibfnamefont {A.}~\bibnamefont
  {F{\"o}rster}}\ and\ \bibinfo {author} {\bibfnamefont {L.}~\bibnamefont
  {Visscher}},\ }\bibfield  {title} {\enquote {\bibinfo {title} {{Quasiparticle
  {{Self-Consistent GW-Bethe}}{\textendash}{{Salpeter Equation Calculations}}
  for {{Large Chromophoric Systems}}}},}\ }\href
  {https://doi.org/10.1021/acs.jctc.2c00531} {\bibfield  {journal} {\bibinfo
  {journal} {J. Chem. Theory Comput.}\ }\textbf {\bibinfo {volume} {18}},\
  \bibinfo {pages} {6779--6793} (\bibinfo {year} {2022})}\BibitemShut {NoStop}%
\bibitem [{\citenamefont {Yu}\ and\ \citenamefont {Govoni}(2022)}]{Yu_2022}%
  \BibitemOpen
  \bibfield  {author} {\bibinfo {author} {\bibfnamefont {V.~W.-z.}\
  \bibnamefont {Yu}}\ and\ \bibinfo {author} {\bibfnamefont {M.}~\bibnamefont
  {Govoni}},\ }\bibfield  {title} {\enquote {\bibinfo {title} {{GPU
  Acceleration of Large-Scale Full-Frequency GW Calculations}},}\ }\href
  {https://doi.org/10.1021/acs.jctc.2c00241} {\bibfield  {journal} {\bibinfo
  {journal} {J. Chem. Theory Comput.}\ }\textbf {\bibinfo {volume} {18}},\
  \bibinfo {pages} {4690--4707} (\bibinfo {year} {2022})}\BibitemShut {NoStop}%
\bibitem [{\citenamefont {T{\"o}lle}, \citenamefont {Niemeyer},\ and\
  \citenamefont {Neugebauer}(2024)}]{Tolle_2024}%
  \BibitemOpen
  \bibfield  {author} {\bibinfo {author} {\bibfnamefont {J.}~\bibnamefont
  {T{\"o}lle}}, \bibinfo {author} {\bibfnamefont {N.}~\bibnamefont
  {Niemeyer}},\ and\ \bibinfo {author} {\bibfnamefont {J.}~\bibnamefont
  {Neugebauer}},\ }\bibfield  {title} {\enquote {\bibinfo {title} {Accelerating
  analytic-continuation gw calculations with a laplace transform and natural
  auxiliary functions},}\ }\href {https://doi.org/10.1021/acs.jctc.3c01264}
  {\bibfield  {journal} {\bibinfo  {journal} {J. Chem. Theory Comput.}\
  }\textbf {\bibinfo {volume} {20}},\ \bibinfo {pages} {2022--2032} (\bibinfo
  {year} {2024})}\BibitemShut {NoStop}%
\bibitem [{\citenamefont {Gallandi}\ \emph {et~al.}(2016)\citenamefont
  {Gallandi}, \citenamefont {Marom}, \citenamefont {Rinke},\ and\ \citenamefont
  {K{\"o}rzd{\"o}rfer}}]{Gallandi_2016}%
  \BibitemOpen
  \bibfield  {author} {\bibinfo {author} {\bibfnamefont {L.}~\bibnamefont
  {Gallandi}}, \bibinfo {author} {\bibfnamefont {N.}~\bibnamefont {Marom}},
  \bibinfo {author} {\bibfnamefont {P.}~\bibnamefont {Rinke}},\ and\ \bibinfo
  {author} {\bibfnamefont {T.}~\bibnamefont {K{\"o}rzd{\"o}rfer}},\ }\bibfield
  {title} {\enquote {\bibinfo {title} {Accurate {{Ionization Potentials}} and
  {{Electron Affinities}} of {{Acceptor Molecules II}}: {{Non}}-{{Empirically
  Tuned Long}}-{{Range Corrected Hybrid Functionals}}},}\ }\href
  {https://doi.org/10.1021/acs.jctc.5b00873} {\bibfield  {journal} {\bibinfo
  {journal} {J. Chem. Theory Comput.}\ }\textbf {\bibinfo {volume} {12}},\
  \bibinfo {pages} {605--614} (\bibinfo {year} {2016})}\BibitemShut {NoStop}%
\bibitem [{\citenamefont {Richard}\ \emph {et~al.}(2016)\citenamefont
  {Richard}, \citenamefont {Marshall}, \citenamefont {Dolgounitcheva},
  \citenamefont {Ortiz}, \citenamefont {Br{\'e}das}, \citenamefont {Marom},\
  and\ \citenamefont {Sherrill}}]{Richard_2016}%
  \BibitemOpen
  \bibfield  {author} {\bibinfo {author} {\bibfnamefont {R.~M.}\ \bibnamefont
  {Richard}}, \bibinfo {author} {\bibfnamefont {M.~S.}\ \bibnamefont
  {Marshall}}, \bibinfo {author} {\bibfnamefont {O.}~\bibnamefont
  {Dolgounitcheva}}, \bibinfo {author} {\bibfnamefont {J.~V.}\ \bibnamefont
  {Ortiz}}, \bibinfo {author} {\bibfnamefont {J.-L.}\ \bibnamefont
  {Br{\'e}das}}, \bibinfo {author} {\bibfnamefont {N.}~\bibnamefont {Marom}},\
  and\ \bibinfo {author} {\bibfnamefont {C.~D.}\ \bibnamefont {Sherrill}},\
  }\bibfield  {title} {\enquote {\bibinfo {title} {Accurate {{Ionization
  Potentials}} and {{Electron Affinities}} of {{Acceptor Molecules I}}.
  {{Reference Data}} at the {{CCSD}}({{T}}) {{Complete Basis Set Limit}}},}\
  }\href {https://doi.org/10.1021/acs.jctc.5b00875} {\bibfield  {journal}
  {\bibinfo  {journal} {J. Chem. Theory Comput.}\ }\textbf {\bibinfo {volume}
  {12}},\ \bibinfo {pages} {595--604} (\bibinfo {year} {2016})}\BibitemShut
  {NoStop}%
\bibitem [{\citenamefont {Knight}\ \emph {et~al.}(2016)\citenamefont {Knight},
  \citenamefont {Wang}, \citenamefont {Gallandi}, \citenamefont
  {Dolgounitcheva}, \citenamefont {Ren}, \citenamefont {Ortiz}, \citenamefont
  {Rinke}, \citenamefont {K{\"o}rzd{\"o}rfer},\ and\ \citenamefont
  {Marom}}]{Knight_2016}%
  \BibitemOpen
  \bibfield  {author} {\bibinfo {author} {\bibfnamefont {J.~W.}\ \bibnamefont
  {Knight}}, \bibinfo {author} {\bibfnamefont {X.}~\bibnamefont {Wang}},
  \bibinfo {author} {\bibfnamefont {L.}~\bibnamefont {Gallandi}}, \bibinfo
  {author} {\bibfnamefont {O.}~\bibnamefont {Dolgounitcheva}}, \bibinfo
  {author} {\bibfnamefont {X.}~\bibnamefont {Ren}}, \bibinfo {author}
  {\bibfnamefont {J.~V.}\ \bibnamefont {Ortiz}}, \bibinfo {author}
  {\bibfnamefont {P.}~\bibnamefont {Rinke}}, \bibinfo {author} {\bibfnamefont
  {T.}~\bibnamefont {K{\"o}rzd{\"o}rfer}},\ and\ \bibinfo {author}
  {\bibfnamefont {N.}~\bibnamefont {Marom}},\ }\bibfield  {title} {\enquote
  {\bibinfo {title} {Accurate {{Ionization Potentials}} and {{Electron
  Affinities}} of {{Acceptor Molecules III}}: {{A Benchmark}} of
  {{{\emph{GW}}}} {{Methods}}},}\ }\href
  {https://doi.org/10.1021/acs.jctc.5b00871} {\bibfield  {journal} {\bibinfo
  {journal} {J. Chem. Theory Comput.}\ }\textbf {\bibinfo {volume} {12}},\
  \bibinfo {pages} {615--626} (\bibinfo {year} {2016})}\BibitemShut {NoStop}%
\bibitem [{\citenamefont {Dolgounitcheva}\ \emph {et~al.}(2016)\citenamefont
  {Dolgounitcheva}, \citenamefont {D{\'\i}az-Tinoco}, \citenamefont
  {Zakrzewski}, \citenamefont {Richard}, \citenamefont {Marom}, \citenamefont
  {Sherrill},\ and\ \citenamefont {Ortiz}}]{Dolgounitcheva_2016}%
  \BibitemOpen
  \bibfield  {author} {\bibinfo {author} {\bibfnamefont {O.}~\bibnamefont
  {Dolgounitcheva}}, \bibinfo {author} {\bibfnamefont {M.}~\bibnamefont
  {D{\'\i}az-Tinoco}}, \bibinfo {author} {\bibfnamefont {V.~G.}\ \bibnamefont
  {Zakrzewski}}, \bibinfo {author} {\bibfnamefont {R.~M.}\ \bibnamefont
  {Richard}}, \bibinfo {author} {\bibfnamefont {N.}~\bibnamefont {Marom}},
  \bibinfo {author} {\bibfnamefont {C.~D.}\ \bibnamefont {Sherrill}},\ and\
  \bibinfo {author} {\bibfnamefont {J.~V.}\ \bibnamefont {Ortiz}},\ }\bibfield
  {title} {\enquote {\bibinfo {title} {Accurate {{Ionization Potentials}} and
  {{Electron Affinities}} of {{Acceptor Molecules IV}}:
  {{Electron}}-{{Propagator Methods}}},}\ }\href
  {https://doi.org/10.1021/acs.jctc.5b00872} {\bibfield  {journal} {\bibinfo
  {journal} {J. Chem. Theory Comput.}\ }\textbf {\bibinfo {volume} {12}},\
  \bibinfo {pages} {627--637} (\bibinfo {year} {2016})}\BibitemShut {NoStop}%
\bibitem [{\citenamefont {Strinati}, \citenamefont {Mattausch},\ and\
  \citenamefont {Hanke}(1980)}]{Strinati_1980}%
  \BibitemOpen
  \bibfield  {author} {\bibinfo {author} {\bibfnamefont {G.}~\bibnamefont
  {Strinati}}, \bibinfo {author} {\bibfnamefont {H.~J.}\ \bibnamefont
  {Mattausch}},\ and\ \bibinfo {author} {\bibfnamefont {W.}~\bibnamefont
  {Hanke}},\ }\bibfield  {title} {\enquote {\bibinfo {title} {Dynamical
  correlation effects on the quasiparticle bloch states of a covalent
  crystal},}\ }\href {https://doi.org/10.1103/PhysRevLett.45.290} {\bibfield
  {journal} {\bibinfo  {journal} {Phys. Rev. Lett.}\ }\textbf {\bibinfo
  {volume} {45}},\ \bibinfo {pages} {290--294} (\bibinfo {year}
  {1980})}\BibitemShut {NoStop}%
\bibitem [{\citenamefont {Hybertsen}\ and\ \citenamefont
  {Louie}(1985)}]{Hybertsen_1985a}%
  \BibitemOpen
  \bibfield  {author} {\bibinfo {author} {\bibfnamefont {M.~S.}\ \bibnamefont
  {Hybertsen}}\ and\ \bibinfo {author} {\bibfnamefont {S.~G.}\ \bibnamefont
  {Louie}},\ }\bibfield  {title} {\enquote {\bibinfo {title}
  {First-{{Principles Theory}} of {{Quasiparticles}}: {{Calculation}} of {{Band
  Gaps}} in {{Semiconductors}} and {{Insulators}}},}\ }\href
  {https://doi.org/10.1103/PhysRevLett.55.1418} {\bibfield  {journal} {\bibinfo
   {journal} {Phys. Rev. Lett.}\ }\textbf {\bibinfo {volume} {55}},\ \bibinfo
  {pages} {1418--1421} (\bibinfo {year} {1985})}\BibitemShut {NoStop}%
\bibitem [{\citenamefont {Godby}, \citenamefont {Schl\"uter},\ and\
  \citenamefont {Sham}(1988)}]{Godby_1988}%
  \BibitemOpen
  \bibfield  {author} {\bibinfo {author} {\bibfnamefont {R.~W.}\ \bibnamefont
  {Godby}}, \bibinfo {author} {\bibfnamefont {M.}~\bibnamefont {Schl\"uter}},\
  and\ \bibinfo {author} {\bibfnamefont {L.~J.}\ \bibnamefont {Sham}},\
  }\bibfield  {title} {\enquote {\bibinfo {title} {Self-energy operators and
  exchange-correlation potentials in semiconductors},}\ }\href
  {https://doi.org/10.1103/PhysRevB.37.10159} {\bibfield  {journal} {\bibinfo
  {journal} {Phys. Rev. B}\ }\textbf {\bibinfo {volume} {37}},\ \bibinfo
  {pages} {10159--10175} (\bibinfo {year} {1988})}\BibitemShut {NoStop}%
\bibitem [{\citenamefont {von~der Linden}\ and\ \citenamefont
  {Horsch}(1988)}]{Linden_1988}%
  \BibitemOpen
  \bibfield  {author} {\bibinfo {author} {\bibfnamefont {W.}~\bibnamefont
  {von~der Linden}}\ and\ \bibinfo {author} {\bibfnamefont {P.}~\bibnamefont
  {Horsch}},\ }\bibfield  {title} {\enquote {\bibinfo {title} {Precise
  quasiparticle energies and hartree-fock bands of semiconductors and
  insulators},}\ }\href {https://doi.org/10.1103/PhysRevB.37.8351} {\bibfield
  {journal} {\bibinfo  {journal} {Phys. Rev. B}\ }\textbf {\bibinfo {volume}
  {37}},\ \bibinfo {pages} {8351--8362} (\bibinfo {year} {1988})}\BibitemShut
  {NoStop}%
\bibitem [{\citenamefont {Northrup}, \citenamefont {Hybertsen},\ and\
  \citenamefont {Louie}(1991)}]{Northrup_1991}%
  \BibitemOpen
  \bibfield  {author} {\bibinfo {author} {\bibfnamefont {J.~E.}\ \bibnamefont
  {Northrup}}, \bibinfo {author} {\bibfnamefont {M.~S.}\ \bibnamefont
  {Hybertsen}},\ and\ \bibinfo {author} {\bibfnamefont {S.~G.}\ \bibnamefont
  {Louie}},\ }\bibfield  {title} {\enquote {\bibinfo {title} {{Many-body
  Calculation of the Surface-State Energies for
  Si(111)2\ifmmode\times\else\texttimes\fi{}1}},}\ }\href
  {https://doi.org/10.1103/PhysRevLett.66.500} {\bibfield  {journal} {\bibinfo
  {journal} {Phys. Rev. Lett.}\ }\textbf {\bibinfo {volume} {66}},\ \bibinfo
  {pages} {500--503} (\bibinfo {year} {1991})}\BibitemShut {NoStop}%
\bibitem [{\citenamefont {Blase}, \citenamefont {Zhu},\ and\ \citenamefont
  {Louie}(1994)}]{Blase_1994}%
  \BibitemOpen
  \bibfield  {author} {\bibinfo {author} {\bibfnamefont {X.}~\bibnamefont
  {Blase}}, \bibinfo {author} {\bibfnamefont {X.}~\bibnamefont {Zhu}},\ and\
  \bibinfo {author} {\bibfnamefont {S.~G.}\ \bibnamefont {Louie}},\ }\bibfield
  {title} {\enquote {\bibinfo {title} {Self-energy effects on the surface-state
  energies of h-si(111)1\ifmmode\times\else\texttimes\fi{}1},}\ }\href
  {https://doi.org/10.1103/PhysRevB.49.4973} {\bibfield  {journal} {\bibinfo
  {journal} {Phys. Rev. B}\ }\textbf {\bibinfo {volume} {49}},\ \bibinfo
  {pages} {4973--4980} (\bibinfo {year} {1994})}\BibitemShut {NoStop}%
\bibitem [{\citenamefont {Rohlfing}, \citenamefont {Kr{\"u}ger},\ and\
  \citenamefont {Pollmann}(1995)}]{Rohlfing_1995}%
  \BibitemOpen
  \bibfield  {author} {\bibinfo {author} {\bibfnamefont {M.}~\bibnamefont
  {Rohlfing}}, \bibinfo {author} {\bibfnamefont {P.}~\bibnamefont
  {Kr{\"u}ger}},\ and\ \bibinfo {author} {\bibfnamefont {J.}~\bibnamefont
  {Pollmann}},\ }\bibfield  {title} {\enquote {\bibinfo {title} {{Efficient
  Scheme for $GW$ Quasiparticle Band-Structure Calculations with Applications
  to Bulk Si and to the Si(001)-(2$\times$1) Surface}},}\ }\href
  {https://doi.org/10.1103/PhysRevB.52.1905} {\bibfield  {journal} {\bibinfo
  {journal} {Phys. Rev. B}\ }\textbf {\bibinfo {volume} {52}},\ \bibinfo
  {pages} {1905--1917} (\bibinfo {year} {1995})}\BibitemShut {NoStop}%
\bibitem [{\citenamefont {Hybertsen}\ and\ \citenamefont
  {Louie}(1986)}]{Hybertsen_1986}%
  \BibitemOpen
  \bibfield  {author} {\bibinfo {author} {\bibfnamefont {M.~S.}\ \bibnamefont
  {Hybertsen}}\ and\ \bibinfo {author} {\bibfnamefont {S.~G.}\ \bibnamefont
  {Louie}},\ }\bibfield  {title} {\enquote {\bibinfo {title} {Electron
  correlation in semiconductors and insulators: {{Band}} gaps and quasiparticle
  energies},}\ }\href {https://doi.org/10.1103/PhysRevB.34.5390} {\bibfield
  {journal} {\bibinfo  {journal} {Phys. Rev. B}\ }\textbf {\bibinfo {volume}
  {34}},\ \bibinfo {pages} {5390--5413} (\bibinfo {year} {1986})}\BibitemShut
  {NoStop}%
\bibitem [{\citenamefont {Shishkin}\ and\ \citenamefont
  {Kresse}(2007)}]{Shishkin_2007a}%
  \BibitemOpen
  \bibfield  {author} {\bibinfo {author} {\bibfnamefont {M.}~\bibnamefont
  {Shishkin}}\ and\ \bibinfo {author} {\bibfnamefont {G.}~\bibnamefont
  {Kresse}},\ }\bibfield  {title} {\enquote {\bibinfo {title} {{Self-Consistent
  $GW$ Calculations for Semiconductors and Insulators}},}\ }\href
  {https://doi.org/10.1103/PhysRevB.75.235102} {\bibfield  {journal} {\bibinfo
  {journal} {Phys. Rev. B}\ }\textbf {\bibinfo {volume} {75}},\ \bibinfo
  {pages} {235102} (\bibinfo {year} {2007})}\BibitemShut {NoStop}%
\bibitem [{\citenamefont {Blase}\ and\ \citenamefont
  {Attaccalite}(2011)}]{Blase_2011a}%
  \BibitemOpen
  \bibfield  {author} {\bibinfo {author} {\bibfnamefont {X.}~\bibnamefont
  {Blase}}\ and\ \bibinfo {author} {\bibfnamefont {C.}~\bibnamefont
  {Attaccalite}},\ }\bibfield  {title} {\enquote {\bibinfo {title}
  {Charge-transfer excitations in molecular donor-acceptor complexes within the
  many-body bethe-salpeter approach},}\ }\href
  {https://doi.org/10.1063/1.3655352} {\bibfield  {journal} {\bibinfo
  {journal} {Appl. Phys. Lett.}\ }\textbf {\bibinfo {volume} {99}},\ \bibinfo
  {pages} {171909} (\bibinfo {year} {2011})}\BibitemShut {NoStop}%
\bibitem [{\citenamefont {Faber}\ \emph {et~al.}(2011)\citenamefont {Faber},
  \citenamefont {Attaccalite}, \citenamefont {Olevano}, \citenamefont {Runge},\
  and\ \citenamefont {Blase}}]{Faber_2011}%
  \BibitemOpen
  \bibfield  {author} {\bibinfo {author} {\bibfnamefont {C.}~\bibnamefont
  {Faber}}, \bibinfo {author} {\bibfnamefont {C.}~\bibnamefont {Attaccalite}},
  \bibinfo {author} {\bibfnamefont {V.}~\bibnamefont {Olevano}}, \bibinfo
  {author} {\bibfnamefont {E.}~\bibnamefont {Runge}},\ and\ \bibinfo {author}
  {\bibfnamefont {X.}~\bibnamefont {Blase}},\ }\bibfield  {title} {\enquote
  {\bibinfo {title} {First-principles {{GW}} calculations for {{DNA}} and
  {{RNA}} nucleobases},}\ }\href {https://doi.org/10.1103/PhysRevB.83.115123}
  {\bibfield  {journal} {\bibinfo  {journal} {Phys. Rev. B}\ }\textbf {\bibinfo
  {volume} {83}},\ \bibinfo {pages} {115123} (\bibinfo {year}
  {2011})}\BibitemShut {NoStop}%
\bibitem [{\citenamefont {Rangel}\ \emph {et~al.}(2016)\citenamefont {Rangel},
  \citenamefont {Hamed}, \citenamefont {Bruneval},\ and\ \citenamefont
  {Neaton}}]{Rangel_2016}%
  \BibitemOpen
  \bibfield  {author} {\bibinfo {author} {\bibfnamefont {T.}~\bibnamefont
  {Rangel}}, \bibinfo {author} {\bibfnamefont {S.~M.}\ \bibnamefont {Hamed}},
  \bibinfo {author} {\bibfnamefont {F.}~\bibnamefont {Bruneval}},\ and\
  \bibinfo {author} {\bibfnamefont {J.~B.}\ \bibnamefont {Neaton}},\ }\bibfield
   {title} {\enquote {\bibinfo {title} {{Evaluating the $GW$ Approximation with
  CCSD(T) for Charged Excitations Across the Oligoacenes}},}\ }\href
  {https://doi.org/10.1021/acs.jctc.6b00163} {\bibfield  {journal} {\bibinfo
  {journal} {J. Chem. Theory Comput.}\ }\textbf {\bibinfo {volume} {12}},\
  \bibinfo {pages} {2834--2842} (\bibinfo {year} {2016})}\BibitemShut {NoStop}%
\bibitem [{\citenamefont {Gui}, \citenamefont {Holzer},\ and\ \citenamefont
  {Klopper}(2018)}]{Gui_2018}%
  \BibitemOpen
  \bibfield  {author} {\bibinfo {author} {\bibfnamefont {X.}~\bibnamefont
  {Gui}}, \bibinfo {author} {\bibfnamefont {C.}~\bibnamefont {Holzer}},\ and\
  \bibinfo {author} {\bibfnamefont {W.}~\bibnamefont {Klopper}},\ }\bibfield
  {title} {\enquote {\bibinfo {title} {Accuracy assessment of gw starting
  points for calculating molecular excitation energies using the
  bethe--salpeter formalism},}\ }\href
  {https://doi.org/10.1021/acs.jctc.8b00014} {\bibfield  {journal} {\bibinfo
  {journal} {J. Chem. Theory Comput.}\ }\textbf {\bibinfo {volume} {14}},\
  \bibinfo {pages} {2127--2136} (\bibinfo {year} {2018})}\BibitemShut {NoStop}%
\bibitem [{\citenamefont {Faleev}, \citenamefont {{van Schilfgaarde}},\ and\
  \citenamefont {Kotani}(2004)}]{Faleev_2004}%
  \BibitemOpen
  \bibfield  {author} {\bibinfo {author} {\bibfnamefont {S.~V.}\ \bibnamefont
  {Faleev}}, \bibinfo {author} {\bibfnamefont {M.}~\bibnamefont {{van
  Schilfgaarde}}},\ and\ \bibinfo {author} {\bibfnamefont {T.}~\bibnamefont
  {Kotani}},\ }\bibfield  {title} {\enquote {\bibinfo {title} {All-{{Electron
  Self}}-{{Consistent G W Approximation}}: {{Application}} to {{Si}}, {{MnO}},
  and {{NiO}}},}\ }\href {https://doi.org/10.1103/PhysRevLett.93.126406}
  {\bibfield  {journal} {\bibinfo  {journal} {Phys. Rev. Lett.}\ }\textbf
  {\bibinfo {volume} {93}},\ \bibinfo {pages} {126406} (\bibinfo {year}
  {2004})}\BibitemShut {NoStop}%
\bibitem [{\citenamefont {{van Schilfgaarde}}, \citenamefont {Kotani},\ and\
  \citenamefont {Faleev}(2006)}]{vanSchilfgaarde_2006}%
  \BibitemOpen
  \bibfield  {author} {\bibinfo {author} {\bibfnamefont {M.}~\bibnamefont {{van
  Schilfgaarde}}}, \bibinfo {author} {\bibfnamefont {T.}~\bibnamefont
  {Kotani}},\ and\ \bibinfo {author} {\bibfnamefont {S.}~\bibnamefont
  {Faleev}},\ }\bibfield  {title} {\enquote {\bibinfo {title} {Quasiparticle
  {{Self}}-{{Consistent G W Theory}}},}\ }\href
  {https://doi.org/10.1103/PhysRevLett.96.226402} {\bibfield  {journal}
  {\bibinfo  {journal} {Phys. Rev. Lett.}\ }\textbf {\bibinfo {volume} {96}},\
  \bibinfo {pages} {226402} (\bibinfo {year} {2006})}\BibitemShut {NoStop}%
\bibitem [{\citenamefont {Kotani}, \citenamefont {{van Schilfgaarde}},\ and\
  \citenamefont {Faleev}(2007)}]{Kotani_2007}%
  \BibitemOpen
  \bibfield  {author} {\bibinfo {author} {\bibfnamefont {T.}~\bibnamefont
  {Kotani}}, \bibinfo {author} {\bibfnamefont {M.}~\bibnamefont {{van
  Schilfgaarde}}},\ and\ \bibinfo {author} {\bibfnamefont {S.~V.}\ \bibnamefont
  {Faleev}},\ }\bibfield  {title} {\enquote {\bibinfo {title} {Quasiparticle
  self-consistent {{G W}} method: {{A}} basis for the independent-particle
  approximation},}\ }\href {https://doi.org/10.1103/PhysRevB.76.165106}
  {\bibfield  {journal} {\bibinfo  {journal} {Phys. Rev. B}\ }\textbf {\bibinfo
  {volume} {76}},\ \bibinfo {pages} {165106} (\bibinfo {year}
  {2007})}\BibitemShut {NoStop}%
\bibitem [{\citenamefont {Ke}(2011)}]{Ke_2011}%
  \BibitemOpen
  \bibfield  {author} {\bibinfo {author} {\bibfnamefont {S.-H.}\ \bibnamefont
  {Ke}},\ }\bibfield  {title} {\enquote {\bibinfo {title} {All-electron {{G W}}
  methods implemented in molecular orbital space: {{Ionization}} energy and
  electron affinity of conjugated molecules},}\ }\href
  {https://doi.org/10.1103/PhysRevB.84.205415} {\bibfield  {journal} {\bibinfo
  {journal} {Phys. Rev. B}\ }\textbf {\bibinfo {volume} {84}},\ \bibinfo
  {pages} {205415} (\bibinfo {year} {2011})}\BibitemShut {NoStop}%
\bibitem [{\citenamefont {Kaplan}\ \emph {et~al.}(2016)\citenamefont {Kaplan},
  \citenamefont {Harding}, \citenamefont {Seiler}, \citenamefont {Weigend},
  \citenamefont {Evers},\ and\ \citenamefont {{van Setten}}}]{Kaplan_2016}%
  \BibitemOpen
  \bibfield  {author} {\bibinfo {author} {\bibfnamefont {F.}~\bibnamefont
  {Kaplan}}, \bibinfo {author} {\bibfnamefont {M.~E.}\ \bibnamefont {Harding}},
  \bibinfo {author} {\bibfnamefont {C.}~\bibnamefont {Seiler}}, \bibinfo
  {author} {\bibfnamefont {F.}~\bibnamefont {Weigend}}, \bibinfo {author}
  {\bibfnamefont {F.}~\bibnamefont {Evers}},\ and\ \bibinfo {author}
  {\bibfnamefont {M.~J.}\ \bibnamefont {{van Setten}}},\ }\bibfield  {title}
  {\enquote {\bibinfo {title} {Quasi-{{Particle Self}}-{{Consistent}}
  {{{\emph{GW}}}} for {{Molecules}}},}\ }\href
  {https://doi.org/10.1021/acs.jctc.5b01238} {\bibfield  {journal} {\bibinfo
  {journal} {J. Chem. Theory Comput.}\ }\textbf {\bibinfo {volume} {12}},\
  \bibinfo {pages} {2528--2541} (\bibinfo {year} {2016})}\BibitemShut {NoStop}%
\bibitem [{\citenamefont {Marie}\ and\ \citenamefont
  {Loos}(2023)}]{Marie_2023}%
  \BibitemOpen
  \bibfield  {author} {\bibinfo {author} {\bibfnamefont {A.}~\bibnamefont
  {Marie}}\ and\ \bibinfo {author} {\bibfnamefont {P.-F.}\ \bibnamefont
  {Loos}},\ }\bibfield  {title} {\enquote {\bibinfo {title} {A {{Similarity
  Renormalization Group Approach}} to {{Green}}'s {{Function Methods}}},}\
  }\href {https://doi.org/10.1021/acs.jctc.3c00281} {\bibfield  {journal}
  {\bibinfo  {journal} {J. Chem. Theory Comput.}\ }\textbf {\bibinfo {volume}
  {19}},\ \bibinfo {pages} {3943--3957} (\bibinfo {year} {2023})}\BibitemShut
  {NoStop}%
\bibitem [{\citenamefont {Moiseyev}, \citenamefont {Certain},\ and\
  \citenamefont {Weinhold}(1978)}]{Moiseyev_1978}%
  \BibitemOpen
  \bibfield  {author} {\bibinfo {author} {\bibfnamefont {N.}~\bibnamefont
  {Moiseyev}}, \bibinfo {author} {\bibfnamefont {P.~R.}\ \bibnamefont
  {Certain}},\ and\ \bibinfo {author} {\bibfnamefont {F.}~\bibnamefont
  {Weinhold}},\ }\bibfield  {title} {\enquote {\bibinfo {title} {{Resonance
  properties of complex-rotated hamiltonians}},}\ }\href
  {https://www.tandfonline.com/doi/abs/10.1080/00268977800102631} {\bibfield
  {journal} {\bibinfo  {journal} {Mol. Phys.}\ }\textbf {\bibinfo {volume}
  {36}},\ \bibinfo {pages} {1613--1630} (\bibinfo {year} {1978})}\BibitemShut
  {NoStop}%
\bibitem [{Note1()}]{Note1}%
  \BibitemOpen
  \bibinfo {note} {Let $\protect \tilde {\protect \boldsymbol {C}}(\eta )$ be
  the matrix of non-orthonormal eigenvectors, i.e., the results of a
  diagonalisation method for non-Hermitian complex matrices. Then, to ensure
  the normalization condition, we perform an adapted Cholesky decomposition to
  get a lower triangular matrix $\protect \boldsymbol {L}$ fulfilling $\protect
  \tilde {\protect \boldsymbol {C}}(\eta )^{\intercal } \cdot \protect
  \boldsymbol {S}\cdot \protect \tilde {\protect \boldsymbol {C}}(\eta ) =
  \protect \boldsymbol {L}\cdot \protect \boldsymbol {L}^{\intercal }$. Then,
  to obtain the orthonormalized eigenvectors, we transform the original
  coefficients matrix using the inverse transpose of the Cholesky factor
  $\protect \boldsymbol {L}$, as follows: $\protect \boldsymbol {C}(\eta ) =
  \protect \tilde {\protect \boldsymbol {C}}(\eta ) \protect \boldsymbol
  {L}^{-\intercal }$.}\BibitemShut {Stop}%
\bibitem [{\citenamefont {Pulay}(1980)}]{Pulay_1980}%
  \BibitemOpen
  \bibfield  {author} {\bibinfo {author} {\bibfnamefont {P.}~\bibnamefont
  {Pulay}},\ }\bibfield  {title} {\enquote {\bibinfo {title} {{Convergence
  Acceleration of Iterative Sequences. The Case of SCF Iteration}},}\ }\href
  {https://doi.org/10.1016/0009-2614(80)80396-4} {\bibfield  {journal}
  {\bibinfo  {journal} {Chem. Phys. Lett.}\ }\textbf {\bibinfo {volume} {73}},\
  \bibinfo {pages} {393--398} (\bibinfo {year} {1980})}\BibitemShut {NoStop}%
\bibitem [{\citenamefont {Pulay}(1982)}]{Pulay_1982}%
  \BibitemOpen
  \bibfield  {author} {\bibinfo {author} {\bibfnamefont {P.}~\bibnamefont
  {Pulay}},\ }\bibfield  {title} {\enquote {\bibinfo {title} {{Improved SCF
  convergence acceleration}},}\ }\href
  {https://doi.org/https://doi.org/10.1002/jcc.540030413} {\bibfield  {journal}
  {\bibinfo  {journal} {J. Comput. Chem.}\ }\textbf {\bibinfo {volume} {3}},\
  \bibinfo {pages} {556--560} (\bibinfo {year} {1982})}\BibitemShut {NoStop}%
\bibitem [{\citenamefont {Santra}, \citenamefont {Cederbaum},\ and\
  \citenamefont {Meyer}(1999)}]{Santra_1999}%
  \BibitemOpen
  \bibfield  {author} {\bibinfo {author} {\bibfnamefont {R.}~\bibnamefont
  {Santra}}, \bibinfo {author} {\bibfnamefont {L.}~\bibnamefont {Cederbaum}},\
  and\ \bibinfo {author} {\bibfnamefont {H.-D.}\ \bibnamefont {Meyer}},\
  }\bibfield  {title} {\enquote {\bibinfo {title} {Electronic decay of
  molecular clusters: non-stationary states computed by standard quantum
  chemistry methods},}\ }\href
  {https://doi.org/https://doi.org/10.1016/S0009-2614(99)00226-2} {\bibfield
  {journal} {\bibinfo  {journal} {Chem. Phys. Lett.}\ }\textbf {\bibinfo
  {volume} {303}},\ \bibinfo {pages} {413--419} (\bibinfo {year}
  {1999})}\BibitemShut {NoStop}%
\bibitem [{\citenamefont {Aryasetiawan}\ and\ \citenamefont
  {Gunnarsson}(1998)}]{Aryasetiawan_1998}%
  \BibitemOpen
  \bibfield  {author} {\bibinfo {author} {\bibfnamefont {F.}~\bibnamefont
  {Aryasetiawan}}\ and\ \bibinfo {author} {\bibfnamefont {O.}~\bibnamefont
  {Gunnarsson}},\ }\bibfield  {title} {\enquote {\bibinfo {title} {{The $GW$
  Method}},}\ }\href {https://doi.org/10.1088/0034-4885/61/3/002} {\bibfield
  {journal} {\bibinfo  {journal} {Rep. Prog. Phys.}\ }\textbf {\bibinfo
  {volume} {61}},\ \bibinfo {pages} {237--312} (\bibinfo {year}
  {1998})}\BibitemShut {NoStop}%
\bibitem [{\citenamefont {Onida}, \citenamefont {Reining},\ and\ \citenamefont
  {Rubio}(2002)}]{Onida_2002}%
  \BibitemOpen
  \bibfield  {author} {\bibinfo {author} {\bibfnamefont {G.}~\bibnamefont
  {Onida}}, \bibinfo {author} {\bibfnamefont {L.}~\bibnamefont {Reining}},\
  and\ \bibinfo {author} {\bibfnamefont {A.}~\bibnamefont {Rubio}},\ }\bibfield
   {title} {\enquote {\bibinfo {title} {Electronic excitations:
  Density-functional versus many-body green's function approaches},}\ }\href
  {https://doi.org/10.1103/RevModPhys.74.601} {\bibfield  {journal} {\bibinfo
  {journal} {Rev. Mod. Phys.}\ }\textbf {\bibinfo {volume} {74}},\ \bibinfo
  {pages} {601--659} (\bibinfo {year} {2002})}\BibitemShut {NoStop}%
\bibitem [{\citenamefont {Reining}(2017)}]{Reining_2017}%
  \BibitemOpen
  \bibfield  {author} {\bibinfo {author} {\bibfnamefont {L.}~\bibnamefont
  {Reining}},\ }\bibfield  {title} {\enquote {\bibinfo {title} {{The $GW$
  Approximation: Content, Successes and Limitations}},}\ }\href
  {https://doi.org/10.1002/wcms.1344} {\bibfield  {journal} {\bibinfo
  {journal} {WIREs Comput. Mol. Sci.}\ }\textbf {\bibinfo {volume} {8}},\
  \bibinfo {pages} {e1344} (\bibinfo {year} {2017})}\BibitemShut {NoStop}%
\bibitem [{\citenamefont {DickHoff}\ and\ \citenamefont
  {Neck}(2008)}]{DickhoffBook}%
  \BibitemOpen
  \bibfield  {author} {\bibinfo {author} {\bibfnamefont {W.~H.}\ \bibnamefont
  {DickHoff}}\ and\ \bibinfo {author} {\bibfnamefont {D.~V.}\ \bibnamefont
  {Neck}},\ }\href@noop {} {\emph {\bibinfo {title} {Many-Body Theory
  Exposed!}}}\ (\bibinfo  {publisher} {{World Sientific}},\ \bibinfo {year}
  {2008})\BibitemShut {NoStop}%
\bibitem [{\citenamefont {Starke}\ and\ \citenamefont
  {Kresse}(2012)}]{Starke_2012}%
  \BibitemOpen
  \bibfield  {author} {\bibinfo {author} {\bibfnamefont {R.}~\bibnamefont
  {Starke}}\ and\ \bibinfo {author} {\bibfnamefont {G.}~\bibnamefont
  {Kresse}},\ }\bibfield  {title} {\enquote {\bibinfo {title} {{Self-consistent
  Green function equations and the hierarchy of approximations for the
  four-point propagator}},}\ }\href
  {https://doi.org/10.1103/PhysRevB.85.075119} {\bibfield  {journal} {\bibinfo
  {journal} {Phys. Rev. B}\ }\textbf {\bibinfo {volume} {85}},\ \bibinfo
  {pages} {075119} (\bibinfo {year} {2012})}\BibitemShut {NoStop}%
\bibitem [{\citenamefont {Orlando}, \citenamefont {Romaniello},\ and\
  \citenamefont {Loos}(2023)}]{Orlando_2023b}%
  \BibitemOpen
  \bibfield  {author} {\bibinfo {author} {\bibfnamefont {R.}~\bibnamefont
  {Orlando}}, \bibinfo {author} {\bibfnamefont {P.}~\bibnamefont
  {Romaniello}},\ and\ \bibinfo {author} {\bibfnamefont {P.-F.}\ \bibnamefont
  {Loos}},\ }\bibfield  {title} {\enquote {\bibinfo {title} {{The three
  channels of many-body perturbation theory: GW, particle--particle, and
  electron--hole T-matrix self-energies}},}\ }\href
  {https://doi.org/10.1063/5.0176898} {\bibfield  {journal} {\bibinfo
  {journal} {J. Chem. Phys.}\ }\textbf {\bibinfo {volume} {159}},\ \bibinfo
  {pages} {184113} (\bibinfo {year} {2023})}\BibitemShut {NoStop}%
\bibitem [{\citenamefont {Marie}, \citenamefont {Romaniello},\ and\
  \citenamefont {Loos}(2024)}]{Marie_2024c}%
  \BibitemOpen
  \bibfield  {author} {\bibinfo {author} {\bibfnamefont {A.}~\bibnamefont
  {Marie}}, \bibinfo {author} {\bibfnamefont {P.}~\bibnamefont {Romaniello}},\
  and\ \bibinfo {author} {\bibfnamefont {P.-F.}\ \bibnamefont {Loos}},\
  }\bibfield  {title} {\enquote {\bibinfo {title} {Anomalous propagators and
  the particle-particle channel: {{Hedin}}'s equations},}\ }\href
  {https://doi.org/10.1103/PhysRevB.110.115155} {\bibfield  {journal} {\bibinfo
   {journal} {Phys. Rev. B}\ }\textbf {\bibinfo {volume} {110}},\ \bibinfo
  {pages} {115155} (\bibinfo {year} {2024})}\BibitemShut {NoStop}%
\bibitem [{\citenamefont {Marie}\ \emph {et~al.}(2025)\citenamefont {Marie},
  \citenamefont {Romaniello}, \citenamefont {Blase},\ and\ \citenamefont
  {Loos}}]{Marie_2025}%
  \BibitemOpen
  \bibfield  {author} {\bibinfo {author} {\bibfnamefont {A.}~\bibnamefont
  {Marie}}, \bibinfo {author} {\bibfnamefont {P.}~\bibnamefont {Romaniello}},
  \bibinfo {author} {\bibfnamefont {X.}~\bibnamefont {Blase}},\ and\ \bibinfo
  {author} {\bibfnamefont {P.-F.}\ \bibnamefont {Loos}},\ }\bibfield  {title}
  {\enquote {\bibinfo {title} {Anomalous propagators and the particle--particle
  channel: {{Bethe}}--{{Salpeter}} equation},}\ }\href
  {https://doi.org/10.1063/5.0250155} {\bibfield  {journal} {\bibinfo
  {journal} {J. Chem. Phys.}\ }\textbf {\bibinfo {volume} {162}},\ \bibinfo
  {pages} {134105} (\bibinfo {year} {2025})}\BibitemShut {NoStop}%
\bibitem [{\citenamefont {Bohm}\ and\ \citenamefont {Pines}(1951)}]{Bohm_1951}%
  \BibitemOpen
  \bibfield  {author} {\bibinfo {author} {\bibfnamefont {D.}~\bibnamefont
  {Bohm}}\ and\ \bibinfo {author} {\bibfnamefont {D.}~\bibnamefont {Pines}},\
  }\bibfield  {title} {\enquote {\bibinfo {title} {{A Collective Description of
  Electron Interactions. I. Magnetic Interactions}},}\ }\href
  {https://doi.org/10.1103/PhysRev.82.625} {\bibfield  {journal} {\bibinfo
  {journal} {Phys. Rev.}\ }\textbf {\bibinfo {volume} {82}},\ \bibinfo {pages}
  {625--634} (\bibinfo {year} {1951})}\BibitemShut {NoStop}%
\bibitem [{\citenamefont {Pines}\ and\ \citenamefont
  {Bohm}(1952)}]{Pines_1952}%
  \BibitemOpen
  \bibfield  {author} {\bibinfo {author} {\bibfnamefont {D.}~\bibnamefont
  {Pines}}\ and\ \bibinfo {author} {\bibfnamefont {D.}~\bibnamefont {Bohm}},\
  }\bibfield  {title} {\enquote {\bibinfo {title} {{A Collective Description of
  Electron Interactions: II. Collective $\mathrm{vs}$ Individual Particle
  Aspects of the Interactions}},}\ }\href
  {https://doi.org/10.1103/PhysRev.85.338} {\bibfield  {journal} {\bibinfo
  {journal} {Phys. Rev.}\ }\textbf {\bibinfo {volume} {85}},\ \bibinfo {pages}
  {338--353} (\bibinfo {year} {1952})}\BibitemShut {NoStop}%
\bibitem [{\citenamefont {Bohm}\ and\ \citenamefont {Pines}(1953)}]{Bohm_1953}%
  \BibitemOpen
  \bibfield  {author} {\bibinfo {author} {\bibfnamefont {D.}~\bibnamefont
  {Bohm}}\ and\ \bibinfo {author} {\bibfnamefont {D.}~\bibnamefont {Pines}},\
  }\bibfield  {title} {\enquote {\bibinfo {title} {{A Collective Description of
  Electron Interactions: III. Coulomb Interactions in a Degenerate Electron
  Gas}},}\ }\href {https://doi.org/10.1103/PhysRev.92.609} {\bibfield
  {journal} {\bibinfo  {journal} {Phys. Rev.}\ }\textbf {\bibinfo {volume}
  {92}},\ \bibinfo {pages} {609--625} (\bibinfo {year} {1953})}\BibitemShut
  {NoStop}%
\bibitem [{\citenamefont {Nozi\`eres}\ and\ \citenamefont
  {Pines}(1958)}]{Nozieres_1958}%
  \BibitemOpen
  \bibfield  {author} {\bibinfo {author} {\bibfnamefont {P.}~\bibnamefont
  {Nozi\`eres}}\ and\ \bibinfo {author} {\bibfnamefont {D.}~\bibnamefont
  {Pines}},\ }\bibfield  {title} {\enquote {\bibinfo {title} {{Correlation
  Energy of a Free Electron Gas}},}\ }\href
  {https://doi.org/10.1103/PhysRev.111.442} {\bibfield  {journal} {\bibinfo
  {journal} {Phys. Rev.}\ }\textbf {\bibinfo {volume} {111}},\ \bibinfo {pages}
  {442--454} (\bibinfo {year} {1958})}\BibitemShut {NoStop}%
\bibitem [{\citenamefont {Holzer}\ \emph {et~al.}(2019)\citenamefont {Holzer},
  \citenamefont {Teale}, \citenamefont {Hampe}, \citenamefont {Stopkowicz},
  \citenamefont {Helgaker},\ and\ \citenamefont {Klopper}}]{Holzer_2019}%
  \BibitemOpen
  \bibfield  {author} {\bibinfo {author} {\bibfnamefont {C.}~\bibnamefont
  {Holzer}}, \bibinfo {author} {\bibfnamefont {A.~M.}\ \bibnamefont {Teale}},
  \bibinfo {author} {\bibfnamefont {F.}~\bibnamefont {Hampe}}, \bibinfo
  {author} {\bibfnamefont {S.}~\bibnamefont {Stopkowicz}}, \bibinfo {author}
  {\bibfnamefont {T.}~\bibnamefont {Helgaker}},\ and\ \bibinfo {author}
  {\bibfnamefont {W.}~\bibnamefont {Klopper}},\ }\bibfield  {title} {\enquote
  {\bibinfo {title} {{$GW$ Quasiparticle Energies of Atoms in Strong Magnetic
  Fields}},}\ }\href {https://doi.org/10.1063/1.5093396} {\bibfield  {journal}
  {\bibinfo  {journal} {J. Chem. Phys.}\ }\textbf {\bibinfo {volume} {150}},\
  \bibinfo {pages} {214112} (\bibinfo {year} {2019})}\BibitemShut {NoStop}%
\bibitem [{Note2()}]{Note2}%
  \BibitemOpen
  \bibinfo {note} {Practically, we ensure the normalization of the RPA
  eigenvectors with the same method as introduced in the HF section, i.e., by a
  Cholesky decomposition of \begin {equation*} \begin {pmatrix} \protect
  \boldsymbol {X}& \protect \boldsymbol {Y}\\ \protect \boldsymbol {Y}&
  \protect \boldsymbol {X}\end {pmatrix}^{\intercal } \cdot \begin {pmatrix}
  \protect \boldsymbol {1}& \protect \boldsymbol {0}\\ \protect \boldsymbol
  {0}& -\protect \boldsymbol {1}\end {pmatrix} \cdot \begin {pmatrix} \protect
  \boldsymbol {X}& \protect \boldsymbol {Y}\\ \protect \boldsymbol {Y}&
  \protect \boldsymbol {X}\end {pmatrix} \end {equation*}}\BibitemShut
  {NoStop}%
\bibitem [{\citenamefont {Szabo}\ and\ \citenamefont
  {Ostlund}(1989)}]{SzaboBook}%
  \BibitemOpen
  \bibfield  {author} {\bibinfo {author} {\bibfnamefont {A.}~\bibnamefont
  {Szabo}}\ and\ \bibinfo {author} {\bibfnamefont {N.~S.}\ \bibnamefont
  {Ostlund}},\ }\href@noop {} {\emph {\bibinfo {title} {Modern quantum
  chemistry}}}\ (\bibinfo  {publisher} {McGraw-Hill},\ \bibinfo {address} {New
  York},\ \bibinfo {year} {1989})\BibitemShut {NoStop}%
\bibitem [{\citenamefont {Loos}(2019)}]{quack}%
  \BibitemOpen
  \bibfield  {author} {\bibinfo {author} {\bibfnamefont {P.~F.}\ \bibnamefont
  {Loos}},\ }\href {https://doi.org/10.5281/zenodo.3745928} {\enquote {\bibinfo
  {title} {{{QuAcK: a software for emerging quantum electronic structure
  methods}}},}\ } (\bibinfo {year} {2019}),\ \bibinfo {note}
  {\url{https://github.com/pfloos/QuAcK}}\BibitemShut {NoStop}%
\bibitem [{\citenamefont {Gayvert}(2024)}]{opencap}%
  \BibitemOpen
  \bibfield  {author} {\bibinfo {author} {\bibfnamefont {J.}~\bibnamefont
  {Gayvert}},\ }\href {https://github.com/gayverjr/opencap} {\enquote {\bibinfo
  {title} {Opencap: An open-source program for studying resonances in
  molecules},}\ } (\bibinfo {year} {2024}),\ \bibinfo {note} {commit: 1562fce.
  Accessed: 2025-04-28}\BibitemShut {NoStop}%
\bibitem [{\citenamefont {V{\'e}ril}\ \emph {et~al.}(2018)\citenamefont
  {V{\'e}ril}, \citenamefont {Romaniello}, \citenamefont {Berger},\ and\
  \citenamefont {Loos}}]{Veril_2018}%
  \BibitemOpen
  \bibfield  {author} {\bibinfo {author} {\bibfnamefont {M.}~\bibnamefont
  {V{\'e}ril}}, \bibinfo {author} {\bibfnamefont {P.}~\bibnamefont
  {Romaniello}}, \bibinfo {author} {\bibfnamefont {J.~A.}\ \bibnamefont
  {Berger}},\ and\ \bibinfo {author} {\bibfnamefont {P.~F.}\ \bibnamefont
  {Loos}},\ }\bibfield  {title} {\enquote {\bibinfo {title} {Unphysical
  discontinuities in gw methods},}\ }\href
  {https://doi.org/10.1021/acs.jctc.8b00745} {\bibfield  {journal} {\bibinfo
  {journal} {J. Chem. Theory Comput.}\ }\textbf {\bibinfo {volume} {14}},\
  \bibinfo {pages} {5220} (\bibinfo {year} {2018})}\BibitemShut {NoStop}%
\bibitem [{\citenamefont {Nestmann}\ and\ \citenamefont
  {Peyerimhoff}(1985)}]{Nestmann_1985b}%
  \BibitemOpen
  \bibfield  {author} {\bibinfo {author} {\bibfnamefont {B.~M.}\ \bibnamefont
  {Nestmann}}\ and\ \bibinfo {author} {\bibfnamefont {S.~D.}\ \bibnamefont
  {Peyerimhoff}},\ }\bibfield  {title} {\enquote {\bibinfo {title} {{CI method
  for determining the location and width of resonances in electron-molecule
  collision processes}},}\ }\href {https://doi.org/10.1088/0022-3700/18/21/017}
  {\bibfield  {journal} {\bibinfo  {journal} {J. Phys. B: At. Mol. Phys.}\
  }\textbf {\bibinfo {volume} {18}},\ \bibinfo {pages} {4309} (\bibinfo {year}
  {1985})}\BibitemShut {NoStop}%
\bibitem [{\citenamefont {V{\'e}ril}\ \emph {et~al.}()\citenamefont
  {V{\'e}ril}, \citenamefont {Scemama}, \citenamefont {Caffarel}, \citenamefont
  {Lipparini}, \citenamefont {Boggio-Pasqua}, \citenamefont {Jacquemin},\ and\
  \citenamefont {Loos}}]{Veril_2021}%
  \BibitemOpen
  \bibfield  {author} {\bibinfo {author} {\bibfnamefont {M.}~\bibnamefont
  {V{\'e}ril}}, \bibinfo {author} {\bibfnamefont {A.}~\bibnamefont {Scemama}},
  \bibinfo {author} {\bibfnamefont {M.}~\bibnamefont {Caffarel}}, \bibinfo
  {author} {\bibfnamefont {F.}~\bibnamefont {Lipparini}}, \bibinfo {author}
  {\bibfnamefont {M.}~\bibnamefont {Boggio-Pasqua}}, \bibinfo {author}
  {\bibfnamefont {D.}~\bibnamefont {Jacquemin}},\ and\ \bibinfo {author}
  {\bibfnamefont {P.-F.}\ \bibnamefont {Loos}},\ }\bibfield  {title} {\enquote
  {\bibinfo {title} {{QUESTDB: A database of highly accurate excitation
  energies for the electronic structure community}},}\ }\href
  {https://doi.org/https://doi.org/10.1002/wcms.1517} {\bibfield  {journal}
  {\bibinfo  {journal} {WIREs Comput. Mol. Sci.}\ }\textbf {\bibinfo {volume}
  {11}},\ \bibinfo {pages} {e1517}}\BibitemShut {NoStop}%
\bibitem [{\citenamefont {Gayvert}\ and\ \citenamefont
  {Bravaya}(2022)}]{Gayvert_2022}%
  \BibitemOpen
  \bibfield  {author} {\bibinfo {author} {\bibfnamefont {J.~R.}\ \bibnamefont
  {Gayvert}}\ and\ \bibinfo {author} {\bibfnamefont {K.~B.}\ \bibnamefont
  {Bravaya}},\ }\bibfield  {title} {\enquote {\bibinfo {title} {{Projected
  CAP-EOM-CCSD method for electronic resonances}},}\ }\href
  {https://doi.org/10.1063/5.0082739} {\bibfield  {journal} {\bibinfo
  {journal} {J. Chem. Phys.}\ }\textbf {\bibinfo {volume} {156}},\ \bibinfo
  {pages} {094108} (\bibinfo {year} {2022})}\BibitemShut {NoStop}%
\bibitem [{\citenamefont {Berman}\ \emph {et~al.}(1983)\citenamefont {Berman},
  \citenamefont {Estrada}, \citenamefont {Cederbaum},\ and\ \citenamefont
  {Domcke}}]{Berman_1983}%
  \BibitemOpen
  \bibfield  {author} {\bibinfo {author} {\bibfnamefont {M.}~\bibnamefont
  {Berman}}, \bibinfo {author} {\bibfnamefont {H.}~\bibnamefont {Estrada}},
  \bibinfo {author} {\bibfnamefont {L.~S.}\ \bibnamefont {Cederbaum}},\ and\
  \bibinfo {author} {\bibfnamefont {W.}~\bibnamefont {Domcke}},\ }\bibfield
  {title} {\enquote {\bibinfo {title} {{Nuclear dynamics in resonant
  electron-molecule scattering beyond the local approximation: The 2.3-eV shape
  resonance in ${\mathrm{N}}_{2}$}},}\ }\href
  {https://doi.org/10.1103/PhysRevA.28.1363} {\bibfield  {journal} {\bibinfo
  {journal} {Phys. Rev. A}\ }\textbf {\bibinfo {volume} {28}},\ \bibinfo
  {pages} {1363--1381} (\bibinfo {year} {1983})}\BibitemShut {NoStop}%
\bibitem [{\citenamefont {Ehrhardt}\ \emph {et~al.}(1968)\citenamefont
  {Ehrhardt}, \citenamefont {Langhans}, \citenamefont {Linder},\ and\
  \citenamefont {Taylor}}]{Ehrhardt_1968}%
  \BibitemOpen
  \bibfield  {author} {\bibinfo {author} {\bibfnamefont {H.}~\bibnamefont
  {Ehrhardt}}, \bibinfo {author} {\bibfnamefont {L.}~\bibnamefont {Langhans}},
  \bibinfo {author} {\bibfnamefont {F.}~\bibnamefont {Linder}},\ and\ \bibinfo
  {author} {\bibfnamefont {H.~S.}\ \bibnamefont {Taylor}},\ }\bibfield  {title}
  {\enquote {\bibinfo {title} {{Resonance Scattering of Slow Electrons from
  ${\mathrm{H}}_{2}$ and CO Angular Distributions}},}\ }\href
  {https://doi.org/10.1103/PhysRev.173.222} {\bibfield  {journal} {\bibinfo
  {journal} {Phys. Rev.}\ }\textbf {\bibinfo {volume} {173}},\ \bibinfo {pages}
  {222--230} (\bibinfo {year} {1968})}\BibitemShut {NoStop}%
\bibitem [{\citenamefont {Zubek}\ and\ \citenamefont
  {Szmytkowski}(1977)}]{Zubek_1977}%
  \BibitemOpen
  \bibfield  {author} {\bibinfo {author} {\bibfnamefont {M.}~\bibnamefont
  {Zubek}}\ and\ \bibinfo {author} {\bibfnamefont {C.}~\bibnamefont
  {Szmytkowski}},\ }\bibfield  {title} {\enquote {\bibinfo {title}
  {{Calculation of resonant vibrational excitation of CO by scattering of
  electrons}},}\ }\href {https://doi.org/10.1088/0022-3700/10/1/006} {\bibfield
   {journal} {\bibinfo  {journal} {J. Phys. B: At. Mol. Phys.}\ }\textbf
  {\bibinfo {volume} {10}},\ \bibinfo {pages} {L27} (\bibinfo {year}
  {1977})}\BibitemShut {NoStop}%
\bibitem [{\citenamefont {Zubek}\ and\ \citenamefont
  {Szmytkowski}(1979)}]{Zubek_1979}%
  \BibitemOpen
  \bibfield  {author} {\bibinfo {author} {\bibfnamefont {M.}~\bibnamefont
  {Zubek}}\ and\ \bibinfo {author} {\bibfnamefont {C.}~\bibnamefont
  {Szmytkowski}},\ }\bibfield  {title} {\enquote {\bibinfo {title} {{Electron
  impact vibrational excitation of CO in the range 1--4 eV}},}\ }\href
  {https://doi.org/https://doi.org/10.1016/0375-9601(79)90583-8} {\bibfield
  {journal} {\bibinfo  {journal} {Phys. Lett. A}\ }\textbf {\bibinfo {volume}
  {74}},\ \bibinfo {pages} {60--62} (\bibinfo {year} {1979})}\BibitemShut
  {NoStop}%
\bibitem [{\citenamefont {Buckman}\ and\ \citenamefont
  {Lohmann}(1986)}]{Buckman_1986}%
  \BibitemOpen
  \bibfield  {author} {\bibinfo {author} {\bibfnamefont {S.~J.}\ \bibnamefont
  {Buckman}}\ and\ \bibinfo {author} {\bibfnamefont {B.}~\bibnamefont
  {Lohmann}},\ }\bibfield  {title} {\enquote {\bibinfo {title} {Electron
  scattering from co in the $^{2}\mathrm{\ensuremath{\Pi}}$ resonance
  region},}\ }\href {https://doi.org/10.1103/PhysRevA.34.1561} {\bibfield
  {journal} {\bibinfo  {journal} {Phys. Rev. A}\ }\textbf {\bibinfo {volume}
  {34}},\ \bibinfo {pages} {1561--1563} (\bibinfo {year} {1986})}\BibitemShut
  {NoStop}%
\bibitem [{\citenamefont {Szmytkowski}, \citenamefont {Maciag},\ and\
  \citenamefont {Karwasz}(1996)}]{Szmytkowski_1996}%
  \BibitemOpen
  \bibfield  {author} {\bibinfo {author} {\bibfnamefont {C.}~\bibnamefont
  {Szmytkowski}}, \bibinfo {author} {\bibfnamefont {K.}~\bibnamefont
  {Maciag}},\ and\ \bibinfo {author} {\bibfnamefont {G.}~\bibnamefont
  {Karwasz}},\ }\bibfield  {title} {\enquote {\bibinfo {title} {Absolute
  electron-scattering total cross section measurements for noble gas atoms and
  diatomic molecules},}\ }\href {https://doi.org/10.1088/0031-8949/54/3/006}
  {\bibfield  {journal} {\bibinfo  {journal} {Phys. Scr.}\ }\textbf {\bibinfo
  {volume} {54}},\ \bibinfo {pages} {271} (\bibinfo {year} {1996})}\BibitemShut
  {NoStop}%
\bibitem [{\citenamefont {Allan}(2010)}]{Allan_2010}%
  \BibitemOpen
  \bibfield  {author} {\bibinfo {author} {\bibfnamefont {M.}~\bibnamefont
  {Allan}},\ }\bibfield  {title} {\enquote {\bibinfo {title} {Electron
  collisions with co: Elastic and vibrational excitation cross sections},}\
  }\href {https://doi.org/10.1103/PhysRevA.81.042706} {\bibfield  {journal}
  {\bibinfo  {journal} {Phys. Rev. A}\ }\textbf {\bibinfo {volume} {81}},\
  \bibinfo {pages} {042706} (\bibinfo {year} {2010})}\BibitemShut {NoStop}%
\bibitem [{\citenamefont {Kochem}\ \emph {et~al.}(1985)\citenamefont {Kochem},
  \citenamefont {Sohn}, \citenamefont {Jung}, \citenamefont {Ehrhardt},\ and\
  \citenamefont {Chang}}]{Kochem_1985}%
  \BibitemOpen
  \bibfield  {author} {\bibinfo {author} {\bibfnamefont {K.~H.}\ \bibnamefont
  {Kochem}}, \bibinfo {author} {\bibfnamefont {W.}~\bibnamefont {Sohn}},
  \bibinfo {author} {\bibfnamefont {K.}~\bibnamefont {Jung}}, \bibinfo {author}
  {\bibfnamefont {H.}~\bibnamefont {Ehrhardt}},\ and\ \bibinfo {author}
  {\bibfnamefont {E.~S.}\ \bibnamefont {Chang}},\ }\bibfield  {title} {\enquote
  {\bibinfo {title} {{Direct and resonant vibrational excitation of C2H2 by
  electron impact from 0 to 3.6 eV}},}\ }\href
  {https://doi.org/10.1088/0022-3700/18/6/025} {\bibfield  {journal} {\bibinfo
  {journal} {J. Phys. B At. Mol. Opt. Phys.}\ }\textbf {\bibinfo {volume}
  {18}},\ \bibinfo {pages} {1253} (\bibinfo {year} {1985})}\BibitemShut
  {NoStop}%
\bibitem [{\citenamefont {Jordan}\ and\ \citenamefont
  {Burrow}(1978)}]{Jordan_1978}%
  \BibitemOpen
  \bibfield  {author} {\bibinfo {author} {\bibfnamefont {K.~D.}\ \bibnamefont
  {Jordan}}\ and\ \bibinfo {author} {\bibfnamefont {P.~D.}\ \bibnamefont
  {Burrow}},\ }\bibfield  {title} {\enquote {\bibinfo {title} {{Studies of the
  Temporary Anion States of Unsaturated Hydrocarbons ny Electron Transmission
  Spectroscopy}},}\ }\href {https://doi.org/10.1021/ar50129a004} {\bibfield
  {journal} {\bibinfo  {journal} {Acc. Chem. Res.}\ }\textbf {\bibinfo {volume}
  {11}},\ \bibinfo {pages} {341--348} (\bibinfo {year} {1978})}\BibitemShut
  {NoStop}%
\bibitem [{\citenamefont {Dressler}\ and\ \citenamefont
  {Allan}(1987)}]{Dressler_1987}%
  \BibitemOpen
  \bibfield  {author} {\bibinfo {author} {\bibfnamefont {R.}~\bibnamefont
  {Dressler}}\ and\ \bibinfo {author} {\bibfnamefont {M.}~\bibnamefont
  {Allan}},\ }\bibfield  {title} {\enquote {\bibinfo {title} {A dissociative
  electron attachment, electron transmission, and electron energy‐loss study
  of the temporary negative ion of acetylene},}\ }\href
  {https://doi.org/10.1063/1.452864} {\bibfield  {journal} {\bibinfo  {journal}
  {J. Chem. Phys.}\ }\textbf {\bibinfo {volume} {87}},\ \bibinfo {pages}
  {4510--4518} (\bibinfo {year} {1987})}\BibitemShut {NoStop}%
\bibitem [{\citenamefont {Andric}\ and\ \citenamefont
  {Hall}(1988)}]{Andric_1988}%
  \BibitemOpen
  \bibfield  {author} {\bibinfo {author} {\bibfnamefont {L.}~\bibnamefont
  {Andric}}\ and\ \bibinfo {author} {\bibfnamefont {R.~I.}\ \bibnamefont
  {Hall}},\ }\bibfield  {title} {\enquote {\bibinfo {title} {Resonance
  phenomena observed in electron scattering from acetylene},}\ }\href
  {https://doi.org/10.1088/0953-4075/21/2/019} {\bibfield  {journal} {\bibinfo
  {journal} {J. Phys. B At. Mol. Opt. Phys.}\ }\textbf {\bibinfo {volume}
  {21}},\ \bibinfo {pages} {355} (\bibinfo {year} {1988})}\BibitemShut
  {NoStop}%
\bibitem [{\citenamefont {Szmytkowski}\ \emph {et~al.}(2014)\citenamefont
  {Szmytkowski}, \citenamefont {Mo\ifmmode~\dot{z}\else \.{z}\fi{}ejko},
  \citenamefont {Zawadzki}, \citenamefont {Macia\ifmmode~\mbox{\c{}}\else
  \c{}\fi{}g},\ and\ \citenamefont {Ptasi\ifmmode \acute{n}\else~\'{n}\fi{}ska
  Denga}}]{Szmytkowski_2014}%
  \BibitemOpen
  \bibfield  {author} {\bibinfo {author} {\bibfnamefont {C.}~\bibnamefont
  {Szmytkowski}}, \bibinfo {author} {\bibfnamefont {P.}~\bibnamefont
  {Mo\ifmmode~\dot{z}\else \.{z}\fi{}ejko}}, \bibinfo {author} {\bibfnamefont
  {M.}~\bibnamefont {Zawadzki}}, \bibinfo {author} {\bibfnamefont
  {K.}~\bibnamefont {Macia\ifmmode~\mbox{\c{}}\else \c{}\fi{}g}},\ and\
  \bibinfo {author} {\bibfnamefont {E.~d.~z.}\ \bibnamefont {Ptasi\ifmmode
  \acute{n}\else~\'{n}\fi{}ska Denga}},\ }\bibfield  {title} {\enquote
  {\bibinfo {title} {Electron-scattering cross sections for selected alkyne
  molecules: Measurements and calculations},}\ }\href
  {https://doi.org/10.1103/PhysRevA.89.052702} {\bibfield  {journal} {\bibinfo
  {journal} {Phys. Rev. A}\ }\textbf {\bibinfo {volume} {89}},\ \bibinfo
  {pages} {052702} (\bibinfo {year} {2014})}\BibitemShut {NoStop}%
\bibitem [{\citenamefont {Sanche}\ and\ \citenamefont
  {Schulz}(1973)}]{Sanche_1973}%
  \BibitemOpen
  \bibfield  {author} {\bibinfo {author} {\bibfnamefont {L.}~\bibnamefont
  {Sanche}}\ and\ \bibinfo {author} {\bibfnamefont {G.~J.}\ \bibnamefont
  {Schulz}},\ }\bibfield  {title} {\enquote {\bibinfo {title} {Electron
  transmission spectroscopy: Resonances in triatomic molecules and
  hydrocarbons},}\ }\href {https://doi.org/10.1063/1.1679228} {\bibfield
  {journal} {\bibinfo  {journal} {J. Chem. Phys.}\ }\textbf {\bibinfo {volume}
  {58}},\ \bibinfo {pages} {479--493} (\bibinfo {year} {1973})}\BibitemShut
  {NoStop}%
\bibitem [{\citenamefont {Walker}, \citenamefont {Stamatovic},\ and\
  \citenamefont {Wong}(1978)}]{Walker_1978}%
  \BibitemOpen
  \bibfield  {author} {\bibinfo {author} {\bibfnamefont {I.~C.}\ \bibnamefont
  {Walker}}, \bibinfo {author} {\bibfnamefont {A.}~\bibnamefont {Stamatovic}},\
  and\ \bibinfo {author} {\bibfnamefont {S.~F.}\ \bibnamefont {Wong}},\
  }\bibfield  {title} {\enquote {\bibinfo {title} {{Vibrational excitation of
  ethylene by electron impact: 1--11 eV}},}\ }\href
  {https://doi.org/10.1063/1.436547} {\bibfield  {journal} {\bibinfo  {journal}
  {J. Chem. Phys.}\ }\textbf {\bibinfo {volume} {69}},\ \bibinfo {pages}
  {5532--5537} (\bibinfo {year} {1978})}\BibitemShut {NoStop}%
\bibitem [{\citenamefont {Lunt}\ \emph {et~al.}(1994)\citenamefont {Lunt},
  \citenamefont {Randell}, \citenamefont {Ziesel}, \citenamefont {Mrotzek},\
  and\ \citenamefont {Field}}]{Lunt_1994}%
  \BibitemOpen
  \bibfield  {author} {\bibinfo {author} {\bibfnamefont {S.~L.}\ \bibnamefont
  {Lunt}}, \bibinfo {author} {\bibfnamefont {J.}~\bibnamefont {Randell}},
  \bibinfo {author} {\bibfnamefont {J.~P.}\ \bibnamefont {Ziesel}}, \bibinfo
  {author} {\bibfnamefont {G.}~\bibnamefont {Mrotzek}},\ and\ \bibinfo {author}
  {\bibfnamefont {D.}~\bibnamefont {Field}},\ }\bibfield  {title} {\enquote
  {\bibinfo {title} {Low-energy electron scattering from ch4, c2h4 and c2h6},}\
  }\href {https://doi.org/10.1088/0953-4075/27/7/016} {\bibfield  {journal}
  {\bibinfo  {journal} {J. Phys. B At. Mol. Opt. Phys.}\ }\textbf {\bibinfo
  {volume} {27}},\ \bibinfo {pages} {1407} (\bibinfo {year}
  {1994})}\BibitemShut {NoStop}%
\bibitem [{\citenamefont {Panajotovic}\ \emph {et~al.}(2003)\citenamefont
  {Panajotovic}, \citenamefont {Kitajima}, \citenamefont {Tanaka},
  \citenamefont {Jelisavcic}, \citenamefont {Lower}, \citenamefont {Campbell},
  \citenamefont {Brunger},\ and\ \citenamefont {Buckman}}]{Panajotovic_2003}%
  \BibitemOpen
  \bibfield  {author} {\bibinfo {author} {\bibfnamefont {R.}~\bibnamefont
  {Panajotovic}}, \bibinfo {author} {\bibfnamefont {M.}~\bibnamefont
  {Kitajima}}, \bibinfo {author} {\bibfnamefont {H.}~\bibnamefont {Tanaka}},
  \bibinfo {author} {\bibfnamefont {M.}~\bibnamefont {Jelisavcic}}, \bibinfo
  {author} {\bibfnamefont {J.}~\bibnamefont {Lower}}, \bibinfo {author}
  {\bibfnamefont {L.}~\bibnamefont {Campbell}}, \bibinfo {author}
  {\bibfnamefont {M.~J.}\ \bibnamefont {Brunger}},\ and\ \bibinfo {author}
  {\bibfnamefont {S.~J.}\ \bibnamefont {Buckman}},\ }\bibfield  {title}
  {\enquote {\bibinfo {title} {Electron collisions with ethylene},}\ }\href
  {https://doi.org/10.1088/0953-4075/36/8/314} {\bibfield  {journal} {\bibinfo
  {journal} {J. Phys. B At. Mol. Opt. Phys.}\ }\textbf {\bibinfo {volume}
  {36}},\ \bibinfo {pages} {1615} (\bibinfo {year} {2003})}\BibitemShut
  {NoStop}%
\bibitem [{\citenamefont {Szmytkowski}, \citenamefont {Kwitnewski},\ and\
  \citenamefont {Ptasi\ifmmode \acute{n}\else~\'{n}\fi{}ska
  Denga}(2003)}]{Szmytkowski_2003}%
  \BibitemOpen
  \bibfield  {author} {\bibinfo {author} {\bibfnamefont {C.}~\bibnamefont
  {Szmytkowski}}, \bibinfo {author} {\bibfnamefont {S.}~\bibnamefont
  {Kwitnewski}},\ and\ \bibinfo {author} {\bibfnamefont {E.~d.~z.}\
  \bibnamefont {Ptasi\ifmmode \acute{n}\else~\'{n}\fi{}ska Denga}},\ }\bibfield
   {title} {\enquote {\bibinfo {title} {Electron collisions with
  tetrafluoroethylene $({\mathrm{c}}_{2}{\mathrm{f}}_{4})$ and ethylene
  $({\mathrm{c}}_{2}{\mathrm{h}}_{4})$ molecules},}\ }\href
  {https://doi.org/10.1103/PhysRevA.68.032715} {\bibfield  {journal} {\bibinfo
  {journal} {Phys. Rev. A}\ }\textbf {\bibinfo {volume} {68}},\ \bibinfo
  {pages} {032715} (\bibinfo {year} {2003})}\BibitemShut {NoStop}%
\bibitem [{\citenamefont {Allan}, \citenamefont {Winstead},\ and\ \citenamefont
  {McKoy}(2008)}]{Allan_2008}%
  \BibitemOpen
  \bibfield  {author} {\bibinfo {author} {\bibfnamefont {M.}~\bibnamefont
  {Allan}}, \bibinfo {author} {\bibfnamefont {C.}~\bibnamefont {Winstead}},\
  and\ \bibinfo {author} {\bibfnamefont {V.}~\bibnamefont {McKoy}},\ }\bibfield
   {title} {\enquote {\bibinfo {title} {Electron scattering in ethene:
  Excitation of the $\stackrel{\ifmmode \tilde{}\else \~{}\fi{}}{a}\text{
  }{^{3}B}_{1u}$ state, elastic scattering, and vibrational excitation},}\
  }\href {https://doi.org/10.1103/PhysRevA.77.042715} {\bibfield  {journal}
  {\bibinfo  {journal} {Phys. Rev. A}\ }\textbf {\bibinfo {volume} {77}},\
  \bibinfo {pages} {042715} (\bibinfo {year} {2008})}\BibitemShut {NoStop}%
\bibitem [{\citenamefont {Khakoo}\ \emph {et~al.}(2016)\citenamefont {Khakoo},
  \citenamefont {Khakoo}, \citenamefont {Sakaamini}, \citenamefont {Hlousek},
  \citenamefont {Hargreaves}, \citenamefont {Lee},\ and\ \citenamefont
  {Murase}}]{Khakoo_2016}%
  \BibitemOpen
  \bibfield  {author} {\bibinfo {author} {\bibfnamefont {M.~A.}\ \bibnamefont
  {Khakoo}}, \bibinfo {author} {\bibfnamefont {S.~M.}\ \bibnamefont {Khakoo}},
  \bibinfo {author} {\bibfnamefont {A.}~\bibnamefont {Sakaamini}}, \bibinfo
  {author} {\bibfnamefont {B.~A.}\ \bibnamefont {Hlousek}}, \bibinfo {author}
  {\bibfnamefont {L.~R.}\ \bibnamefont {Hargreaves}}, \bibinfo {author}
  {\bibfnamefont {J.}~\bibnamefont {Lee}},\ and\ \bibinfo {author}
  {\bibfnamefont {R.}~\bibnamefont {Murase}},\ }\bibfield  {title} {\enquote
  {\bibinfo {title} {Low-energy elastic electron scattering from ethylene:
  Elastic scattering and vibrational excitation},}\ }\href
  {https://doi.org/10.1103/PhysRevA.93.012710} {\bibfield  {journal} {\bibinfo
  {journal} {Phys. Rev. A}\ }\textbf {\bibinfo {volume} {93}},\ \bibinfo
  {pages} {012710} (\bibinfo {year} {2016})}\BibitemShut {NoStop}%
\bibitem [{\citenamefont {Burrow}\ and\ \citenamefont
  {Michejda}(1976)}]{Burrow_1976}%
  \BibitemOpen
  \bibfield  {author} {\bibinfo {author} {\bibfnamefont {P.}~\bibnamefont
  {Burrow}}\ and\ \bibinfo {author} {\bibfnamefont {J.}~\bibnamefont
  {Michejda}},\ }\bibfield  {title} {\enquote {\bibinfo {title} {Electron
  transmission study of the formaldehyde electron affinity},}\ }\href
  {https://doi.org/https://doi.org/10.1016/0009-2614(76)80351-X} {\bibfield
  {journal} {\bibinfo  {journal} {Chem. Phys. Lett.}\ }\textbf {\bibinfo
  {volume} {42}},\ \bibinfo {pages} {223--226} (\bibinfo {year}
  {1976})}\BibitemShut {NoStop}%
\bibitem [{\citenamefont {{Van Veen}}, \citenamefont {{Van Dijk}},\ and\
  \citenamefont {Brongersma}(1976)}]{VanVeen_1976}%
  \BibitemOpen
  \bibfield  {author} {\bibinfo {author} {\bibfnamefont {E.}~\bibnamefont {{Van
  Veen}}}, \bibinfo {author} {\bibfnamefont {W.}~\bibnamefont {{Van Dijk}}},\
  and\ \bibinfo {author} {\bibfnamefont {H.}~\bibnamefont {Brongersma}},\
  }\bibfield  {title} {\enquote {\bibinfo {title} {Low-energy electron-impact
  excitation spectra of formaldehyde, acetaldehyde and acetone},}\ }\href
  {https://doi.org/https://doi.org/10.1016/0301-0104(76)87029-2} {\bibfield
  {journal} {\bibinfo  {journal} {Chem. Phys.}\ }\textbf {\bibinfo {volume}
  {16}},\ \bibinfo {pages} {337--345} (\bibinfo {year} {1976})}\BibitemShut
  {NoStop}%
\bibitem [{\citenamefont {Benoit}\ and\ \citenamefont
  {Abouaf}(1986)}]{Benoit_1986}%
  \BibitemOpen
  \bibfield  {author} {\bibinfo {author} {\bibfnamefont {C.}~\bibnamefont
  {Benoit}}\ and\ \bibinfo {author} {\bibfnamefont {R.}~\bibnamefont
  {Abouaf}},\ }\bibfield  {title} {\enquote {\bibinfo {title} {Low-energy
  electron collisions with formaldehyde: interference phenomena in the
  differential vibrational excitation cross section},}\ }\href
  {https://doi.org/https://doi.org/10.1016/0009-2614(86)87028-2} {\bibfield
  {journal} {\bibinfo  {journal} {Chem. Phys. Lett.}\ }\textbf {\bibinfo
  {volume} {123}},\ \bibinfo {pages} {134--138} (\bibinfo {year}
  {1986})}\BibitemShut {NoStop}%
\bibitem [{\citenamefont {Loos}\ \emph {et~al.}(2020)\citenamefont {Loos},
  \citenamefont {Scemama}, \citenamefont {Duchemin}, \citenamefont
  {Jacquemin},\ and\ \citenamefont {Blase}}]{Loos_2020e}%
  \BibitemOpen
  \bibfield  {author} {\bibinfo {author} {\bibfnamefont {P.-F.}\ \bibnamefont
  {Loos}}, \bibinfo {author} {\bibfnamefont {A.}~\bibnamefont {Scemama}},
  \bibinfo {author} {\bibfnamefont {I.}~\bibnamefont {Duchemin}}, \bibinfo
  {author} {\bibfnamefont {D.}~\bibnamefont {Jacquemin}},\ and\ \bibinfo
  {author} {\bibfnamefont {X.}~\bibnamefont {Blase}},\ }\bibfield  {title}
  {\enquote {\bibinfo {title} {{Pros and Cons of the Bethe--Salpeter Formalism
  for Ground-State Energies}},}\ }\href
  {https://doi.org/10.1021/acs.jpclett.0c00460} {\bibfield  {journal} {\bibinfo
   {journal} {J. Phys. Chem. Lett.}\ }\textbf {\bibinfo {volume} {11}},\
  \bibinfo {pages} {3536--3545} (\bibinfo {year} {2020})}\BibitemShut {NoStop}%
\bibitem [{\citenamefont {Monino}\ and\ \citenamefont
  {Loos}(2022)}]{Monino_2022}%
  \BibitemOpen
  \bibfield  {author} {\bibinfo {author} {\bibfnamefont {E.}~\bibnamefont
  {Monino}}\ and\ \bibinfo {author} {\bibfnamefont {P.-F.}\ \bibnamefont
  {Loos}},\ }\bibfield  {title} {\enquote {\bibinfo {title} {{Unphysical
  discontinuities, intruder states and regularization in $GW$ methods}},}\
  }\href {https://doi.org/10.1063/5.0089317} {\bibfield  {journal} {\bibinfo
  {journal} {J. Chem. Phys.}\ }\textbf {\bibinfo {volume} {156}},\ \bibinfo
  {pages} {231101} (\bibinfo {year} {2022})}\BibitemShut {NoStop}%
\bibitem [{\citenamefont {Burrow}\ and\ \citenamefont
  {Sanche}(1972)}]{Burrow_1972}%
  \BibitemOpen
  \bibfield  {author} {\bibinfo {author} {\bibfnamefont {P.~D.}\ \bibnamefont
  {Burrow}}\ and\ \bibinfo {author} {\bibfnamefont {L.}~\bibnamefont
  {Sanche}},\ }\bibfield  {title} {\enquote {\bibinfo {title} {{Elastic
  Scattering of Low-Energy Electrons at 180\ifmmode^\circ\else\textdegree\fi{}
  in C${\mathrm{O}}_{2}$}},}\ }\href
  {https://doi.org/10.1103/PhysRevLett.28.333} {\bibfield  {journal} {\bibinfo
  {journal} {Phys. Rev. Lett.}\ }\textbf {\bibinfo {volume} {28}},\ \bibinfo
  {pages} {333--336} (\bibinfo {year} {1972})}\BibitemShut {NoStop}%
\bibitem [{\citenamefont {Boness}\ and\ \citenamefont
  {Schulz}(1974)}]{Boness_1974}%
  \BibitemOpen
  \bibfield  {author} {\bibinfo {author} {\bibfnamefont {M.~J.~W.}\
  \bibnamefont {Boness}}\ and\ \bibinfo {author} {\bibfnamefont {G.~J.}\
  \bibnamefont {Schulz}},\ }\bibfield  {title} {\enquote {\bibinfo {title}
  {Vibrational excitation in c${\mathrm{o}}_{2}$ via the 3.8-ev resonance},}\
  }\href {https://doi.org/10.1103/PhysRevA.9.1969} {\bibfield  {journal}
  {\bibinfo  {journal} {Phys. Rev. A}\ }\textbf {\bibinfo {volume} {9}},\
  \bibinfo {pages} {1969--1979} (\bibinfo {year} {1974})}\BibitemShut {NoStop}%
\bibitem [{\citenamefont {Allan}(2001)}]{Allan_2001}%
  \BibitemOpen
  \bibfield  {author} {\bibinfo {author} {\bibfnamefont {M.}~\bibnamefont
  {Allan}},\ }\bibfield  {title} {\enquote {\bibinfo {title} {Selectivity in
  the excitation of fermi-coupled vibrations in ${\mathrm{co}}_{2}$ by impact
  of slow electrons},}\ }\href {https://doi.org/10.1103/PhysRevLett.87.033201}
  {\bibfield  {journal} {\bibinfo  {journal} {Phys. Rev. Lett.}\ }\textbf
  {\bibinfo {volume} {87}},\ \bibinfo {pages} {033201} (\bibinfo {year}
  {2001})}\BibitemShut {NoStop}%
\bibitem [{\citenamefont {Lozano}\ \emph {et~al.}(2022)\citenamefont {Lozano},
  \citenamefont {Garc{\'\i}a-Abenza}, \citenamefont {Blanco~Ramos},
  \citenamefont {Hasan}, \citenamefont {Slaughter}, \citenamefont {Weber},
  \citenamefont {McEachran}, \citenamefont {White}, \citenamefont {Brunger},
  \citenamefont {Lim{\~a}o-Vieira},\ and\ \citenamefont {Garc{\'\i}a
  G{\'o}mez-Tejedor}}]{Lozano_2022}%
  \BibitemOpen
  \bibfield  {author} {\bibinfo {author} {\bibfnamefont {A.~I.}\ \bibnamefont
  {Lozano}}, \bibinfo {author} {\bibfnamefont {A.}~\bibnamefont
  {Garc{\'\i}a-Abenza}}, \bibinfo {author} {\bibfnamefont {F.}~\bibnamefont
  {Blanco~Ramos}}, \bibinfo {author} {\bibfnamefont {M.}~\bibnamefont {Hasan}},
  \bibinfo {author} {\bibfnamefont {D.~S.}\ \bibnamefont {Slaughter}}, \bibinfo
  {author} {\bibfnamefont {T.}~\bibnamefont {Weber}}, \bibinfo {author}
  {\bibfnamefont {R.~P.}\ \bibnamefont {McEachran}}, \bibinfo {author}
  {\bibfnamefont {R.~D.}\ \bibnamefont {White}}, \bibinfo {author}
  {\bibfnamefont {M.~J.}\ \bibnamefont {Brunger}}, \bibinfo {author}
  {\bibfnamefont {P.}~\bibnamefont {Lim{\~a}o-Vieira}},\ and\ \bibinfo {author}
  {\bibfnamefont {G.}~\bibnamefont {Garc{\'\i}a G{\'o}mez-Tejedor}},\
  }\bibfield  {title} {\enquote {\bibinfo {title} {{Electron and Positron
  Scattering Cross Sections from CO2: A Comparative Study over a Broad Energy
  Range (0.1--5000 eV)}},}\ }\href {https://doi.org/10.1021/acs.jpca.2c05005}
  {\bibfield  {journal} {\bibinfo  {journal} {J. Phys. Chem. A}\ }\textbf
  {\bibinfo {volume} {126}},\ \bibinfo {pages} {6032--6046} (\bibinfo {year}
  {2022})}\BibitemShut {NoStop}%
\bibitem [{\citenamefont {Bruneval}\ and\ \citenamefont
  {Marques}(2013)}]{Bruneval_2013}%
  \BibitemOpen
  \bibfield  {author} {\bibinfo {author} {\bibfnamefont {F.}~\bibnamefont
  {Bruneval}}\ and\ \bibinfo {author} {\bibfnamefont {M.~A.~L.}\ \bibnamefont
  {Marques}},\ }\bibfield  {title} {\enquote {\bibinfo {title} {{Benchmarking
  the {{Starting Points}} of the {{$GW$ Approximation}} for {{Molecules}}}},}\
  }\href {https://doi.org/10.1021/ct300835h} {\bibfield  {journal} {\bibinfo
  {journal} {J. Chem. Theory Comput.}\ }\textbf {\bibinfo {volume} {9}},\
  \bibinfo {pages} {324--329} (\bibinfo {year} {2013})}\BibitemShut {NoStop}%
\bibitem [{\citenamefont {Suhai}(1983)}]{Suhai_1983}%
  \BibitemOpen
  \bibfield  {author} {\bibinfo {author} {\bibfnamefont {S.}~\bibnamefont
  {Suhai}},\ }\bibfield  {title} {\enquote {\bibinfo {title} {Quasiparticle
  energy-band structures in semiconducting polymers: {{Correlation}} effects on
  the band gap in polyacetylene},}\ }\href
  {https://doi.org/10.1103/PhysRevB.27.3506} {\bibfield  {journal} {\bibinfo
  {journal} {Phys. Rev. B}\ }\textbf {\bibinfo {volume} {27}},\ \bibinfo
  {pages} {3506--3518} (\bibinfo {year} {1983})}\BibitemShut {NoStop}%
\bibitem [{\citenamefont {Holleboom}\ and\ \citenamefont
  {Snijders}(1990)}]{Holleboom_1990}%
  \BibitemOpen
  \bibfield  {author} {\bibinfo {author} {\bibfnamefont {L.~J.}\ \bibnamefont
  {Holleboom}}\ and\ \bibinfo {author} {\bibfnamefont {J.~G.}\ \bibnamefont
  {Snijders}},\ }\bibfield  {title} {\enquote {\bibinfo {title} {{A Comparison
  between the Moller-Plesset and {{Green}}'s Function Perturbative Approaches
  to the Calculation of the Correlation Energy in the Many-electron
  Problem}},}\ }\href {https://doi.org/10.1063/1.459578} {\bibfield  {journal}
  {\bibinfo  {journal} {J. Chem. Phys.}\ }\textbf {\bibinfo {volume} {93}},\
  \bibinfo {pages} {5826--5837} (\bibinfo {year} {1990})}\BibitemShut {NoStop}%
\bibitem [{\citenamefont {Casida}\ and\ \citenamefont
  {Chong}(1989)}]{Casida_1989}%
  \BibitemOpen
  \bibfield  {author} {\bibinfo {author} {\bibfnamefont {M.~E.}\ \bibnamefont
  {Casida}}\ and\ \bibinfo {author} {\bibfnamefont {D.~P.}\ \bibnamefont
  {Chong}},\ }\bibfield  {title} {\enquote {\bibinfo {title} {Physical
  interpretation and assessment of the {{Coulomb}}-hole and screened-exchange
  approximation for molecules},}\ }\href
  {https://doi.org/10.1103/PhysRevA.40.4837} {\bibfield  {journal} {\bibinfo
  {journal} {Phys. Rev. A}\ }\textbf {\bibinfo {volume} {40}},\ \bibinfo
  {pages} {4837--4848} (\bibinfo {year} {1989})}\BibitemShut {NoStop}%
\bibitem [{\citenamefont {Casida}\ and\ \citenamefont
  {Chong}(1991)}]{Casida_1991}%
  \BibitemOpen
  \bibfield  {author} {\bibinfo {author} {\bibfnamefont {M.~E.}\ \bibnamefont
  {Casida}}\ and\ \bibinfo {author} {\bibfnamefont {D.~P.}\ \bibnamefont
  {Chong}},\ }\bibfield  {title} {\enquote {\bibinfo {title} {Simplified
  {{Green}}-function approximations: {{Further}} assessment of a polarization
  model for second-order calculation of outer-valence ionization potentials in
  molecules},}\ }\href {https://doi.org/10.1103/PhysRevA.44.5773} {\bibfield
  {journal} {\bibinfo  {journal} {Phys. Rev. A}\ }\textbf {\bibinfo {volume}
  {44}},\ \bibinfo {pages} {5773--5783} (\bibinfo {year} {1991})}\BibitemShut
  {NoStop}%
\bibitem [{\citenamefont {Stefanucci}\ and\ \citenamefont {van
  Leeuwen}(2013)}]{Stefanucci_2013}%
  \BibitemOpen
  \bibfield  {author} {\bibinfo {author} {\bibfnamefont {G.}~\bibnamefont
  {Stefanucci}}\ and\ \bibinfo {author} {\bibfnamefont {R.}~\bibnamefont {van
  Leeuwen}},\ }\href@noop {} {\emph {\bibinfo {title} {Nonequilibrium Many-Body
  Theory of Quantum Systems: A Modern Introduction}}}\ (\bibinfo  {publisher}
  {{Cambridge University Press}},\ \bibinfo {address} {Cambridge},\ \bibinfo
  {year} {2013})\BibitemShut {NoStop}%
\bibitem [{\citenamefont {Ortiz}(2013)}]{Ortiz_2013}%
  \BibitemOpen
  \bibfield  {author} {\bibinfo {author} {\bibfnamefont {J.~V.}\ \bibnamefont
  {Ortiz}},\ }\bibfield  {title} {\enquote {\bibinfo {title} {Electron
  propagator theory: An approach to prediction and interpretation in quantum
  chemistry: {{Electron}} propagator theory},}\ }\href
  {https://doi.org/10.1002/wcms.1116} {\bibfield  {journal} {\bibinfo
  {journal} {WIREs Comput. Mol. Sci.}\ }\textbf {\bibinfo {volume} {3}},\
  \bibinfo {pages} {123--142} (\bibinfo {year} {2013})}\BibitemShut {NoStop}%
\bibitem [{\citenamefont {Phillips}\ and\ \citenamefont
  {Zgid}(2014)}]{Phillips_2014}%
  \BibitemOpen
  \bibfield  {author} {\bibinfo {author} {\bibfnamefont {J.~J.}\ \bibnamefont
  {Phillips}}\ and\ \bibinfo {author} {\bibfnamefont {D.}~\bibnamefont
  {Zgid}},\ }\bibfield  {title} {\enquote {\bibinfo {title} {{Communication:
  The Description of Strong Correlation within Self-Consistent Green's Function
  Second-Order Perturbation Theory}},}\ }\href
  {https://doi.org/10.1063/1.4884951} {\bibfield  {journal} {\bibinfo
  {journal} {J. Chem. Phys.}\ }\textbf {\bibinfo {volume} {140}},\ \bibinfo
  {pages} {241101} (\bibinfo {year} {2014})}\BibitemShut {NoStop}%
\bibitem [{\citenamefont {Phillips}, \citenamefont {Kananenka},\ and\
  \citenamefont {Zgid}(2015)}]{Phillips_2015}%
  \BibitemOpen
  \bibfield  {author} {\bibinfo {author} {\bibfnamefont {J.~J.}\ \bibnamefont
  {Phillips}}, \bibinfo {author} {\bibfnamefont {A.~A.}\ \bibnamefont
  {Kananenka}},\ and\ \bibinfo {author} {\bibfnamefont {D.}~\bibnamefont
  {Zgid}},\ }\bibfield  {title} {\enquote {\bibinfo {title} {{Fractional Charge
  and Spin Errors in Self-Consistent Green's Function Theory}},}\ }\href
  {https://doi.org/10.1063/1.4921259} {\bibfield  {journal} {\bibinfo
  {journal} {J. Chem. Phys.}\ }\textbf {\bibinfo {volume} {142}},\ \bibinfo
  {pages} {194108} (\bibinfo {year} {2015})}\BibitemShut {NoStop}%
\bibitem [{\citenamefont {Rusakov}, \citenamefont {Phillips},\ and\
  \citenamefont {Zgid}(2014)}]{Rusakov_2014}%
  \BibitemOpen
  \bibfield  {author} {\bibinfo {author} {\bibfnamefont {A.~A.}\ \bibnamefont
  {Rusakov}}, \bibinfo {author} {\bibfnamefont {J.~J.}\ \bibnamefont
  {Phillips}},\ and\ \bibinfo {author} {\bibfnamefont {D.}~\bibnamefont
  {Zgid}},\ }\bibfield  {title} {\enquote {\bibinfo {title} {{Local
  {{Hamiltonians}} for Quantitative {{Green}}'s Function Embedding Methods}},}\
  }\href {https://doi.org/10.1063/1.4901432} {\bibfield  {journal} {\bibinfo
  {journal} {J. Chem. Phys.}\ }\textbf {\bibinfo {volume} {141}},\ \bibinfo
  {pages} {194105} (\bibinfo {year} {2014})}\BibitemShut {NoStop}%
\bibitem [{\citenamefont {Rusakov}\ and\ \citenamefont
  {Zgid}(2016)}]{Rusakov_2016}%
  \BibitemOpen
  \bibfield  {author} {\bibinfo {author} {\bibfnamefont {A.~A.}\ \bibnamefont
  {Rusakov}}\ and\ \bibinfo {author} {\bibfnamefont {D.}~\bibnamefont {Zgid}},\
  }\bibfield  {title} {\enquote {\bibinfo {title} {{Self-Consistent
  Second-Order Green's Function Perturbation Theory for Periodic Systems}},}\
  }\href {https://doi.org/10.1063/1.4940900} {\bibfield  {journal} {\bibinfo
  {journal} {J. Chem. Phys.}\ }\textbf {\bibinfo {volume} {144}},\ \bibinfo
  {pages} {054106} (\bibinfo {year} {2016})}\BibitemShut {NoStop}%
\bibitem [{\citenamefont {Hirata}\ \emph {et~al.}(2015)\citenamefont {Hirata},
  \citenamefont {Hermes}, \citenamefont {Simons},\ and\ \citenamefont
  {Ortiz}}]{Hirata_2015}%
  \BibitemOpen
  \bibfield  {author} {\bibinfo {author} {\bibfnamefont {S.}~\bibnamefont
  {Hirata}}, \bibinfo {author} {\bibfnamefont {M.~R.}\ \bibnamefont {Hermes}},
  \bibinfo {author} {\bibfnamefont {J.}~\bibnamefont {Simons}},\ and\ \bibinfo
  {author} {\bibfnamefont {J.~V.}\ \bibnamefont {Ortiz}},\ }\bibfield  {title}
  {\enquote {\bibinfo {title} {General-{{Order Many}}-{{Body Green}}'s
  {{Function Method}}},}\ }\href {https://doi.org/10.1021/acs.jctc.5b00005}
  {\bibfield  {journal} {\bibinfo  {journal} {J. Chem. Theory Comput.}\
  }\textbf {\bibinfo {volume} {11}},\ \bibinfo {pages} {1595--1606} (\bibinfo
  {year} {2015})}\BibitemShut {NoStop}%
\bibitem [{\citenamefont {Hirata}\ \emph {et~al.}(2017)\citenamefont {Hirata},
  \citenamefont {Doran}, \citenamefont {Knowles},\ and\ \citenamefont
  {Ortiz}}]{Hirata_2017}%
  \BibitemOpen
  \bibfield  {author} {\bibinfo {author} {\bibfnamefont {S.}~\bibnamefont
  {Hirata}}, \bibinfo {author} {\bibfnamefont {A.~E.}\ \bibnamefont {Doran}},
  \bibinfo {author} {\bibfnamefont {P.~J.}\ \bibnamefont {Knowles}},\ and\
  \bibinfo {author} {\bibfnamefont {J.~V.}\ \bibnamefont {Ortiz}},\ }\bibfield
  {title} {\enquote {\bibinfo {title} {One-particle many-body {{Green}}'s
  function theory: {{Algebraic}} recursive definitions, linked-diagram theorem,
  irreducible-diagram theorem, and general-order algorithms},}\ }\href
  {https://doi.org/10.1063/1.4994837} {\bibfield  {journal} {\bibinfo
  {journal} {J. Chem. Phys.}\ }\textbf {\bibinfo {volume} {147}},\ \bibinfo
  {pages} {044108} (\bibinfo {year} {2017})}\BibitemShut {NoStop}%
\bibitem [{\citenamefont {Backhouse}, \citenamefont {Santana-Bonilla},\ and\
  \citenamefont {Booth}(2021)}]{Backhouse_2021}%
  \BibitemOpen
  \bibfield  {author} {\bibinfo {author} {\bibfnamefont {O.~J.}\ \bibnamefont
  {Backhouse}}, \bibinfo {author} {\bibfnamefont {A.}~\bibnamefont
  {Santana-Bonilla}},\ and\ \bibinfo {author} {\bibfnamefont {G.~H.}\
  \bibnamefont {Booth}},\ }\bibfield  {title} {\enquote {\bibinfo {title}
  {Scalable and predictive spectra of correlated molecules with moment
  truncated iterated perturbation theory},}\ }\href
  {https://doi.org/10.1021/acs.jpclett.1c02383} {\bibfield  {journal} {\bibinfo
   {journal} {J. Phys. Chem. Lett.}\ }\textbf {\bibinfo {volume} {12}},\
  \bibinfo {pages} {7650--7658} (\bibinfo {year} {2021})}\BibitemShut {NoStop}%
\bibitem [{\citenamefont {Backhouse}\ and\ \citenamefont
  {Booth}(2020)}]{Backhouse_2020b}%
  \BibitemOpen
  \bibfield  {author} {\bibinfo {author} {\bibfnamefont {O.~J.}\ \bibnamefont
  {Backhouse}}\ and\ \bibinfo {author} {\bibfnamefont {G.~H.}\ \bibnamefont
  {Booth}},\ }\bibfield  {title} {\enquote {\bibinfo {title} {Efficient
  excitations and spectra within a perturbative renormalization approach},}\
  }\href {https://doi.org/10.1021/acs.jctc.0c00701} {\bibfield  {journal}
  {\bibinfo  {journal} {J. Chem. Theory Comput.}\ }\textbf {\bibinfo {volume}
  {16}},\ \bibinfo {pages} {6294--6304} (\bibinfo {year} {2020})}\BibitemShut
  {NoStop}%
\bibitem [{\citenamefont {Backhouse}, \citenamefont {Nusspickel},\ and\
  \citenamefont {Booth}(2020)}]{Backhouse_2020a}%
  \BibitemOpen
  \bibfield  {author} {\bibinfo {author} {\bibfnamefont {O.~J.}\ \bibnamefont
  {Backhouse}}, \bibinfo {author} {\bibfnamefont {M.}~\bibnamefont
  {Nusspickel}},\ and\ \bibinfo {author} {\bibfnamefont {G.~H.}\ \bibnamefont
  {Booth}},\ }\bibfield  {title} {\enquote {\bibinfo {title} {Wave function
  perspective and efficient truncation of renormalized second-order
  perturbation theory},}\ }\href {https://doi.org/10.1021/acs.jctc.9b01182}
  {\bibfield  {journal} {\bibinfo  {journal} {J. Chem. Theory Comput.}\
  }\textbf {\bibinfo {volume} {16}},\ \bibinfo {pages} {1090--1104} (\bibinfo
  {year} {2020})}\BibitemShut {NoStop}%
\bibitem [{\citenamefont {Pokhilko}\ and\ \citenamefont
  {Zgid}(2021)}]{Pokhilko_2021a}%
  \BibitemOpen
  \bibfield  {author} {\bibinfo {author} {\bibfnamefont {P.}~\bibnamefont
  {Pokhilko}}\ and\ \bibinfo {author} {\bibfnamefont {D.}~\bibnamefont
  {Zgid}},\ }\bibfield  {title} {\enquote {\bibinfo {title} {{Interpretation of
  multiple solutions in fully iterative GF2 and GW schemes using local analysis
  of two-particle density matrices}},}\ }\href
  {https://doi.org/10.1063/5.0055191} {\bibfield  {journal} {\bibinfo
  {journal} {J. Chem. Phys.}\ }\textbf {\bibinfo {volume} {155}},\ \bibinfo
  {pages} {024101} (\bibinfo {year} {2021})}\BibitemShut {NoStop}%
\bibitem [{\citenamefont {Pokhilko}\ \emph {et~al.}(2021)\citenamefont
  {Pokhilko}, \citenamefont {Iskakov}, \citenamefont {Yeh},\ and\ \citenamefont
  {Zgid}}]{Pokhilko_2021b}%
  \BibitemOpen
  \bibfield  {author} {\bibinfo {author} {\bibfnamefont {P.}~\bibnamefont
  {Pokhilko}}, \bibinfo {author} {\bibfnamefont {S.}~\bibnamefont {Iskakov}},
  \bibinfo {author} {\bibfnamefont {C.-N.}\ \bibnamefont {Yeh}},\ and\ \bibinfo
  {author} {\bibfnamefont {D.}~\bibnamefont {Zgid}},\ }\bibfield  {title}
  {\enquote {\bibinfo {title} {{Evaluation of two-particle properties within
  finite-temperature self-consistent one-particle Green's function methods:
  Theory and application to $GW$ and GF2}},}\ }\href
  {https://doi.org/10.1063/5.0054661} {\bibfield  {journal} {\bibinfo
  {journal} {J. Chem. Phys.}\ }\textbf {\bibinfo {volume} {155}},\ \bibinfo
  {pages} {024119} (\bibinfo {year} {2021})}\BibitemShut {NoStop}%
\bibitem [{\citenamefont {Pokhilko}, \citenamefont {Yeh},\ and\ \citenamefont
  {Zgid}(2022)}]{Pokhilko_2022}%
  \BibitemOpen
  \bibfield  {author} {\bibinfo {author} {\bibfnamefont {P.}~\bibnamefont
  {Pokhilko}}, \bibinfo {author} {\bibfnamefont {C.-N.}\ \bibnamefont {Yeh}},\
  and\ \bibinfo {author} {\bibfnamefont {D.}~\bibnamefont {Zgid}},\ }\bibfield
  {title} {\enquote {\bibinfo {title} {{Iterative subspace algorithms for
  finite-temperature solution of Dyson equation}},}\ }\href
  {https://doi.org/10.1063/5.0082586} {\bibfield  {journal} {\bibinfo
  {journal} {J. Chem. Phys.}\ }\textbf {\bibinfo {volume} {156}},\ \bibinfo
  {pages} {094101} (\bibinfo {year} {2022})}\BibitemShut {NoStop}%
\bibitem [{\citenamefont {Loos}, \citenamefont {Marie},\ and\ \citenamefont
  {Ammar}(2024)}]{Loos_2024}%
  \BibitemOpen
  \bibfield  {author} {\bibinfo {author} {\bibfnamefont {P.-F.}\ \bibnamefont
  {Loos}}, \bibinfo {author} {\bibfnamefont {A.}~\bibnamefont {Marie}},\ and\
  \bibinfo {author} {\bibfnamefont {A.}~\bibnamefont {Ammar}},\ }\bibfield
  {title} {\enquote {\bibinfo {title} {{Cumulant Green's function methods for
  molecules}},}\ }\href {https://doi.org/10.1039/D4FD00037D} {\bibfield
  {journal} {\bibinfo  {journal} {Faraday Discuss.}\ }\textbf {\bibinfo
  {volume} {254}},\ \bibinfo {pages} {240--260} (\bibinfo {year}
  {2024})}\BibitemShut {NoStop}%
\end{thebibliography}%
%%%%%%%%%%%%%%%%%%%%%%%%

\end{document}